\documentclass[a4paper,fleqn]{cas-sc}
\usepackage{graphicx}
\usepackage{epstopdf}
\usepackage[numbers]{natbib}

\def\tsc#1{\csdef{#1}{\textsc{\lowercase{#1}}\xspace}}
\tsc{WGM}
\tsc{QE}
\tsc{EP}
\tsc{PMS}
\tsc{BEC}
\tsc{DE}

\begin{document}
\let\WriteBookmarks\relax
\def\floatpagepagefraction{1}
\def\textpagefraction{.001}
\shorttitle{Effects of chain resolution on the configurational and rheological predictions from Brownian dynamics simulations of an isolated polymer chain in flow}
\shortauthors{Praphul Kumar et~al.}

\title [mode = title]{Effects of chain resolution on the configurational and rheological predictions from Brownian dynamics simulations of an isolated polymer chain in flow}                      

\author{Praphul Kumar}[type=,
                        auid=000,bioid=,
                        prefix=,
                        role=, orcid=]

\author{Indranil Saha Dalal}[type=,
                        auid=000,bioid=,
                        prefix=,
                        role=, orcid=0000-0002-3136-4697]
\cormark[1]
\ead{indrasd@iitk.ac.in}

\address{Department of Chemical Engineering, Indian Institute of Technology
Kanpur, Kanpur-208016, India}

\cortext[cor1]{Corresponding author}

\begin{abstract}
A reasonably accurate representation of a polymer chain is provided by beads connected with rods, or stiff, inextensible springs that mimic a single Kuhn step. Due to high computational cost, coarse-grained bead-spring models are used in typical applications, where each spring is supposed to replace several Kuhn steps. Earlier investigations indicate that the BD simulation predictions of the steady state in different flows, with these different levels of discretization, are largely qualitatively similar. However, subtle quantitative differences exist even for the steady states. In this study, we perform a detailed analysis of the behavioral differences arising out of the varying degrees of chain discretization, ranging from one to several hundred Kuhn steps. We compare the transient and steady behavior of both configuration and rheological properties for a single chain in uniaxial extension and steady shear flow. Our analysis highlights differences, particularly in the stress and viscosity values, obtained at intermediate and high flow rates, between the bead-rod and bead-spring models. Such a thorough understanding helps to provide an estimate of the best possible bead-spring representation for an underlying polymer chain in a given application. Additionally, we also investigate the limit of break-down of the spring laws i.e. the minimum number of Kuhn steps that a spring can mimic faithfully.

\end{abstract}




\ExplSyntaxOn
\keys_set:nn { stm / mktitle } { nologo }
\ExplSyntaxOff
\maketitle

   \section{Introduction}

Over the past few decades, Brownian dynamics (BD) simulations have emerged as an efficient tool to investigate various aspects of the dynamics of polymer solutions. In the dilute regime, the interactions between different chains can be neglected and a single chain may be considered for all investigations. Before BD simulations, the analytical foundations were set by the seminal work by Rouse \cite{rouse1953theory}, who proposed the decomposition of the chain dynamics into normal modes. This is further extended by Zimm \cite{zimm1956dynamics}, who introduced the effects of hydrodynamic interactions, which is highly relevant for dilute solutions. However, the Rouse and Zimm models considered beads connected by Hookean springs and can’t be applied, without any further adjustments, to understand the dynamics of chains in flow rates that are sufficiently strong to significantly deform the chain.

However, in BD simulations, finitely extensible chain models can be considered, which makes it efficient to study the behavior even in strong flows. The most accurate representation of a flexible polymer chain is a freely jointed chain, that consists of a series of beads connected by freely jointed links. These links can usually be rigid rods, but recent simulations have also used stiff Fraenkel \cite{dalal2012multiple,saha2012tumbling,dalal2014effects} and the FENE-Fraenkel spring \cite{hsieh2006brownian}. In all such representations, the chain is discretized to the level of a single Kuhn step. Thus, BD simulations of the dynamics of even a relatively short chain, say 100 or more Kuhn steps, becomes computationally prohibitive, especially when mechanisms like excluded volume and hydrodynamic interactions are active \cite{dalal2014effects}.

For computational efficiency, further course-grained representations of a polymer chain are used in BD simulations. In these, beads are connected by springs, whose spring law is adjusted to mimic the dynamics of several Kuhn steps. Examples include the Cohen Padé approximation  \cite{cohen1991pade,larson2005rheology} and the more recent Underhill-Doyle spring law \cite{underhill2004coarse}. In principle, in such models, any number of Kuhn steps can be mimicked by a single spring, which provides a highly attractive option to reduce the simulation cost by keeping the number of springs to a minimum. In several earlier simulations, the number of beads indeed have been chosen arbitrarily to fit experimental results \cite{hsieh2003modeling,hsieh2004modeling,li2000brownian,jendrejack2002stochastic,larson1999brownian}. This arbitrariness is not present for the chain model by Hsieh et al. \cite{hsieh2006brownian}, where FENE-Fraenkel springs are used. However, this implied chain discretization is at the level of a single Kuhn step, which is computationally prohibitive, as discussed earlier. On the other hand, Ravi Prakash and coworkers, in a series of investigations, have proposed a successive fine graining scheme to overcome this problem \cite{prabhakar2004successive,sunthar2004prediction,sunthar2005parameter,PHAM20089}. In this method, the properties of the bead-rod chain are predicted by extrapolating the corresponding values obtained from bead-spring models, where the number of springs is successively increased, while still remaining much lower than the number of Kuhn steps. This scheme has shown promise for the prediction of rheological properties of a bead-rod chain, without actually performing simulations of the same, thereby circumventing the problem of high computational cost. However, despite these studies, the ability of coarser bead-spring chains, to provide accurate predictions of the corresponding chain resolved to a single Kuhn step, is still relatively uncertain. This is particularly true for the transient behavior at startup of flows, or relaxation processes starting from a stretched chain.  In the studies involving successive fine graining, mostly the prediction of steady state rheological properties is investigated \cite{prabhakar2004successive,PHAM20089}.  A direct comparison of the transient startup predictions with actual bead-rod simulations is still missing.However, we note that a few comparisons have been performed with experimental results in the presence of HI \cite{prabhakar2004successive}.   Even if the steady state agrees well, it is not clear if the entire transient behavior leading to the steady state can be predicted by such an extrapolation scheme. Also, quite interestingly, even a mismatch in the steady state has been reported at higher flow rates \cite{PHAM20089}. Additionally, note that direct comparisons have been carried out only for very short bead-rod chains, of the order of 10-20 Kuhn steps. This owes to the prohibitive computational cost when the chain is resolved to a single Kuhn step.

Due to massive progress in computing power, the last decade has observed quite a few simulations of highly resolved chains (i.e. to the level of a single Kuhn step) \cite{dalal2012multiple,saha2012tumbling,dalal2014effects,soh2018untying}. These mostly mimic a Kuhn step by a stiff, almost inextensible, Fraenkel spring. As such, some more direct comparisons of bead-rod (where the “rod” refers to a nearly inextensible spring) and different bead-spring representations are now available. However, these are mostly for steady states of chain configuration in shear flows and the corresponding end-over-end tumbling dynamics \cite{dalal2012multiple,saha2012tumbling}. As observed earlier for viscosity at high shear rates, differences also exist between the predictions of the tumbling period and average chain thickness from bead-rod and bead-spring representations. These results cast further doubts on the ability of bead-spring models, even with reasonably large number of springs, to mimic the behavior of chains discretized to a single Kuhn step.

In this study, we perform a detailed analysis between BD simulations of the bead-rod and bead-spring representations under imposed flow fields. The “rod” in this manuscript refers to stiff Fraenkel springs, which represent nearly inextensible units.  In the manuscript, the same has also been referred to by “bead-KS” model in various places, to note that it is resolved to a single Kuhn step.  We compare both the transient behavior (startup of flow) and steady state under uniaxial extension and steady shear flow fields. For our investigations, we consider a reasonably large chain of 500 rods ( or Kuhn steps) and various bead-spring representations of the same. However, we stay  well above the limit of very short springs, mimicking the dynamics of few rods, where the spring laws might break. This is much larger than all the earlier comparisons with bead-rod chains considered in the earlier studies, including those towards the successive fine graining scheme. We compare the dynamics in terms of both the evolution of chain configurations and rheological properties. Our simulations clearly indicate that, the bead-spring models, in most cases, match progressively better as the number of springs are increased. However, important quantitative differences exist in terms of the rheological properties between the two representations, even though qualitative trends are similar. The steady state values of the first normal stress difference obtained from bead-rod model in uniaxial extension  do not  agree with that from bead-spring models. Similar mismatch exists for the steady state shear stress and shear viscosity in shear flow.  We also observe that, when the SFG procedure is used, the extrapolated steady state values mostly agree well with the bead-KS model. However, significant differences exist between the overall transient startup behavior between the SFG predictions and bead-KS results.  Thus, not only our study provides a comprehensive analysis of the differences that exist between these representations (and thereby showing the information that is lost due to coarse graining), but our results also provide a foundation to select a particular bead-spring representation to mimic the dynamics of a given underlying bead-rod chain. We hope that this detailed investigation provides researchers with sufficient information to select an appropriate number of springs to represent a chain for a given application, by trading off between accuracy and computational cost. Such applications include, but is not limited to, recent attempts at flow simulations that combine computational fluid mechanics with the dynamics of actual polymer chains, modelled by a few springs \cite{abedijaberi2012sedimentation,koppol2009anomalous}. Additionally, we investigate the limit at which the spring laws break down, thereby providing inaccurate results at equilibrium. Although these laws have been used extensively earlier, this limit, in terms of the minimum number of Kuhn steps that the spring law can mimic faithfully, has not been reported earlier, to the best of our knowledge.  

Our BD simulations in this manuscript are without additional mechanisms of excluded volume (EV) and hydrodynamic interactions (HI). Since the main focus of this study is to understand the differences between bead-KS and bead-spring representations in details, for a reasonably long chain, thereby ascertaining the accuracy of such coarse-graining procedures, we have not considered EV and HI for this article. The addition of these mechanisms requires further details to be incorporated in BD simulations, thereby increasing the complexity of the analysis. Hence, these details for a similar problem will be considered in a follow-up study, that will further complete the analysis presented here.

This article is organised in the following manner. In section 2, we briefly discuss the Brownian dynamics simulation method for dilute polymer solutions under imposed flow field. The simulation results for the bead-KS as well as the coarse-grained bead-spring models for uniaxial extension and steady shear flow are discussed in section 3. Finally, the key findings are summarized in section 4.

\section{Simulation methodology}

In this study, the chain is modelled by $N + 1$ beads (labelled from $0$ to $N$) connected by $N$ springs (labelled from $1$ to $N$). The equation of motion for the $i^{th}$ bead in the polymer chain, in the absence of excluded volume and hydrodynamic interactions, is given as:

\begin{equation}\label{eq:1}
\frac{d{\vec{r}}_i^\ast}{dt^\ast}-\overleftrightarrow{\kappa}^\ast\cdot{\vec{r}}_i^\ast-\left[{\vec{F}}_{i+1}^{s\ast}-{\vec{F}}_i^{s\ast}\right]-\left(\frac{6}{\Delta t^\ast}\right)^{{\frac{1}{2}}}{\vec{n}}_i=0    
\end{equation}

where ‘$\ast$’ in the superscript denotes that the variables are made dimensionless. Here, ${\vec{r}}_i^\ast$ is the normalized position vector of the $i^{th}$ bead, ${\vec{F}}_i^{s\ast}$ is the normalized force due to the $i^{th}$ spring and $\Delta t^\ast$ is the dimensionless time step size. The first and second terms, in Eq. \ref{eq:1}, together, represents the total drag force acting on the $i^{th}$ bead. The third term accounts for the total spring force acting on the $i^{th}$ bead. The last term constitutes the Brownian force, in dimensionless form.  All of these are summarized in the earlier studies using BD simulations \cite{dalal2012multiple}.  Here, the variables are made dimensionless using a length scale of $b_k$ (length of one Kuhn step), a time scale of $\zeta b_k^2/k_BT$ (where $\zeta$ is the drag coefficient) and using a scale of $k_BT/b_k$ (where $k_B$ is the Boltzmann’s constant and $T$ is the temperature) for the force. $\overleftrightarrow{\kappa}^\ast$ is the dimensionless transpose of the velocity gradient tensor and ${\vec{n}}_i$ is random vector generated uniformly in the interval $[-1, 1]$ \cite{dalal2012multiple,larson2005rheology}.

In this article, to represent the same chain (having the same total number of Kuhn steps) in different levels of resolution, we use the “fine-grained” bead-rod and the “coarse-grained” bead-spring models, similar to the earlier study \cite{dalal2012multiple}. The bead-rod model discretizes the chain at the level of a single Kuhn step, while the bead-spring models provide coarser representations of the same, where several Kuhn steps are mimicked by a single spring. For the bead-rod model, following earlier work \cite{dalal2012multiple}, each nearly inextensible ‘rod’ is modelled by a stiff Fraenkel spring with the spring law given by:

\begin{equation}
    {\vec{F}}_i^{s\ast}=K^\ast\left(\left|{\vec{Q}}_i^\ast\right|-1\right)\frac{{\vec{Q}}_i^\ast}{\left|{\vec{Q}}_i^\ast\right|}
\end{equation}

where $K^\ast=Kb_k^2/k_BT$ is the dimensionless spring constant that adjusts the stiffness of the Fraenkel spring. The value of $K^\ast$ is fixed at $10^4$ even in the absence of flow, to prevent the overstretching of the springs. Here, ${\vec{Q}}_i^\ast={\vec{r}}_i^\ast-{\vec{r}}_{i-1}^\ast$ is the $i^{th}$ dimensionless connector vector and $\left|{\vec{Q}}_i^\ast\right|$ denotes its magnitude.  As mentioned earlier, since we have actually used a stiff Fraenkel spring to mimic a Kuhn step and not a rigid rod, henceforth the terms “bead-rod” and “bead-KS” model will be used interchangeably. Here, “KS” denotes a Kuhn step. However, note that the stiff Fraenkel spring has been observed to be a good approximation to rigid rods for such BD simulations\cite{praphul2022}.  For the bead-spring model, we consider two spring laws – the Cohen-Padé \cite{cohen1991pade} approximation to the inverse Langevin function \cite{bird1987dynamics,larson2013constitutive} and the spring potential developed more recently by Underhill and Doyle \cite{underhill2004coarse}. The magnitude of spring force for these are given by the following equation:

\begin{equation}
\left|{\vec{F}}_i^{s\ast}\right|=\frac{\left|{\vec{F}}_i^s\right|b_k}{k_BT}=\frac{\alpha\hat{r}-\beta{\hat{r}}^3}{1-{\hat{r}}^2} 
\end{equation}

where $\hat{r}$ is the fractional extension of the spring (i.e. spring length scaled by $\nu b_k$, where $\nu$ denotes the number of Kuhn steps that each spring is supposed to mimic), and $\alpha$, $\beta$ are the parameters that depend on the spring law. For the Cohen-Padé approximation, the values of $\alpha$ and $\beta$ are 3 and 1, respectively. However, for the Underhill and Doyle spring law, the parameters depend on the level of discretization   and are given as:

\begin{eqnarray}
    \alpha=3-\frac{10}{3\nu}+\frac{10}{27\nu^2} \\
    \beta=1+\frac{2}{3\nu}+\frac{10}{27\nu^2}  
\end{eqnarray}

In this work, we consider two rheologically important flow fields - uniaxial extension and steady shear flow. The value of $\overleftrightarrow{\kappa}^\ast$ for uniaxial extension is given as:

\begin{equation}
\overleftrightarrow{\kappa}^\ast=\left(\begin{matrix}\dot{\epsilon}&0&0\\0&-\dot{\epsilon}/2&0\\0&0&-\dot{\epsilon}/2\\\end{matrix}\right)
\end{equation}

where $\dot{\epsilon}$  represents the dimensionless extensional rate. The same for steady shear is given as
\begin{equation}
\overleftrightarrow{\kappa}^\ast=\left(\begin{matrix}0&\dot{\gamma}&0\\0&0&0\\0&0&0\\\end{matrix}\right)    
\end{equation}
where $\dot{\gamma}$ is the dimensionless shear rate.

   \section{Results and discussion }   
   
The size of the polymer chain is measured in terms of the root mean square end-to-end distance, $R_{end-end}$ or the radius of gyration $R_{g}$.  The end-to-end distance of a polymer chain is defined as 

\begin{equation}
  R_{end-end}  =\left\langle \left( {\vec{r}}_N^\ast-{\vec{r}}_0^\ast\right) ^{2}\right\rangle ^{1/2}  
\end{equation}

where ${\vec{r}}_N^\ast-{\vec{r}}_0^\ast$ denotes the end-to-end vector of the chain and $\left\langle ... \right\rangle$ represents an ensemble average. Correspondingly, the components of the same in different directions can be identically defined. The radius of gyration of a polymer chain is given by the following expression \cite{larson2005rheology}

\begin{equation}
R_{g}  = \sqrt{ \sum_{i=0}^{N}  {\frac{\left\langle \left( {\vec{r}}_i^\ast-{\vec{r}}_{cm}\right) ^{2}\right\rangle}{N+1}}}
\end{equation}

where  ${\vec{r}}_i^\ast$ and ${\vec{r}}_{cm}$ are the position vector of the $i^{th}$ bead and the centre of mass of the chain, respectively. The centre of the mass of the chain is defined by
\begin{equation}
{\vec{r}}_{cm}  = \frac{1}{N+1}\sum_{i=0}^{N} {\vec{r}}_i^\ast
\end{equation}

The component of the radius of gyration in the x-direction can be written as

\begin{equation}
R_{gx}  = \sqrt{ \sum_{i=0}^{i=N}  {\frac{\left\langle \left( x_i^\ast-x_{cm}\right) ^{2}\right\rangle}{N+1}}        }
\end{equation}

where $x_{cm}$ and $x_i^\ast$ are the centre of mass and the coordinate of the $i^{th}$ bead along the x-axis, respectively. Similar expressions can be written for y and z components of the radius of gyration.

When a flow field is imposed on a polymer solution, the deformation in the chain depends on the strength and type of the applied field. The strength of the shear and extensional flow is characterized by the Weissenberg number (Wi) \cite{larson2005rheology}, which is defined as 
\begin{equation}
   \text{Wi}=\dot{\gamma}\tau 
\end{equation}
and
\begin{equation}
   \text{Wi}=\dot{\epsilon}\tau
\end{equation}

where $\tau$ is the longest relaxation time obtained from the autocorrelation function of the end-to-end vector of the chain. $\dot{\gamma}$ and $\dot{\epsilon}$ are the shear and extensional rates, respectively. The relaxation time is calculated by fitting an exponential decay to the last $70\%$ of the autocorrelation function of the end-to-end vector of a polymer chain at equilibrium [3].

The dimensionless polymeric contribution to the stress $\overleftrightarrow{\sigma}$, (normalixed by $n_pk_BT$, where $n_p$ is the number density of polymer chains), is calculated using the Kramer’s expression \cite{bird1987dynamics,larson2013constitutive}, which is given by

\begin{equation}
\overleftrightarrow{\sigma}= N\overleftrightarrow{\delta}+ \sum_{i=1}^{N} \left\langle {\vec{F}}_i^{s\ast}{\vec{Q}}_i^\ast \right\rangle                            
\end{equation}               

where $N\overleftrightarrow{\delta}$ is the isotropic contribution to the stress, ${\vec{F}}_i^{s\ast}$ is the spring force acting on the $i^{th}$ link and ${\vec{Q}}_i^\ast$ is the connector vector.

We also measure the relative shear viscosity (the ratio of shear viscosity $\eta$ and the zero-shear viscosity $\eta_0$). The shear viscosity, $\eta\left(\dot{\gamma}\right)=\sigma_{12}/\dot{\gamma\ }$, is the ratio of the shear stress $\sigma_{12}$ and the shear rate $\dot{\gamma}$.  The zero-shear viscosity $\eta_0$ is defined as the $\eta\left(\dot{\gamma}\right)$ measured in the limit of $\dot{\gamma}\rightarrow 0$.
The first normal stress difference $N_1$ is defined as
\begin{equation}
N_1=\sigma_{11}-\ \sigma_{22}
\end{equation}  

 Here, we also make a note on the figure legends used throughout in this manuscript. The bead-KS model is denoted by “BR”, followed by the number of Kuhn steps in the chain considered. The bead-spring representations are represented by “BS”, followed by two numbers. The first gives the number of springs used, while the second one is $\nu$.  

   \subsection{Extensional flow }

For all simulations in this study, we have used a chain of 500 Kuhn steps. We have performed simulations for different values of Weissenberg numbers for extensional flow to compare the transient behaviour obtained from bead-rod and bead-spring models. Our simulations encompass a wide range of chain discretization levels, ranging from one Kuhn step (bead-KS model) to one single spring (a dumbbell mimicking the entire chain of 500 Kuhn steps). In our bead-spring models, we have used both the Cohen-Padé approximation and the Underhill-Doyle model. We compare the temporal behaviour of both the chain size and stress. However, note that we have limited the usage of bead-spring models to $\nu\geq 10$ for the Cohen-Padé approximation and $\nu\geq 5$ for Underhill-Doyle spring laws, which is above the limit of very low values of $\nu$ where the spring law may be expected to fail \cite{dalal2012multiple}. Note, the value of $\nu$ below which the bead-spring model predictions become unreliable, for both the spring laws used here, has not been investigated earlier, to the best of our knowledge.


\begin{table}[width=.9\linewidth,cols=6,pos=h]
\caption{Square of $R_{end-end}$ of a polymer chain with 500 Kuhn steps at different discretization levels, obtained from BD simulations, in the absence of flow. The values are measured with the Cohe-Padé and Unerhill-Doyle spring laws, each with two different time step sizes $\Delta t^\ast$. }\label{tbl1}

\begin{tabular}{cccccc}
\hline
&     & \begin{tabular}[c]{@{}c@{}}Cohen-Padé\\ spring law\\ ($\Delta t^\ast\ =\ {10}^{-3}$)\end{tabular} & \begin{tabular}[c]{@{}c@{}}Underhill-Doyle\\ spring law\\ ($\Delta t^\ast\ =\ {10}^{-3}$)\end{tabular} & \begin{tabular}[c]{@{}c@{}}Cohen-Padé\\ spring law\\ ($\Delta t^\ast\ =\ {10}^{-4}$)\end{tabular} & \begin{tabular}[c]{@{}c@{}}Underhill-Doyle \\ spring law\\ ($\Delta t^\ast\ =\ {10}^{-4}$)\end{tabular} \\ \hline

$N$   & $\nu$   & $R_{end-end}^2$    & $R_{end-end}^2$   & $R_{end-end}^2$   & $R_{end-end}^2$  \\ \hline

1   & 500 & 499.7  & 499.9   & 499.9    & 499.9  \\
2   & 250 & 500.2  & 500.1   & 500.1    & 500.1  \\
5   & 100 & 499.6  & 499.8   & 499.8    & 499.8  \\
10  & 50  & 499.1  & 499.7   & 499.7    & 499.7  \\
20  & 25  & 499.2  & 500.3   & 500.1    & 499.1  \\
50  & 10  & 482.1  & 500.9   & 485.1    & 500.3  \\
100 & 5   & 383.2  & 502.2   & 393.2    & 502.5  \\
125 & 4   & 376.5  & 556.9   & 380.5    & 550.9  \\ \hline
\end{tabular}
\end{table}

Table \ref{tbl1} shows the square of $R_{end-end}$ of a bead-spring chain at varying levels of chain discretization, for two different time step sizes for integration. At equilibrium, the theoretical value of  $R_{end-end}^2$ is equal to the contour length of the chain (which is 500 Kuhn steps in this case). Clearly, for the Cohen-Padé approximation, the chain size at equilibrium shows considerable error for  $\nu\le 10$. The predictions are much better for the Underhill-Doyle spring law, where the  equilibrium chain size is predicted correctly for $\nu\geq 5$. Henceforth, throughout this study, we have not performed simulations with  $\nu < 10$ for the Cohen-Padé approximation and $\nu< 5 $ for the Underhill-Doyle spring law. Additionally, note that, for high flow rates, the simulations produce erroneous results for $\nu=5$ , even while using  very low  time step sizes. Hence, for some cases of high Weissenberg numbers, we limit ourselves with using $\nu\geq 10$ for the Unerhill-Doyle spring law. To the best of our knowledge, these limits have never been reported earlier.

\begin{figure}[hbt!]
\centering
	\includegraphics[width=\textwidth]{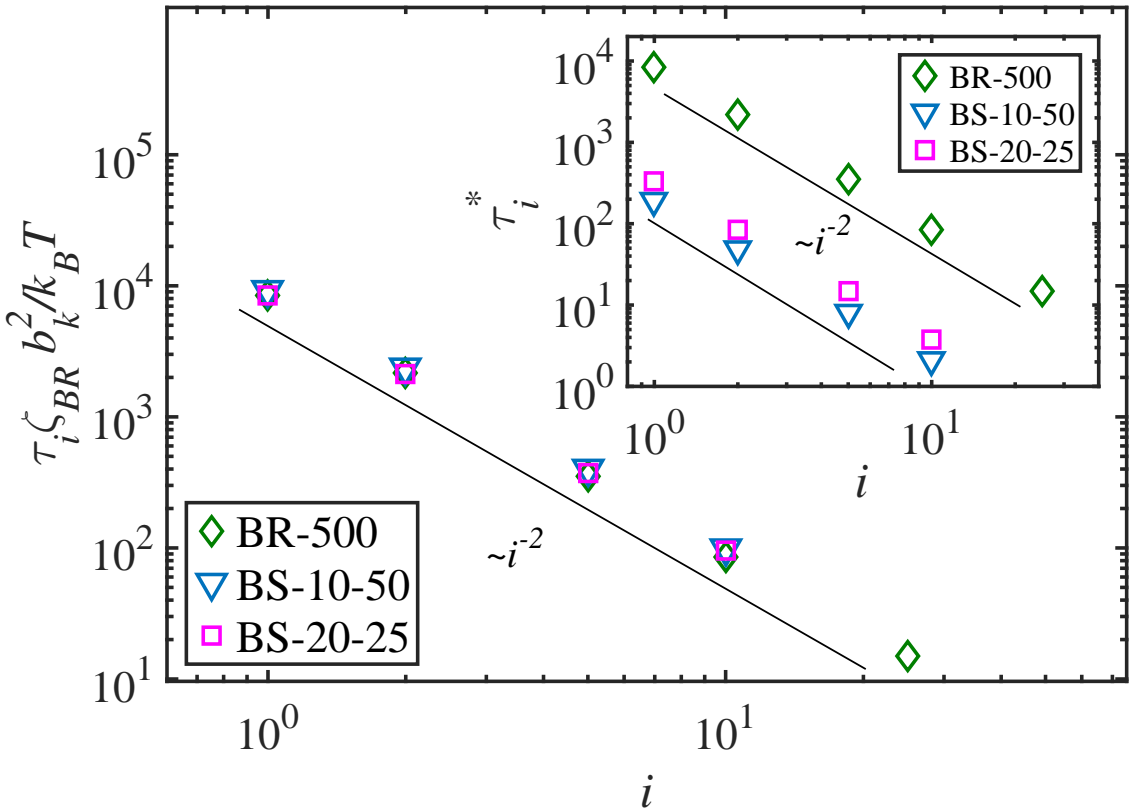}

	\caption {Variation of the unscaled relaxation time with number $i$ for a chain of 500 Kuhn steps in different representations. “BR-500” denotes the bead-KS (or bead-rod) model where 500 stiff Franekel springs are used. “BS-10-50” and “BS-20-25” denotes the bead-spring representations of the same, with 10 and 20 springs, respectively. The inset shows the same in the dimensionless representation.} 
	\label{fig:2a}
\end{figure}

\begin{figure}[hbt!]
\centering
	\includegraphics[width=0.7\textwidth]{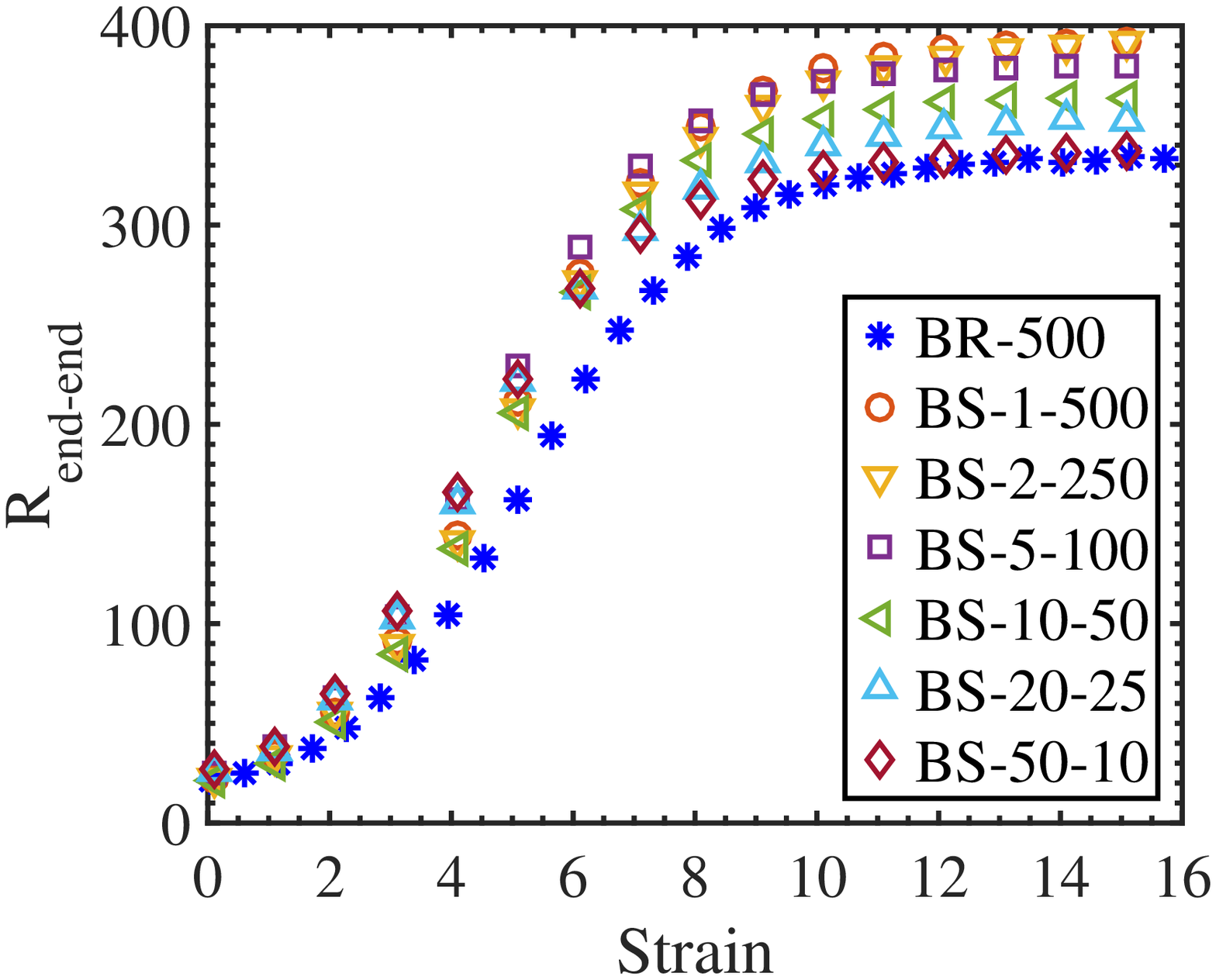}
	\put (-250,210){\LARGE($a$)}
	
	\includegraphics[width=0.7\textwidth]{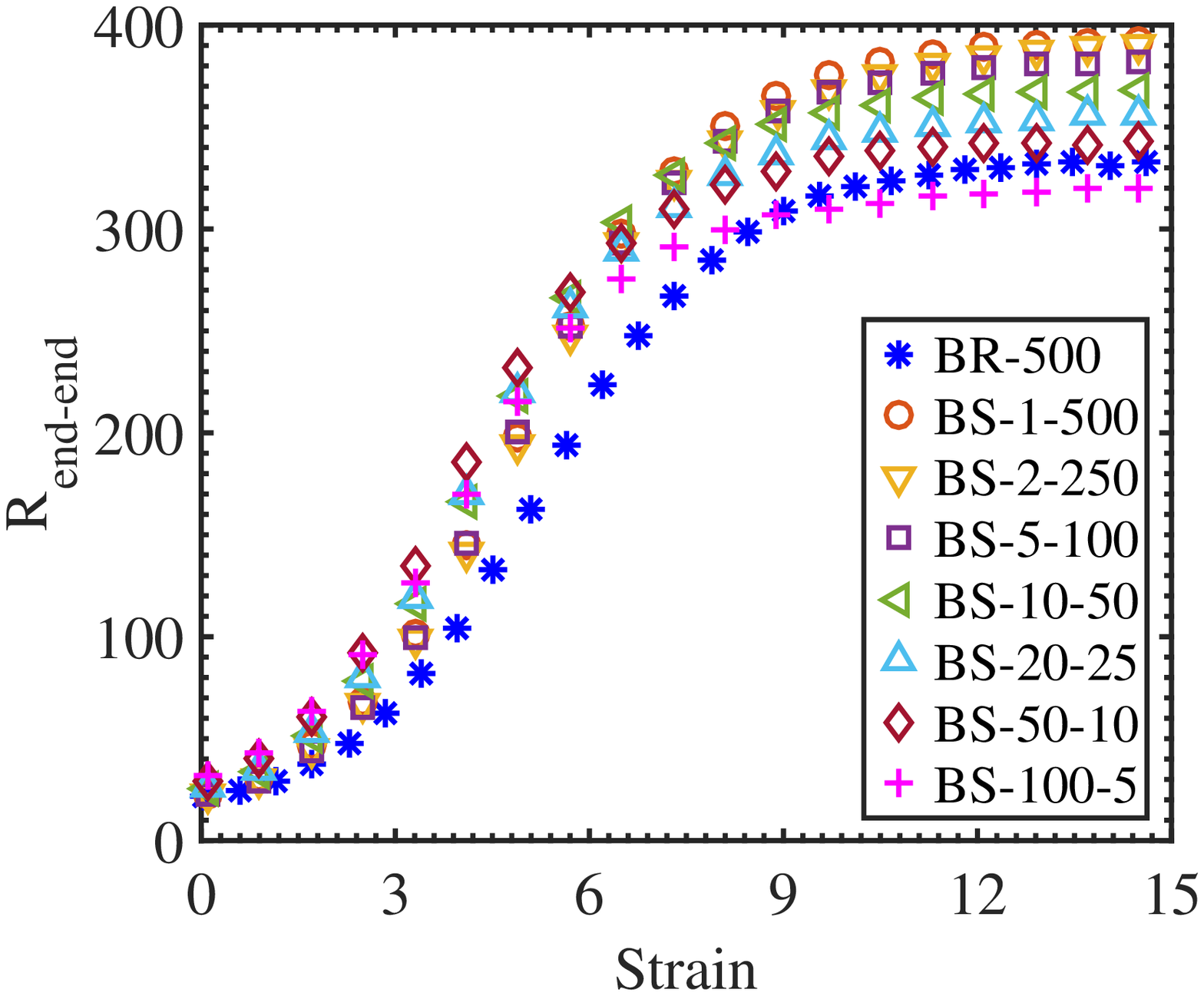}
	\put (-250,210){\LARGE($b$)}
	
	\caption {Variation of $R_{end-end}$ of a polymer chain with strain in uniaxial extensional flow for Wi = 2. BR-500 denotes a bead-rod chain of 500 Kuhn steps, whereas “BS” represents bead-spring chains, using spring laws of ($a$) Cohen-Padé approximation and ($b$) Underhill-Doyle spring law. The first number following “BS” denotes the number of springs, while the second gives the value of $\nu$. The results shown here are averaged over 300 cases, for each model.}
	\label{fig:1}
\end{figure}

\begin{figure}[hbt!]
\centering
	\includegraphics[width=0.7\textwidth]{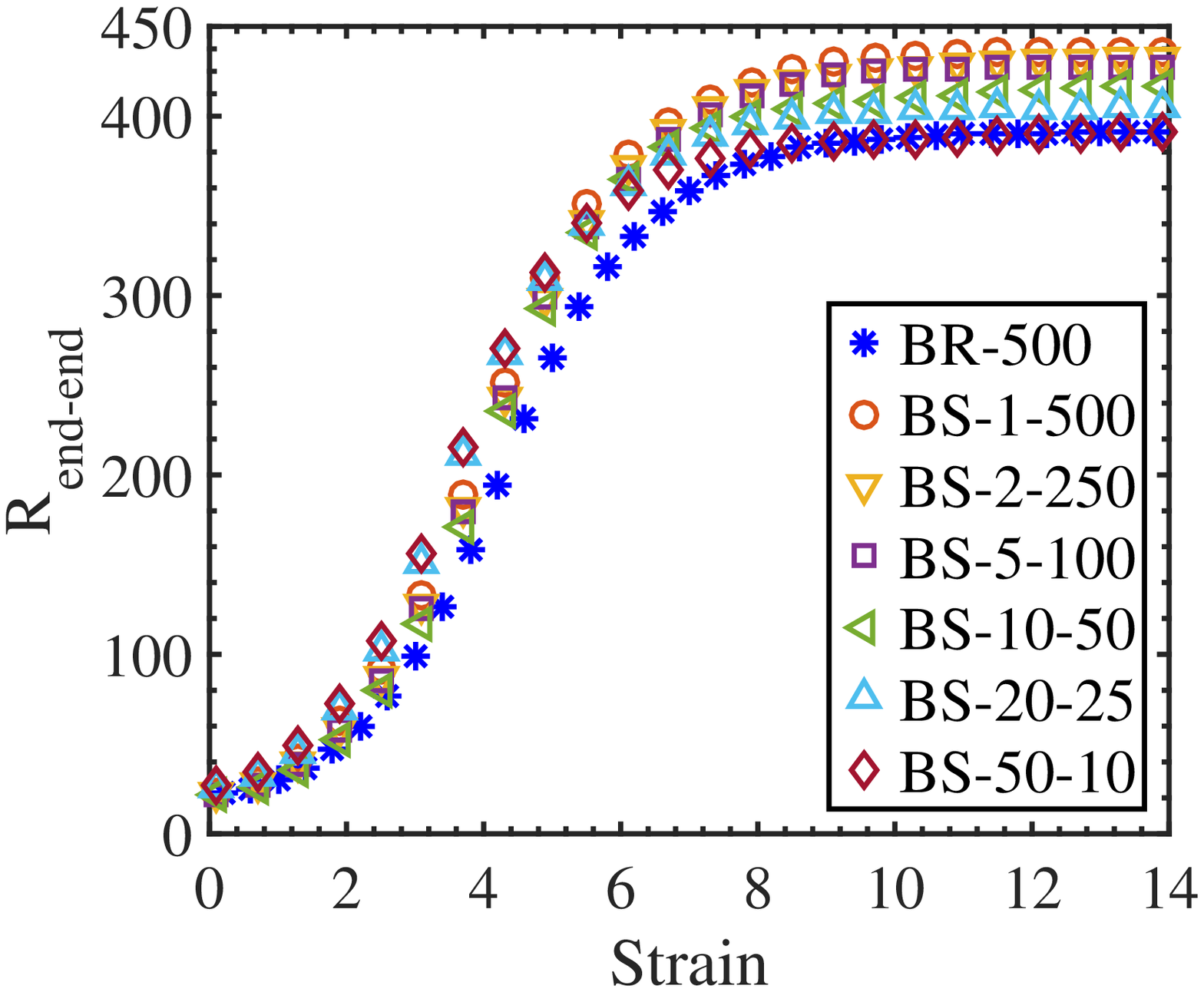}
	\put (-250,210){\LARGE($a$)}
	
	\includegraphics[width=0.7\textwidth]{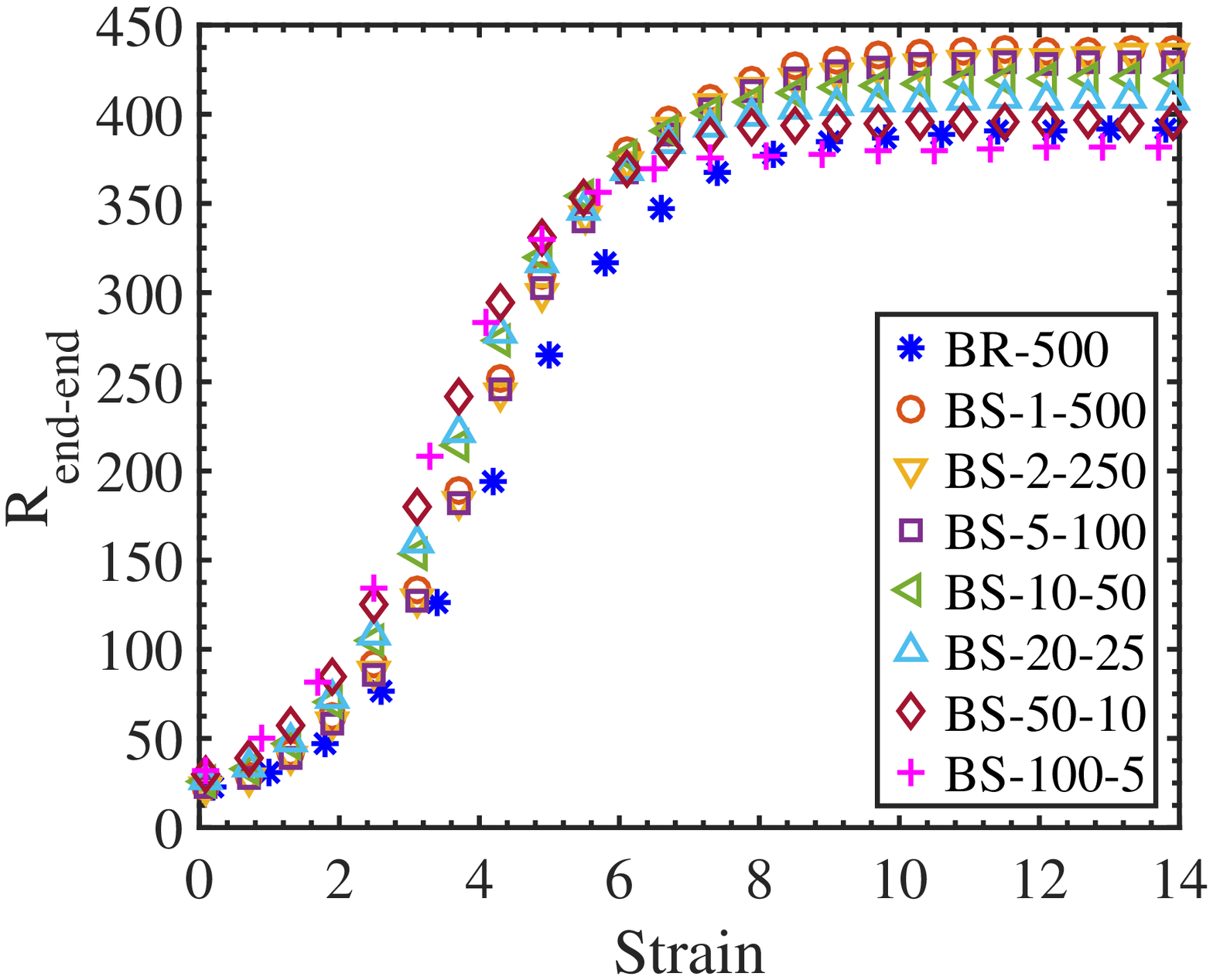}
	\put (-250,210){\LARGE($b$)}
	
	\caption {Variation of $R_{end-end}$ of a polymer chain with strain in uniaxial extensional flow for Wi = 3. BR-500 denotes a bead-rod chain of 500 Kuhn steps, whereas “BS” represents bead-spring chains, using spring laws of ($a$) Cohen-Padé approximation and ($b$) Underhill-Doyle spring law. The first number following “BS” denotes the number of springs, while the second gives the value of $\nu$. The results shown here are averaged over 300 cases, for each model.}
		\label{fig:2}
\end{figure}


\begin{figure}[hbt!]
\centering
	\includegraphics[width=0.7\textwidth]{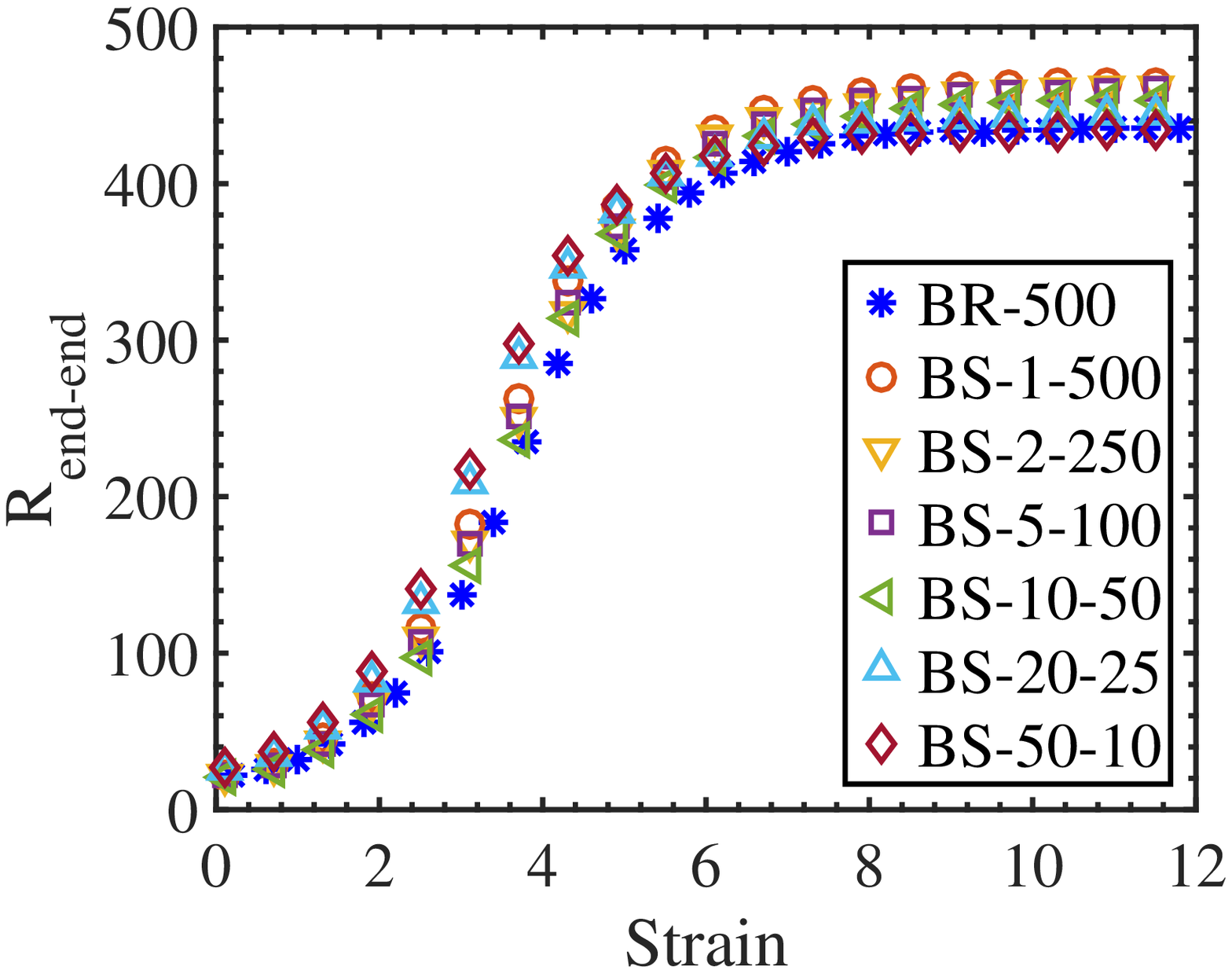}
    \put (-250,210){\LARGE($a$)}
	
	\includegraphics[width=0.7\textwidth]{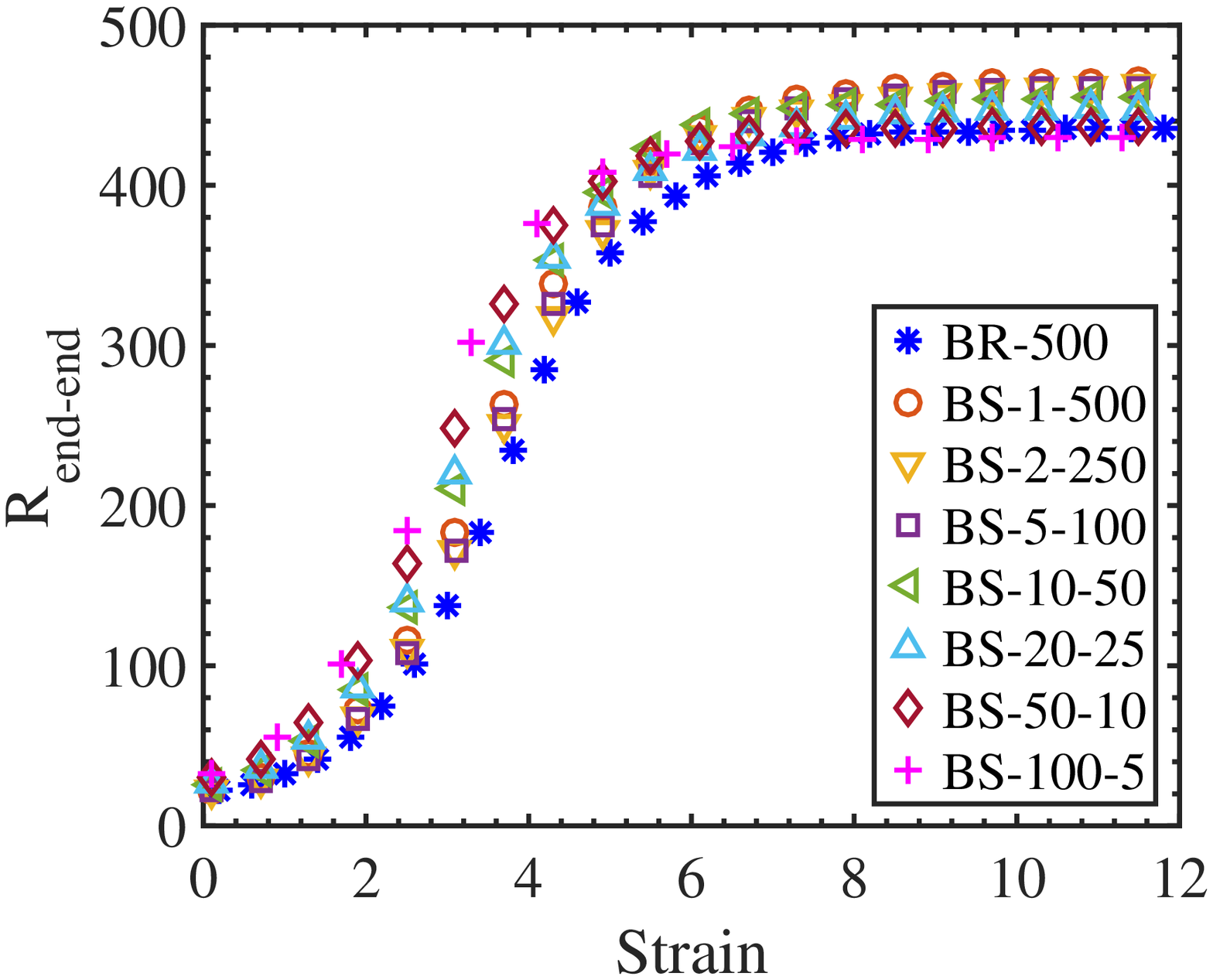}
	\put (-250,210){\LARGE($b$)}
	
	\caption {Variation of $R_{end-end}$ of a polymer chain with strain in uniaxial extensional flow for Wi = 5. BR-500 denotes a bead-rod chain of 500 Kuhn steps, whereas “BS” represents bead-spring chains, using spring laws of ($a$) Cohen-Padé approximation and ($b$) Underhill-Doyle spring law. The first number following “BS” denotes the number of springs, while the second gives the value of $\nu$. The results shown here are averaged over 300 cases, for each model.}
		\label{fig:3}
\end{figure}



\begin{figure}[hbt!]
\centering
	\includegraphics[width=0.7\textwidth]{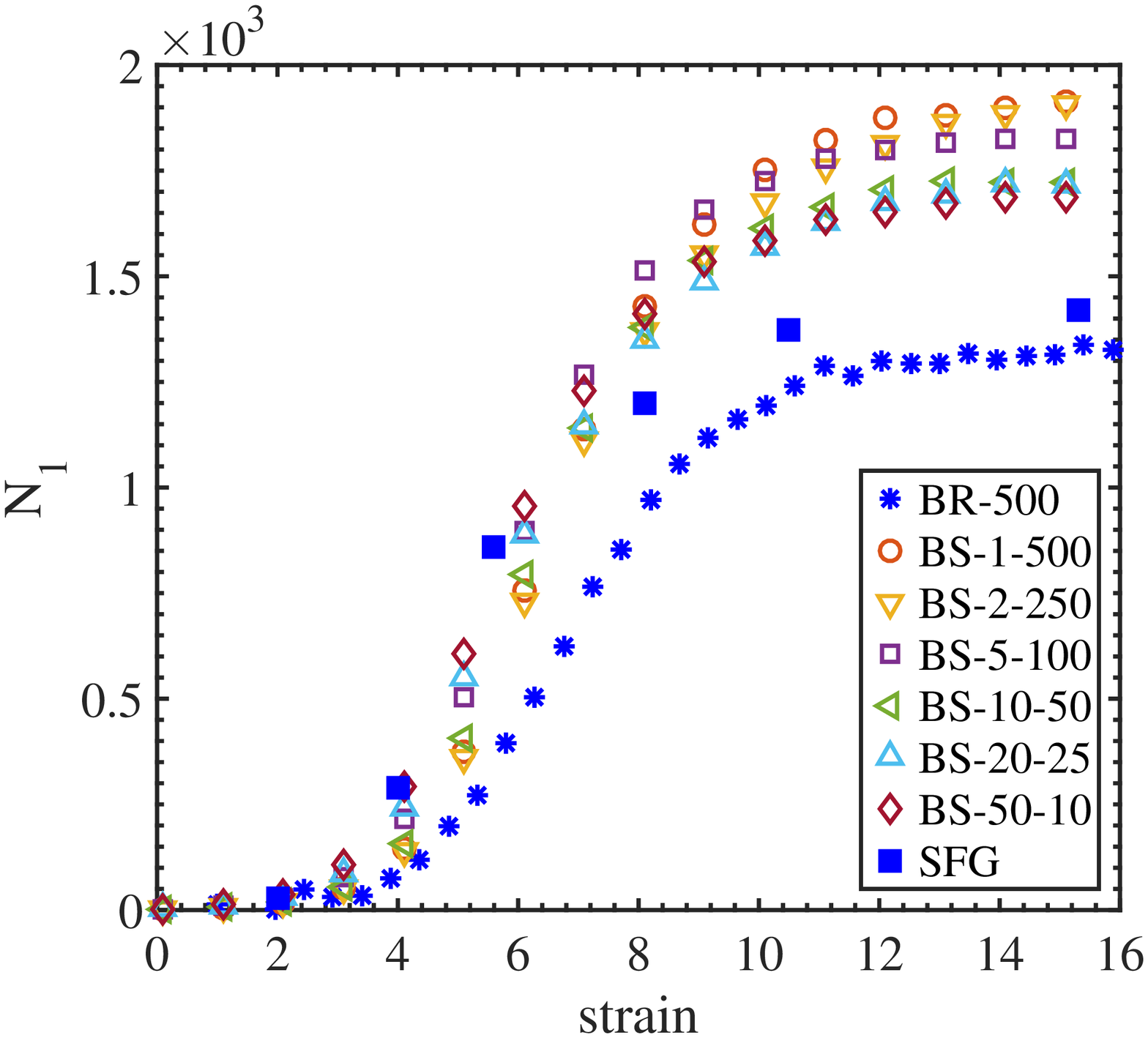}
	\put (-180,230){\LARGE($a$)}
	
	\includegraphics[width=0.7\textwidth]{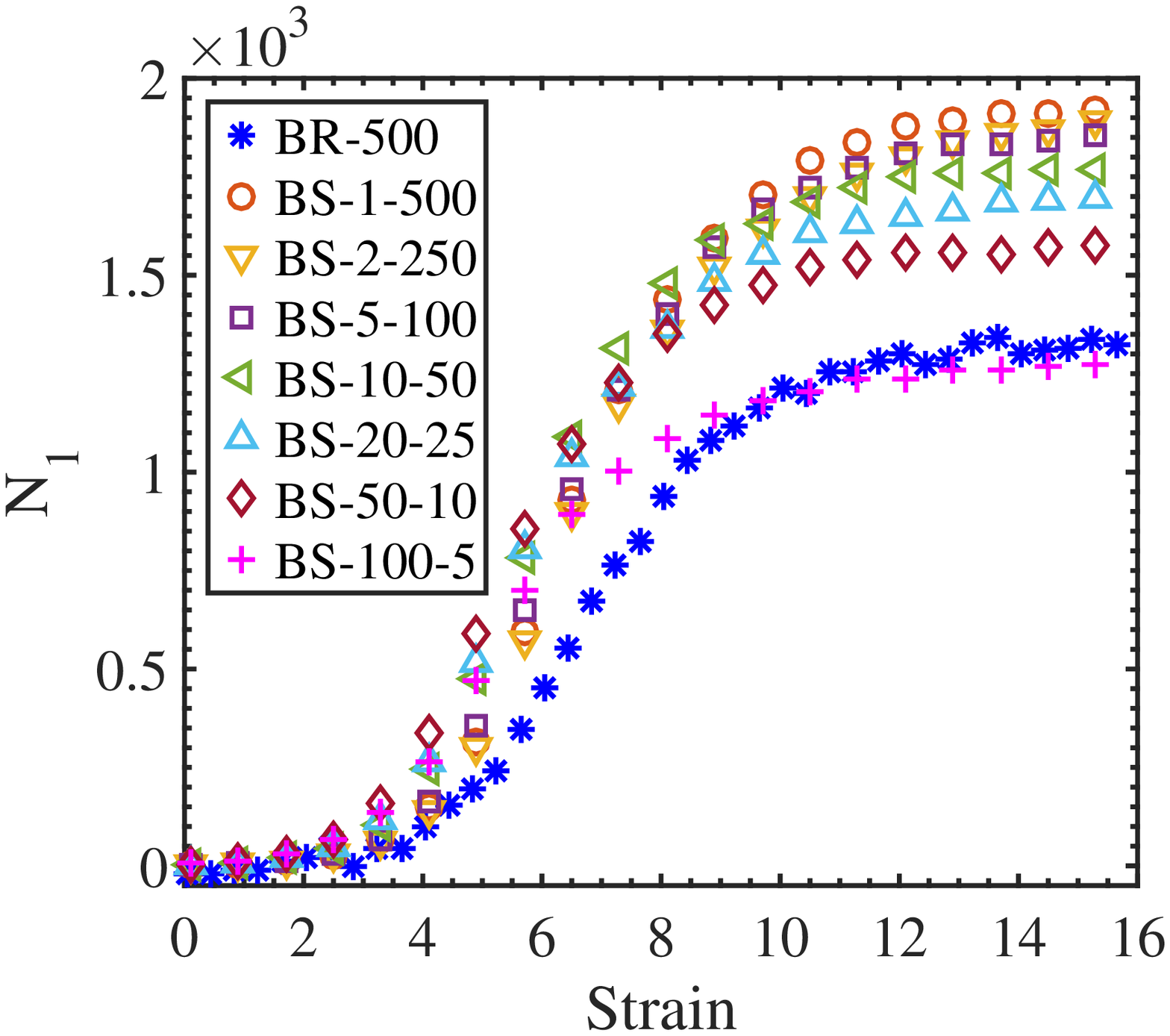}
	\put (-180,230){\LARGE($b$)}
	
	\caption {Variation of $N_1$ of a polymer chain with strain in uniaxial extensional flow for Wi = 2. BR-500 denotes a bead-rod chain of 500 Kuhn steps, whereas “BS” represents bead-spring chains, using spring laws of ($a$) Cohen-Padé approximation and ($b$) Underhill-Doyle spring law. The first number following “BS” denotes the number of springs, while the second gives the value of $\nu$. All results shown here are averaged over 300 cases.}
		\label{fig:4}
\end{figure}

\begin{figure}[hbt!]
\centering
	\includegraphics[width=0.7\textwidth]{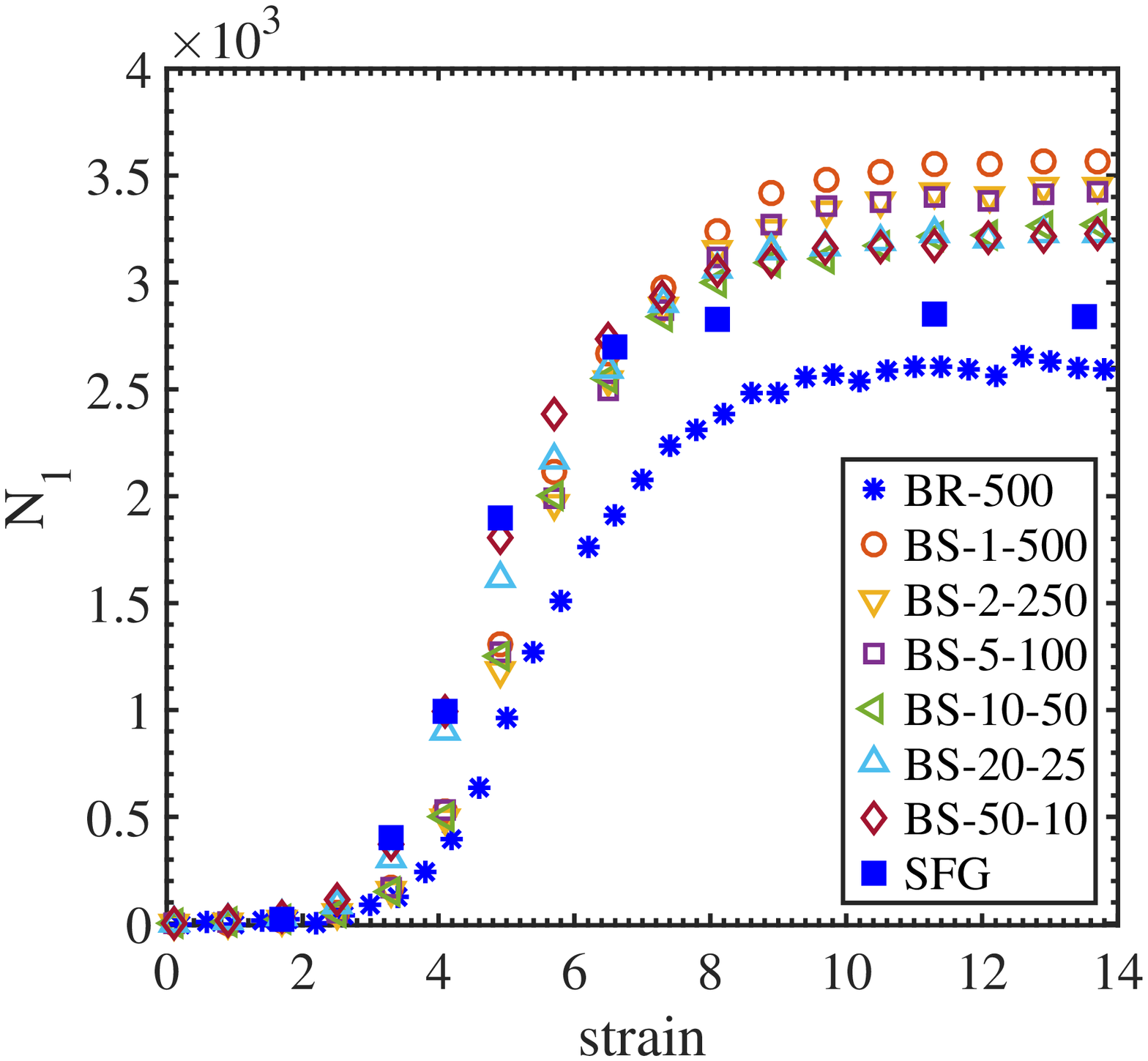}
	\put (-180,230){\LARGE($a$)}
	
	\includegraphics[width=0.7\textwidth]{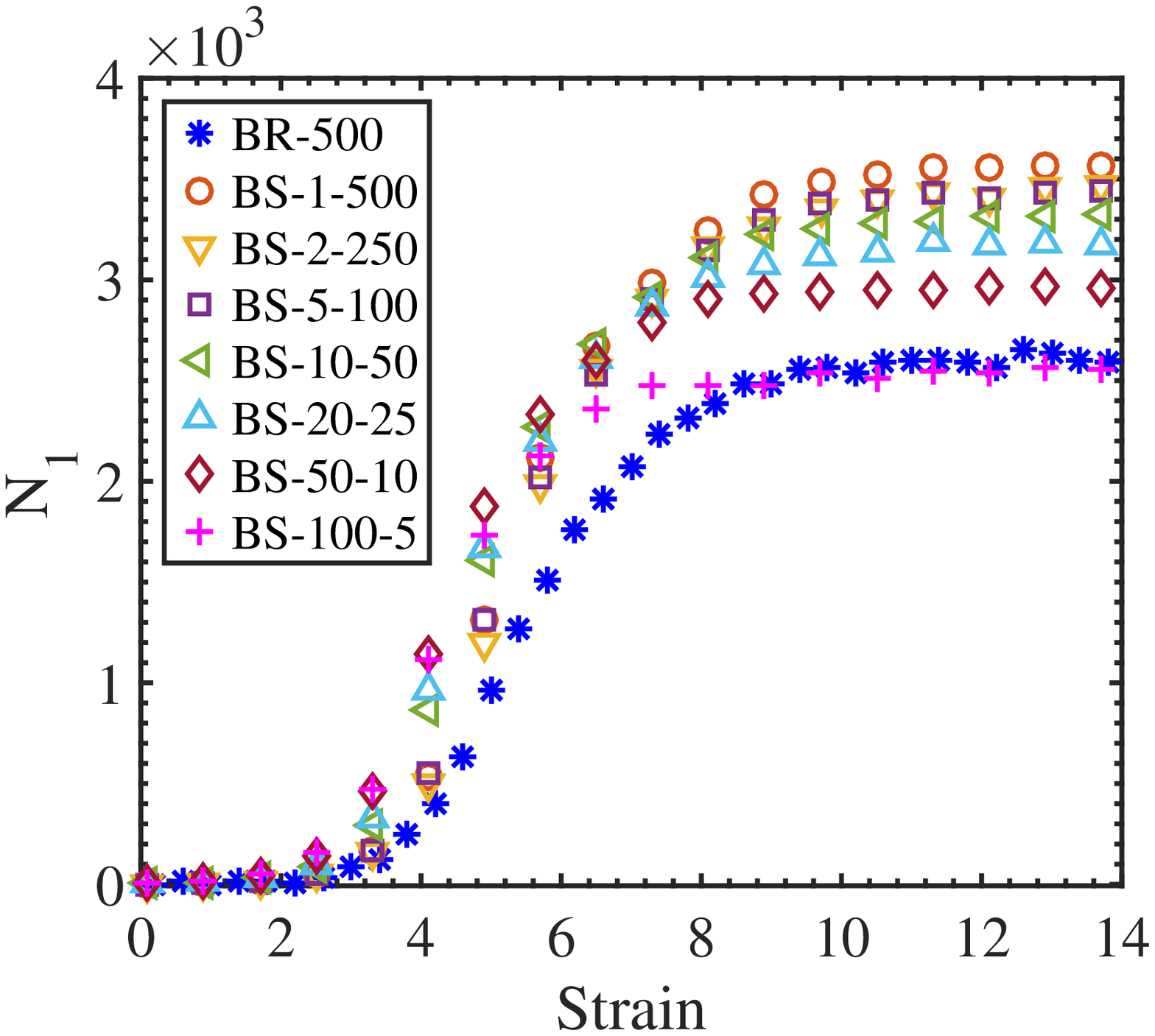}
	\put (-180,230){\LARGE($b$)}
	
	\caption {Variation of $N_1$ of a polymer chain with strain in uniaxial extensional flow for Wi = 3. BR-500 denotes a bead-rod chain of 500 Kuhn steps, whereas “BS” represents bead-spring chains, using spring laws of ($a$) Cohen-Padé approximation and ($b$) Underhill-Doyle spring law. The first number following “BS” denotes the number of springs, while the second gives the value of $\nu$. All results shown here are averaged over 300 cases.}
		\label{fig:5}
\end{figure}

\begin{figure}[hbt!]
\centering
	\includegraphics[width=0.7\textwidth]{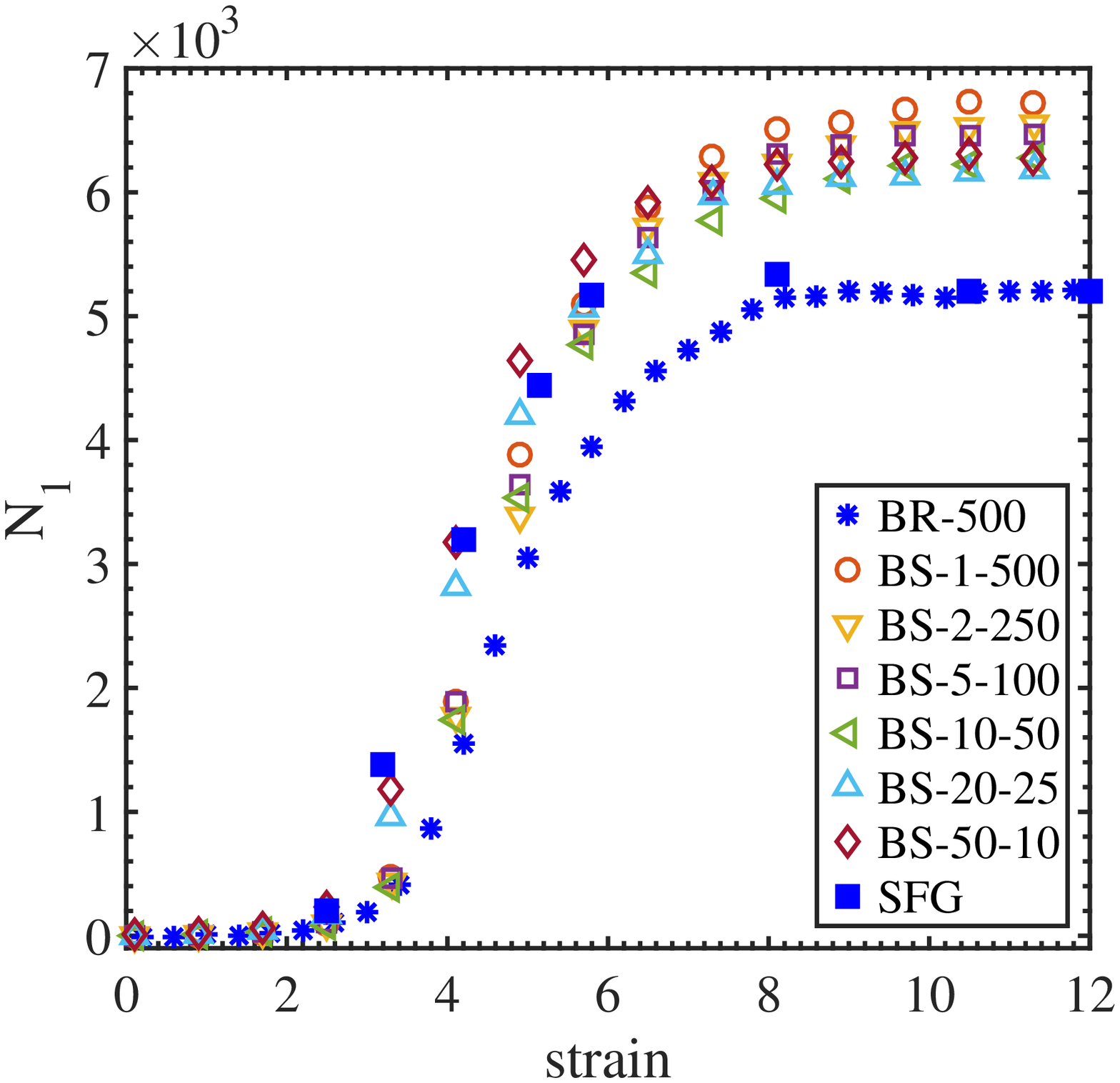}
	\put (-200,220){\LARGE($a$)}
	
	\includegraphics[width=0.7\textwidth]{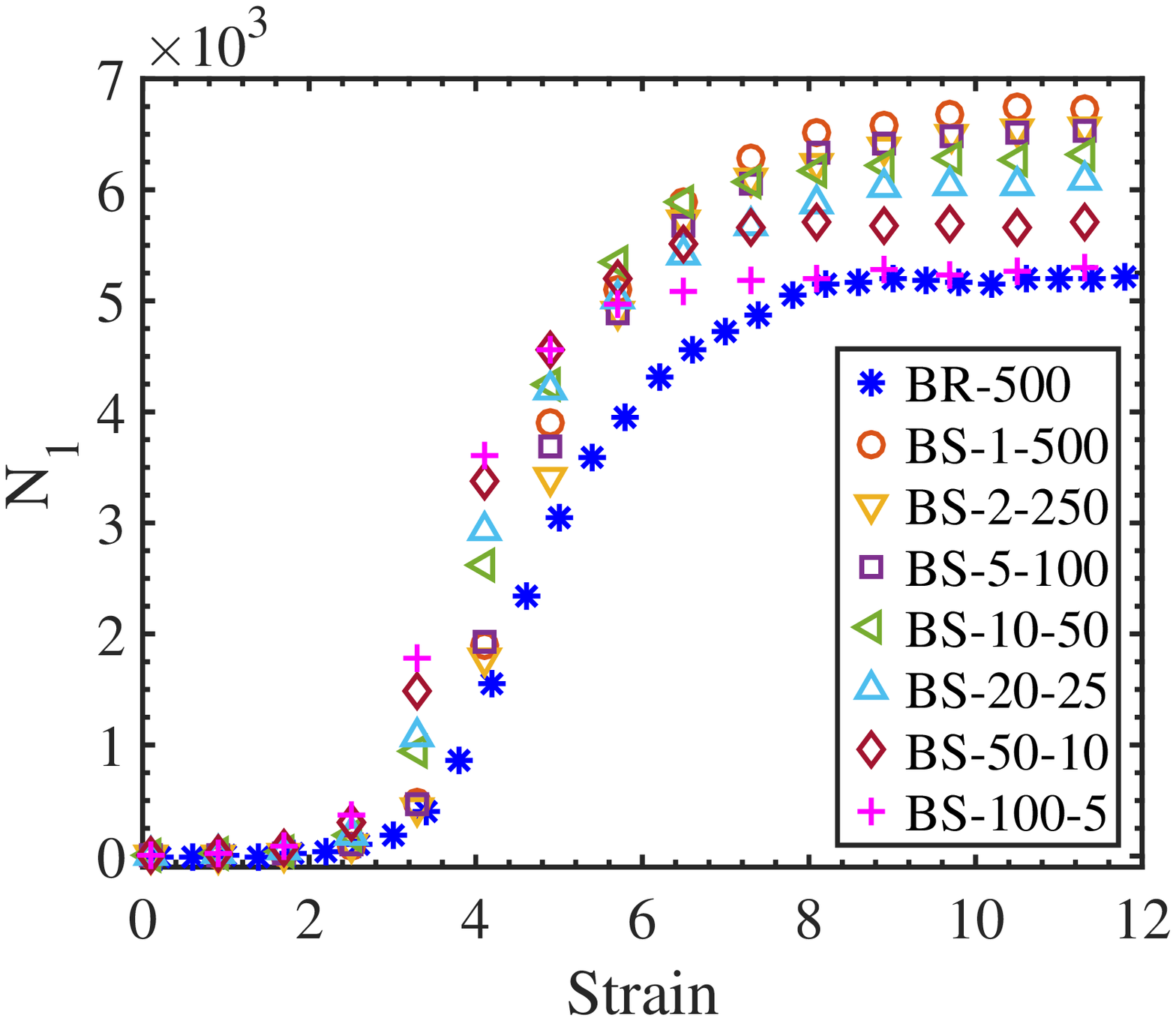}
	\put (-200,210){\LARGE($b$)}
	
	\caption {Variation of $N_1$ of a polymer chain with strain in uniaxial extensional flow for Wi = 5. BR-500 denotes a bead-rod chain of 500 Kuhn steps, whereas “BS” represents bead-spring chains, using spring laws of ($a$) Cohen-Padé approximation and ($b$) Underhill-Doyle spring law. The first number following “BS” denotes the number of springs, while the second gives the value of $\nu$. All results shown here are averaged over 300 cases.}
		\label{fig:6}
\end{figure}



\begin{figure}[hbt!]
\centering
	\includegraphics[width=0.7\textwidth]{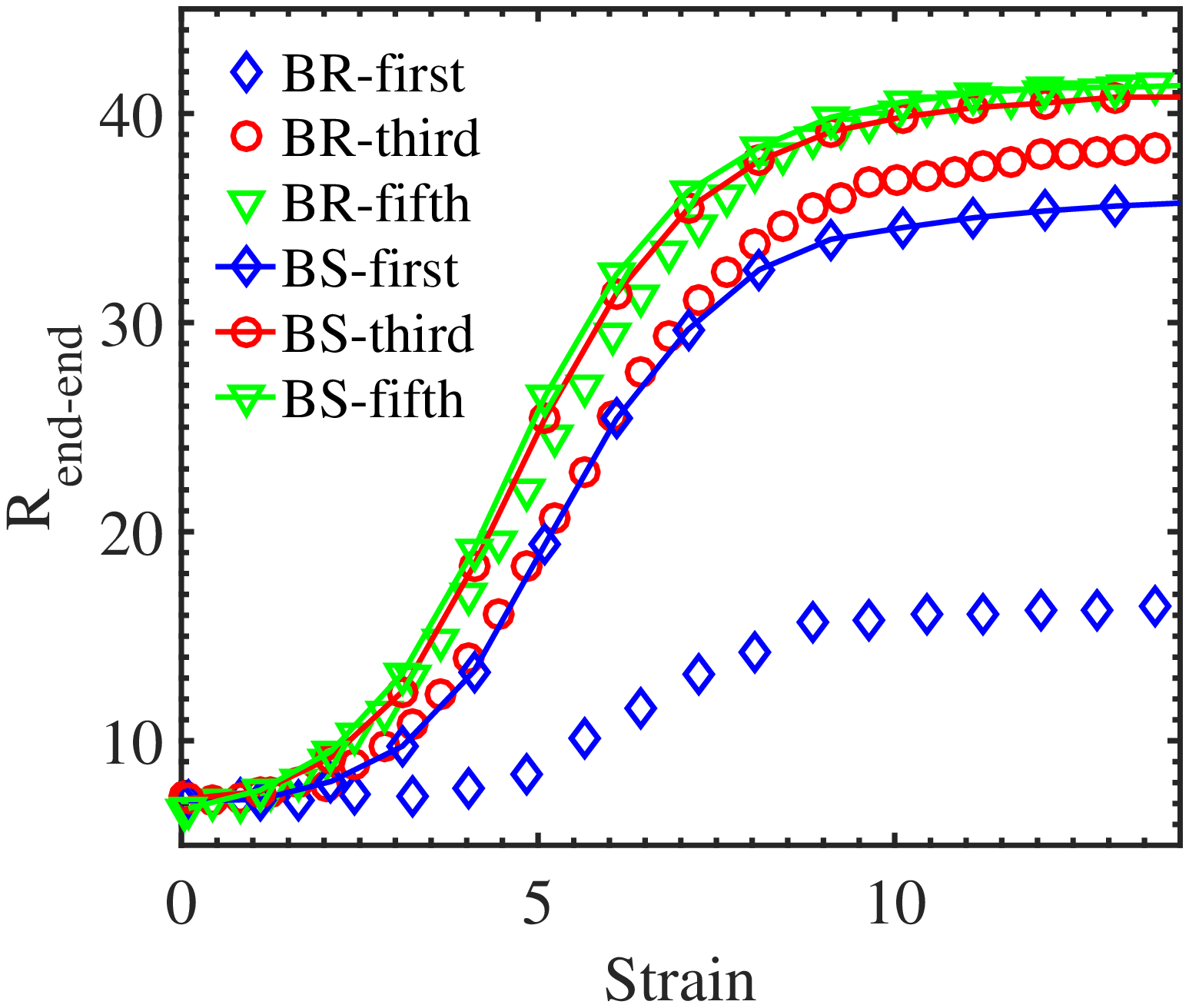}
	\put (-180,230){\LARGE($a$)}
	
	\includegraphics[width=0.7\textwidth]{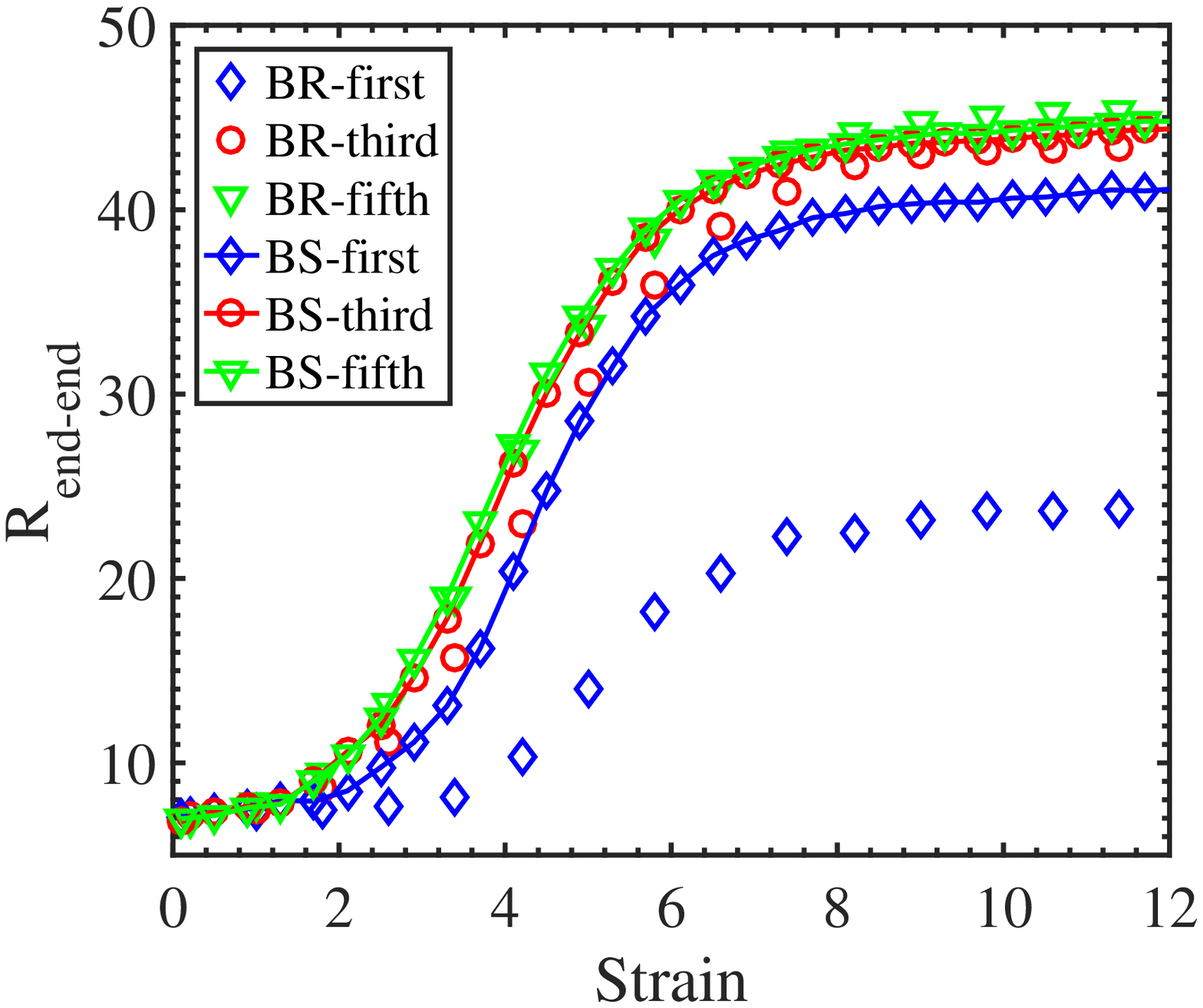}
	\put (-180,230){\LARGE($b$)}
	
	\caption {Variation of $R_{end-end}$ of a segment of polymer chain with strain in uniaxial extensional flow for ($a$) Wi = 2 and ($b$) Wi = 3. BR-first  and BS-first denote the first segment of the bead-rod chain  and the first spring of an equivalent bead-spring chain, respectively. Similarly, third and fifth segment of the bead-rod chain corresponds  to the third and fifth spring of the bead-spring chain, respectively. Bead-spring chain contains 10 springs with each spring mimicking 50 Kuhn steps. The results shown here are averaged over 300 cases, for each model.}
		\label{fig:7}
\end{figure}


\begin{figure}[hbt!]
\centering
	\includegraphics[width=0.7\textwidth]{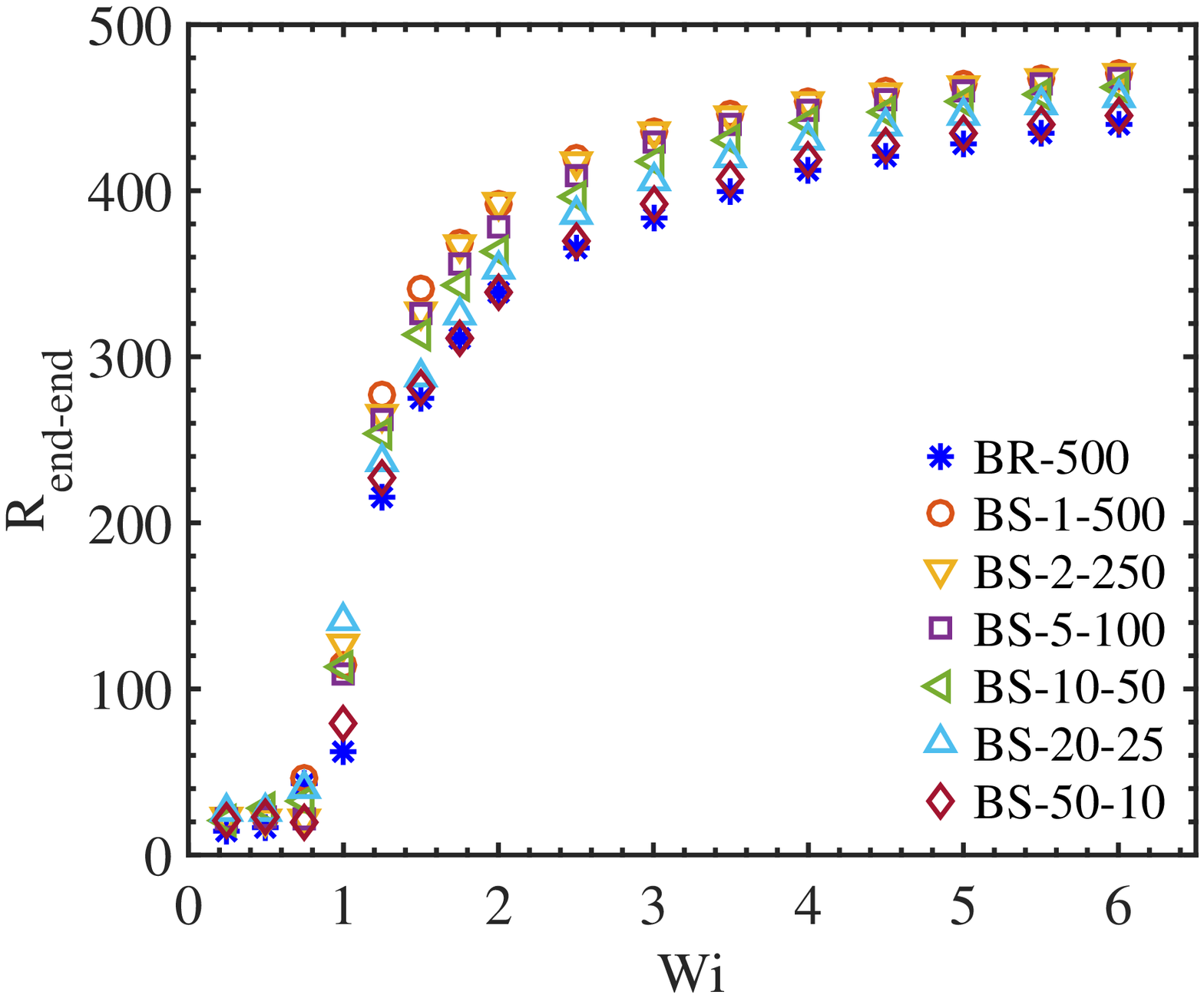}
	\put (-270,220){\LARGE($a$)}
	
	\includegraphics[width=0.7\textwidth]{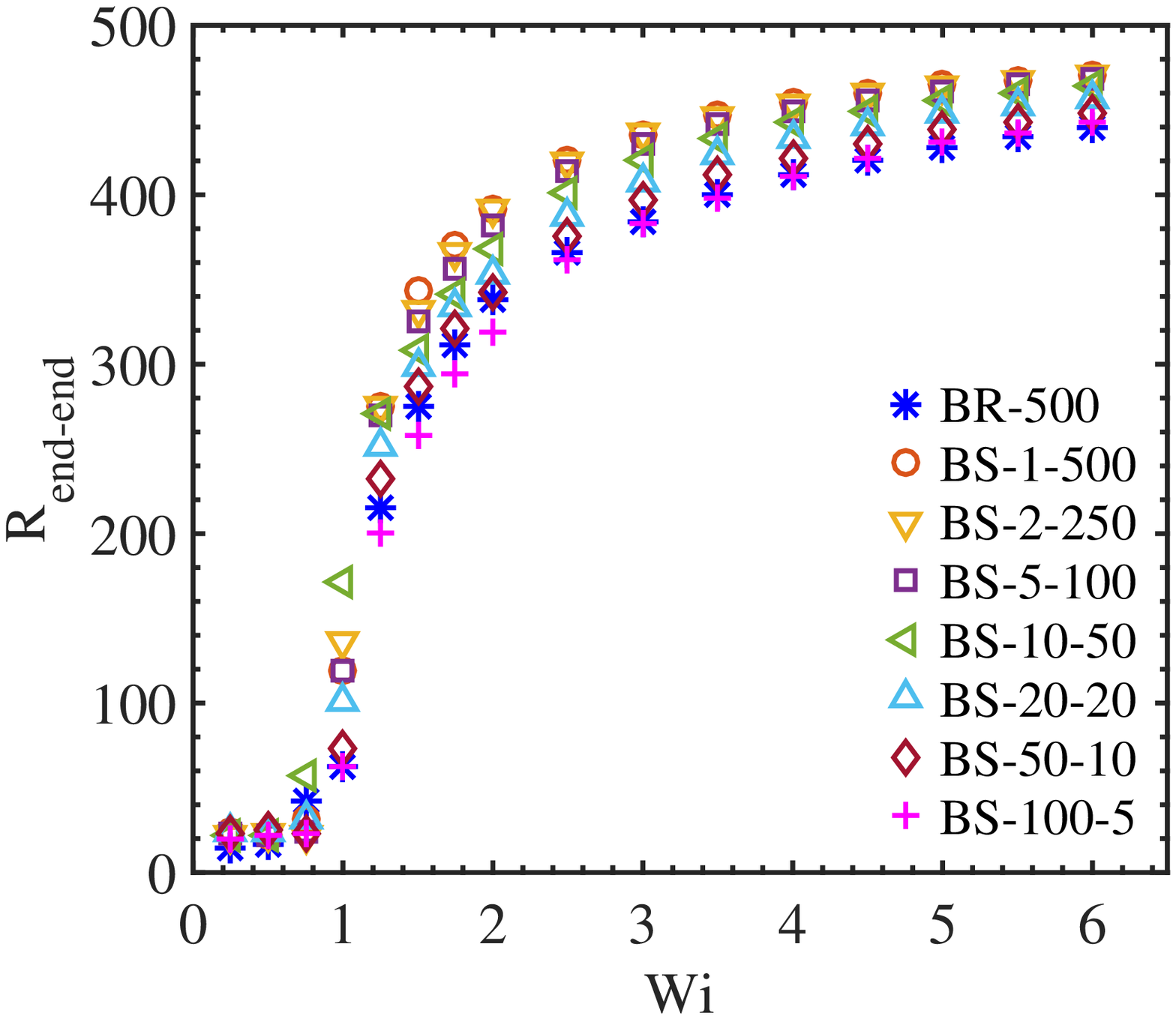}
	\put (-270,230){\LARGE($b$)}
	
	\caption {The variation of the final steady state value of $R_{end-end}$ with the flow rate (measured by Wi) in uniaxial extensional flow. The legends have similar meaning as in Fig. \ref{fig:1}. Comparisons are shown for springs using the ($a$) Cohen-Padé approximation and ($b$) Underhill-Doyle spring law. }
		\label{fig:8}
\end{figure}

\begin{figure}[hbt!]
\centering
	\includegraphics[width=0.7\textwidth]{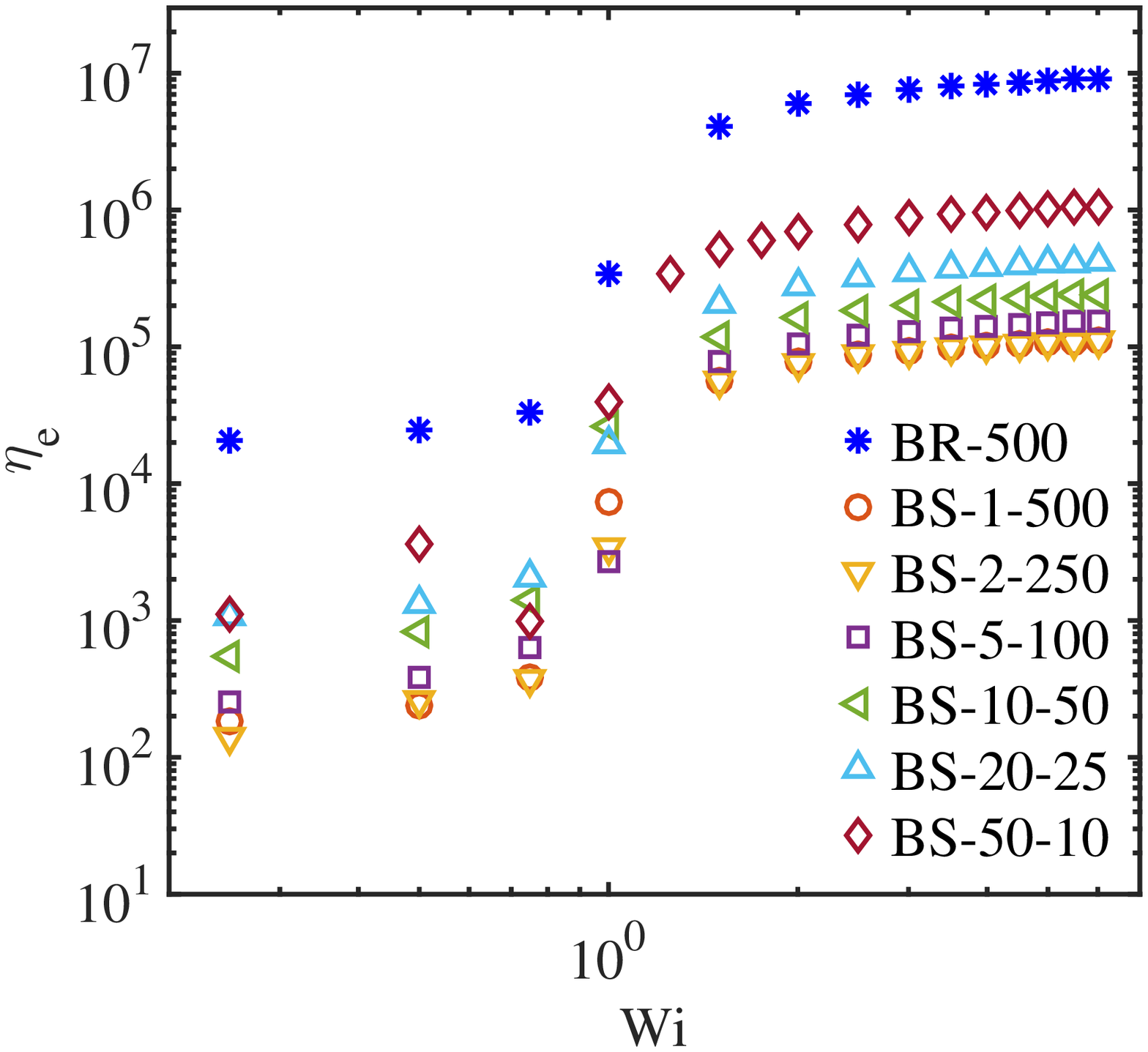}
	\put (-270,230){\LARGE($a$)}
	
	\includegraphics[width=0.7\textwidth]{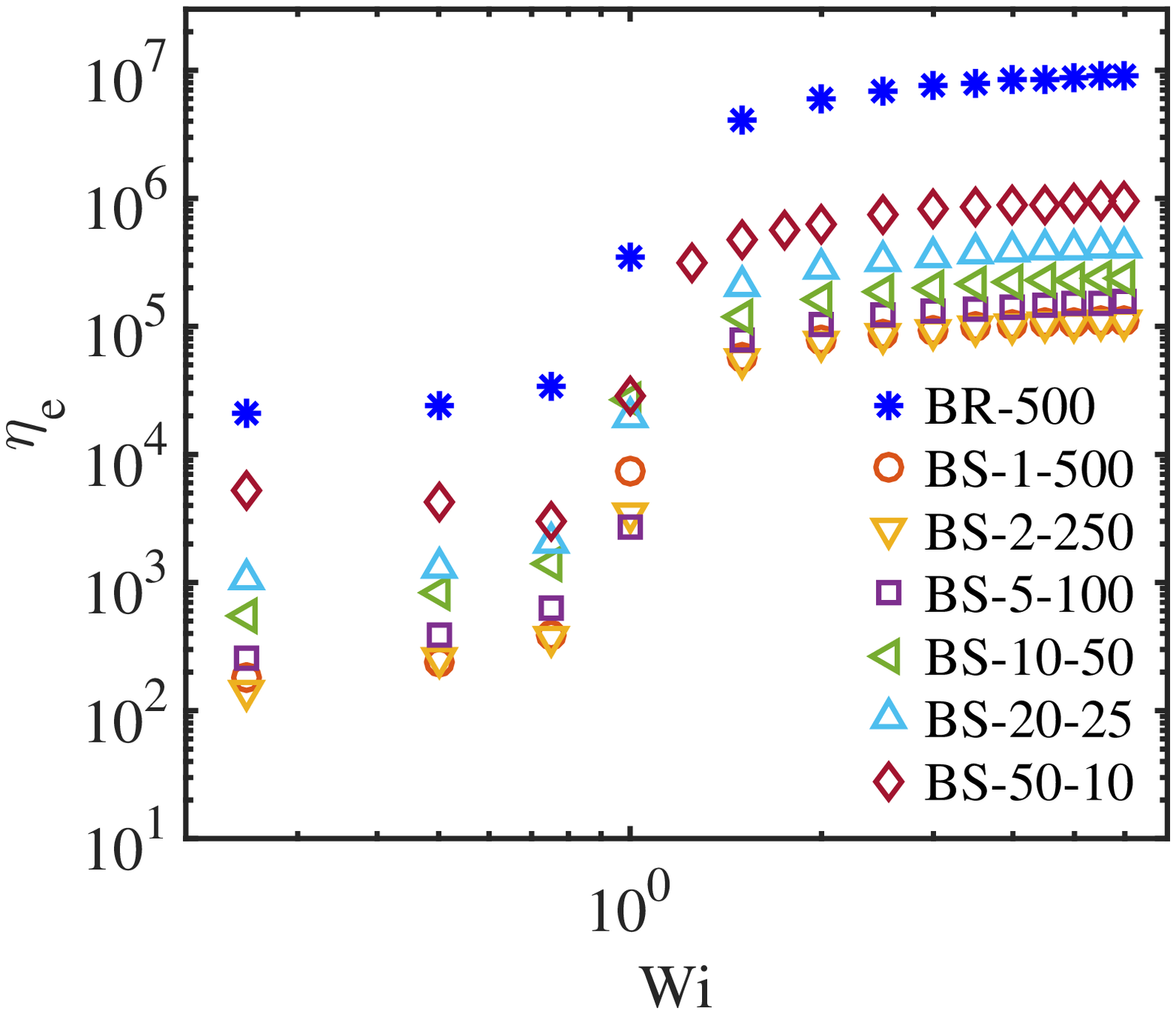}
	\put (-270,230){\LARGE($b$)}
	
	\caption {The variation of extensional viscosity $\eta_e$ with the flow rate (measured by Wi) in uniaxial extensional flow. The legends have similar meaning as in Fig. \ref{fig:1}. Comparisons are shown for springs using the ($a$) Cohen-Padé approximation and ($b$) Underhill-Doyle spring law. }
		\label{fig:9}
\end{figure}

\begin{figure}[hbt!]
\centering
	\includegraphics[width=0.7\textwidth]{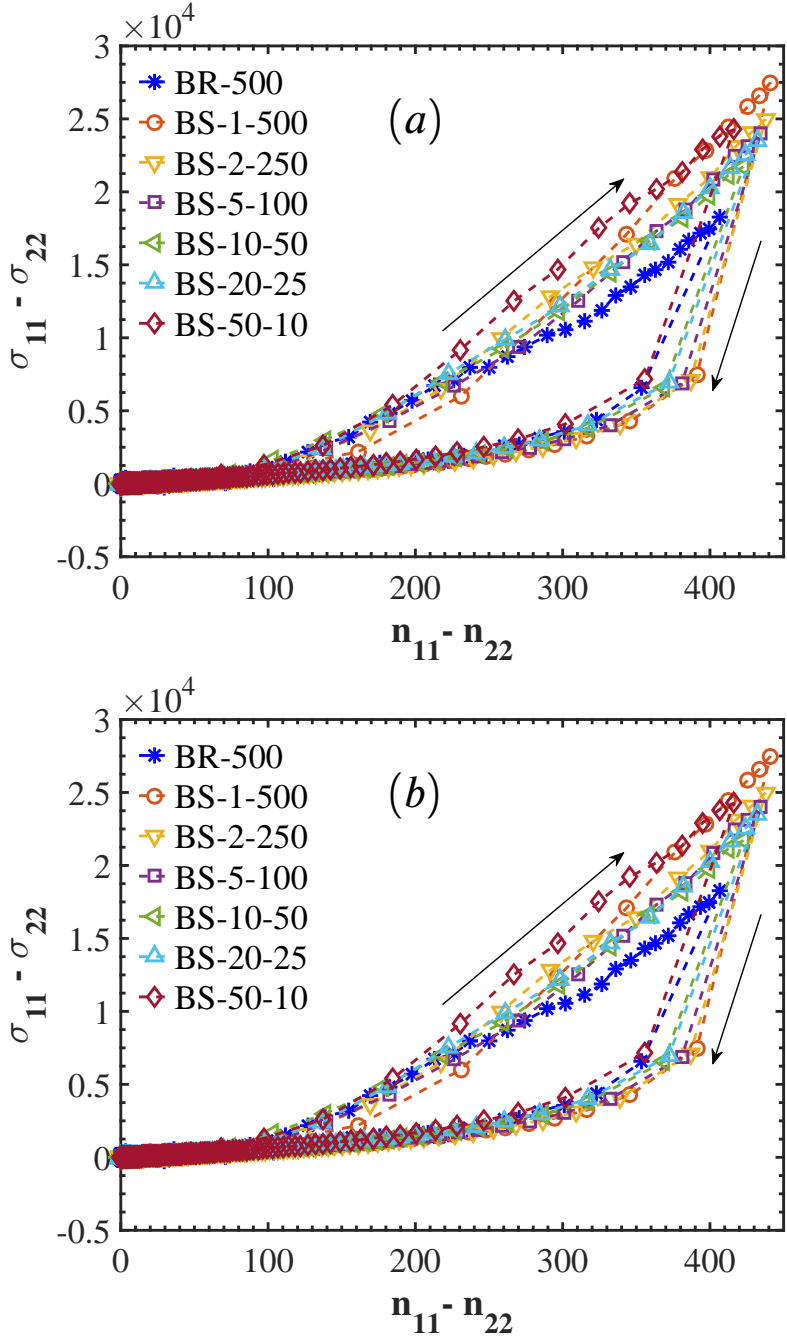}
	\put (-180,200){\LARGE($a$)}
	
	\includegraphics[width=0.7\textwidth]{allimages1/A-WHI-BRBS500-Cohen-Extn-Switched-off-Wi20-Birefringence-Strain-5.eps}
	\put (-180,200){\LARGE($b$)}
	
	\caption {Stress-conformation hysteresis loops for an uniaxial extension flow of Wi = 20. The arrows indicate the forward and backward loop (after cessation of flow). The legends have similar meaning as in Fig. \ref{fig:1}. Dashed-lines are to guide the eye. Comparisons are shown for springs using the ($a$) Cohen-Padé approximation and ($b$) Underhill-Doyle spring law. Further details are provided in the text.}
		\label{fig:10}
\end{figure}

\begin{figure}[hbt!]
\centering
	\includegraphics[width=0.7\textwidth]{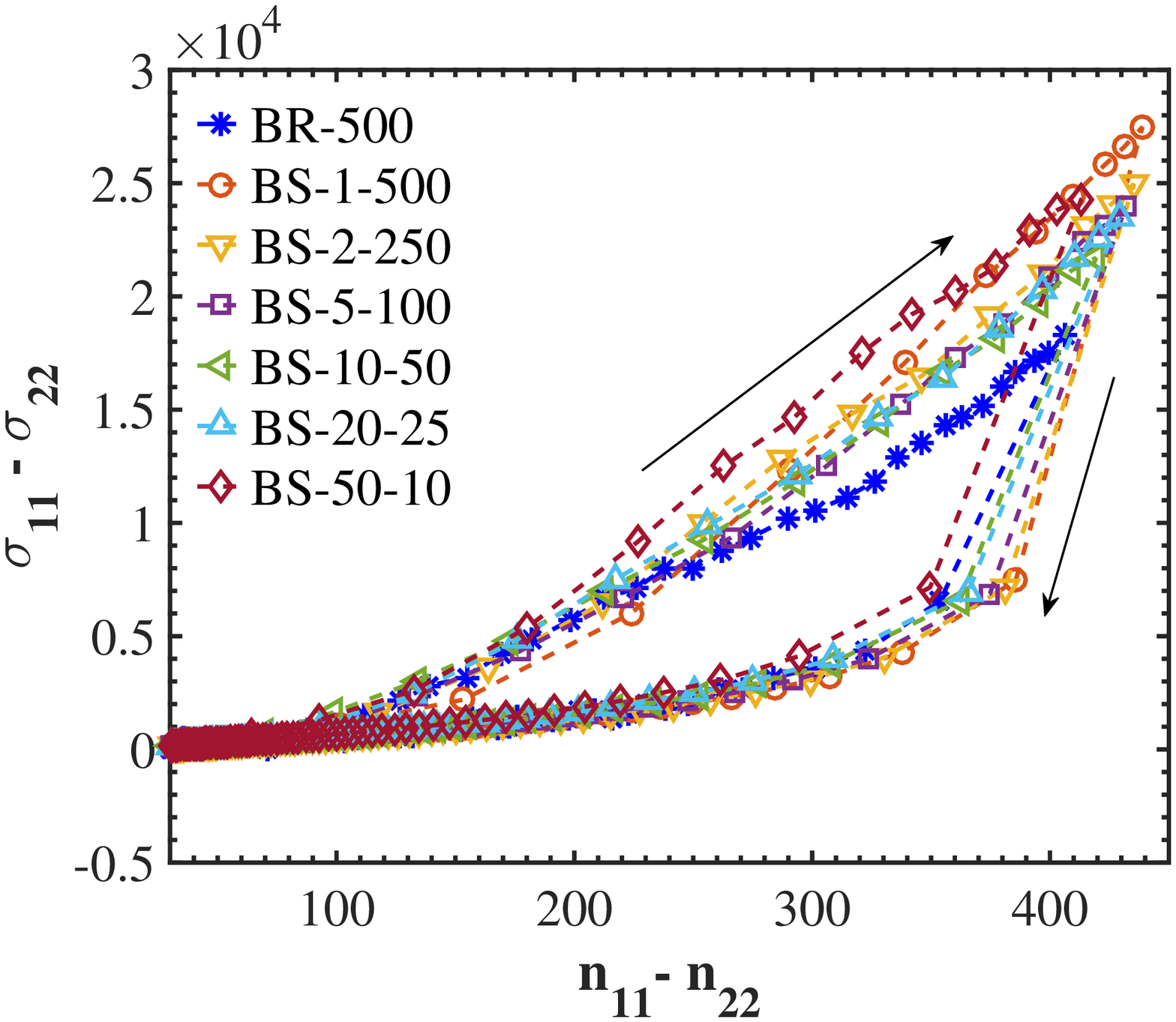}
	\put (-180,200){\LARGE($a$)}
	
	\includegraphics[width=0.7\textwidth]{allimages1/A-WHI-BRBS500-Cohen-Extn-Switched-off-Wi20-Birefringence-Monte-Strain-5.eps}
	\put (-180,200){\LARGE($b$)}
	
	\caption {Stress-conformation hysteresis loops for an uniaxial extension flow of Wi = 20 using the Monte Carlo estimate (Eqn. 21) provided in earlier study \cite{li2000excluded}. The arrows indicate the forward and backward loop (after cessation of flow). The legends have similar meaning as in Fig. \ref{fig:1}. Dashed-lines are to guide the eye. Comparisons are shown for springs using the ($a$) Cohen-Padé approximation and ($b$) Underhill-Doyle spring law. Further details are provided in the text. except that }
		\label{fig:11}
\end{figure}

Here, we also make a note of the equivalence of the bead-KS and the various bead-spring representations of the same. Note that, we have non-dimensionalized the time scales with $\zeta$ for all representations. However, the $\zeta$ for the bead-KS and any corresponding bead-spring representation differ by a factor of $\nu$ as $\zeta_{BS}=\nu \zeta_{BR}$, where $\zeta_{BR}$ and $\zeta_{BS}$ are the drag coefficients of the bead-KS and bead-spring model, respectively. Using this, we can show that the unscaled longest relaxation time of the bead-KS and all bead-spring representations are equal as follows:
 
 The dimensionless end-to-end relaxation times for the bead-rod chain is  $\tau_{BR}^\ast=N_k^2/3\pi^2$, and for the bead-spring chain is $\tau_{BS}^\ast={\nu N}_s^2/3\pi^2$, where $N_k$ is the number of Kuhn-step in the bead-rod chain and $N_s$ is the number of springs in the bead-spring chain. In the unscaled form, the longest relaxation time can be written as
 \begin{equation}\label{eq:1a}
 \tau_{BR}=\tau_{BR}^\ast\frac{\zeta_{BR}b_k^2}{k_BT} = \frac{N_k^2}{3\pi^2}\frac{\zeta_{BR}b_k^2}{k_BT}
 \end{equation}
and,
\begin{equation}\label{eq:2a}
 \tau_{BS}=\tau_{BS}^\ast\frac{\zeta_{BS}b_k^2}{k_BT}=\frac{{\nu N}_s^2}{3\pi^2}\frac{\zeta_{BS}b_k^2}{k_BT} 	   
\end{equation}
		
for the bead-KS and bead-spring models. Now, since $\zeta_{BS}={\nu\zeta}_{BR}$, hence, Eq. \ref{eq:2a}   becomes
\begin{equation}
 \tau_{BS}=\frac{{\nu N}_s^2}{3\pi^2}\frac{{\nu\zeta}_{BR}b_k^2}{k_BT}=\frac{\nu^2N_s^2}{3\pi^2}\frac{\zeta_{BR}b_k^2}{k_BT}=\frac{\left(\nu N_s\right)^2\zeta_{BR}b_k^2}{3\pi^2k_BT}=\frac{N_k^2}{3\pi^2}\frac{\zeta_{BR}b_k^2}{k_BT}=\tau_{BR}   
\end{equation}
 
Thus, throughout in this manuscript, we are comparing the behavior of the same chain in various representations (using different number of springs), exposed to the same flow rate (since all our comparisons are for the same Wi). Further, we highlight this point by showing that the  relaxation time spectrum (scaled using $\zeta_{BR}$ for all representations) for a given chain and its different bead-spring representations are equal in Figure 1. Also, the scaling law obtained agrees with theoretical expectations in the absence of HI.

For the start-up of extensional flows, we consider flow rates of Wi $\le5$. Fig. \ref{fig:1} shows the temporal variation of the average end-to-end distance ($R_{end-end}$) of a polymer chain with strain (dimensionless time, defined as $\dot{\epsilon}t^\ast$, where $t^\ast$ is the simulation time) for different levels of chain discretization, for Wi = 2. The averaging is performed over 300 cases for all models. The comparison with the Cohen-Padé approximation and Underhill-Doyle spring law are shown in Fig. \ref{fig:1}($a$) and \ref{fig:1}($b$), respectively. Qualitatively, the trends are similar for all models considered.  However, as evident, the dynamics is indeed strongly affected by the level of chain discretization. For Wi = 2, there is a significant quantitative difference between the bead-KS chain and the dumbbell (with $\nu=500$). This difference shows a systematic decrease with increasing levels of chain discretization i.e. decreasing $\nu$. However, a considerable quantitative difference remains even when 20 springs are used (with $\nu=25$). As the strength of the flow is increased to Wi = 3 (Fig. \ref{fig:2}) and Wi = 5 (Fig. \ref{fig:3}), these differences in the temporal evolution of the chain size, for the various representations, progressively reduce. The behaviour remains extremely similar for the two different spring laws at the same value of $\nu$ , for all Wi values.

A comparison of the temporal variation of the first normal stress difference $N_1$ is presented in Figs.  \ref{fig:4}-\ref{fig:6} for Wi = 2, 3 and 5. As observed earlier for the chain size (Figs.  \ref{fig:1}-\ref{fig:3}), the trends are qualitatively similar for all models considered. However, quite significantly, there exists a significant difference between the bead-KS and bead-spring models, for all flow rates considered. While the bead-spring models are converging as the value of $\nu$ is increased, the quantitative agreement with the bead-KS model is poor in general. Interestingly, the steady state values obtained with $\nu=5$ for the Underhill-Doyle spring law is consistent with those of the bead-KS model, even though the agreement remains poor in the intermediate strains leading to the steady state. For all other values of $\nu$, the behaviour is quite similar for the two spring laws.

Here, we also use the SFG technique to predict the behavior of the bead-KS chain. In the premise of SFG, if local equilibrium exists for the springs, then we can extrapolate the behavior of different bead-spring representations by smooth polynomials to obtain a prediction for the bead-KS chain. At different strains, we extrapolate the results from different bead-spring representations to obtain a SFG prediction that is compared with the actual bead-KS result. The steady state agree reasonably well for all cases, but the temporal behavior obtained from the SFG is different. The SFG predicts a faster rise in the stretch (in accordance  with the fact that all the bead-spring models predict a faster rise), relative to the actual bead-KS simulation.

To understand the origins of this considerable discrepancy in the values of $N_1$ (Figs. \ref{fig:4} - \ref{fig:6}), we compare the average stretch in the various segments of the chain predicted by the bead-KS and the bead-spring model. Here, the chain is divided in 10 segments, each mimicked by a spring. Fig. \ref{fig:7} shows the temporal variation of various segments of the chain for Wi = 2 (Fig. \ref{fig:7}$a$) and 3 (Fig. \ref{fig:7}$b$). Here, we have shown the behaviour of three segments of the chain. Each segment contains 50 rods, represented by one spring in the equivalent bead-spring model containing a total of 10 springs ($\nu=50$). Thus, the first segment of the bead-KS chain is compared with the first spring of the bead-spring chain. Similarly, intermediate and middle segments of the chain are compared with the third and fifth springs, respectively. Our results, for both Wi =2 and 3, show a significant difference in the stretch of the first segment of the bead-KS model with the first spring of the bead-spring model. The stretch in the other segments agree well. Thus, the contribution of the first segment (or the terminal spring) to the stress is significantly different for the bead-KS and the bead-spring model. This considerably lower stretch of the terminal segment in the bead-KS model leads to lower predictions of the stress, relative to the bead-spring model.

The variation of the final steady-state values obtained for the end-to-end distance and the extensional viscosity $\eta_e$ of a polymer chain with Wi is presented in Fig. \ref{fig:8} and \ref{fig:9}, respectively. The end-to-end distance of a polymer chain increases with an increase in the Weissenberg number and almost attains the fully extended state at Wi = 5. The trends clearly indicate a systematic variation with chain discretization, as also observed also in the transient results in Figs.  \ref{fig:1}-\ref{fig:6}. Again, as observed earlier, the results from the two spring models for same values of $\nu$ agree well with each other. However, there exists a significant quantitative difference between the predictions of the extensional viscosity from the bead-KS and the most resolved bead-spring model.

Figs. \ref{fig:10}-\ref{fig:11} show the stress-conformation hysteresis loops, obtained with the bead-KS and bead-spring models of varying resolutions. For the bead-KS model, the “conformation” is defined as 

\begin{equation}
    \Delta n = n_{11} - n_{22}
\end{equation}

where the tensor $\overleftrightarrow{n}$  is defined as $\overleftrightarrow{n}=\left\langle{\vec{Q}}_i^\ast{\vec{Q}}_i^\ast\right\rangle$. Note that the conformation is proportional to the birefringence [18]. For the bead-spring model, the conformation is estimated as \cite{li2000excluded}

\begin{equation}
    \Delta n = \nu \sum_{i=1}^{N_s} \left\langle \left( \frac{3}{2}C^2 - \frac{1}{2}\right) \left(f_{xx}-f_{yy}\right)\right\rangle_i
\end{equation}

Where the summation is over all springs in the model. For the $i^{th}$ spring, $C=\frac{\left|\vec{Q}_{i,1}^\ast \right|}{\left|\vec{Q}_{i}^\ast\right|}$ , where  $\vec{Q}_{i,1}^\ast$ is the component of the $i^{th}$ spring in the flow direction. Here, as discussed in the study by Li and Larson \cite{li2000excluded}, $f_{xx}$ and $f_{yy}$ are given by:

\begin{eqnarray}
    f_{xx} =  1-2 \frac{\hat{r}}{L^{-1} \left(\hat{r}\right)} \\
    f_{yy} =  \frac{\hat{r}}{L^{-1} \left(\hat{r}\right)} 
\end{eqnarray}

Where $L^{-1}$  denotes the inverse Langevin function and  $\hat{r}$ is the fractional extension for the $i^{th}$ spring. As discussed by Li and Larson \cite{li2000excluded}, the inverse Langevin function can be estimated by the Cohen-Padé approximation, which is also used as the spring law in our simulations:

\begin{equation}
    L^{-1}\left(\hat{r}\right) =   \frac{3\hat{r} -\hat{r}^3 }{1-\hat{r}^2}
\end{equation}

In Fig. \ref{fig:10}($a$), this approximation is used for the inverse Langevin function for bead-spring models. Note that, equivalently, the expression for the Underhill-Doyle spring law can also be used. The results with this are shown in Fig. \ref{fig:10}($b$). The bead-spring simulations in Figs. \ref{fig:10}($a$) and ($b$) are performed using the Cohen-Padé approximation and the Underhill-Doyle spring law, respectively. 
Li and Larson also provided another estimate of  $f_{xx}$ and $f_{yy}$ , obtained from Monte Carlo simulations:

\begin{equation}
   f_{xx}-f_{yy} =   \frac{1}{5} \left[ (\hat{r}^2)^3 + (\hat{r}^2)^2 +  3\hat{r}^2 \right] 
\end{equation}

In Fig. \ref{fig:11}, we use this estimate for the bead-spring models and compare with the bead-KS model. The bead-spring simulations are performed using the Cohen-Padé approximation and the Underhill-Doyle model in Figs. \ref{fig:11}($a$) and ($b$), respectively. 
To the best of our knowledge, no direct comparison of such stress-conformation loops exists in literature for such a long chain (500 Kuhn steps). Overall, the loops obtained are similar, with the bead-spring results converging to the bead-KS ones with increasing number of springs. However, the agreement with 50 springs gets worse relative to the other representations, perhaps since the springs get short enough to be at the verge of failure of the spring laws.

   \subsection{Shear flow}


\begin{figure}[hbt!]
\centering
	\includegraphics[width=0.7\textwidth]{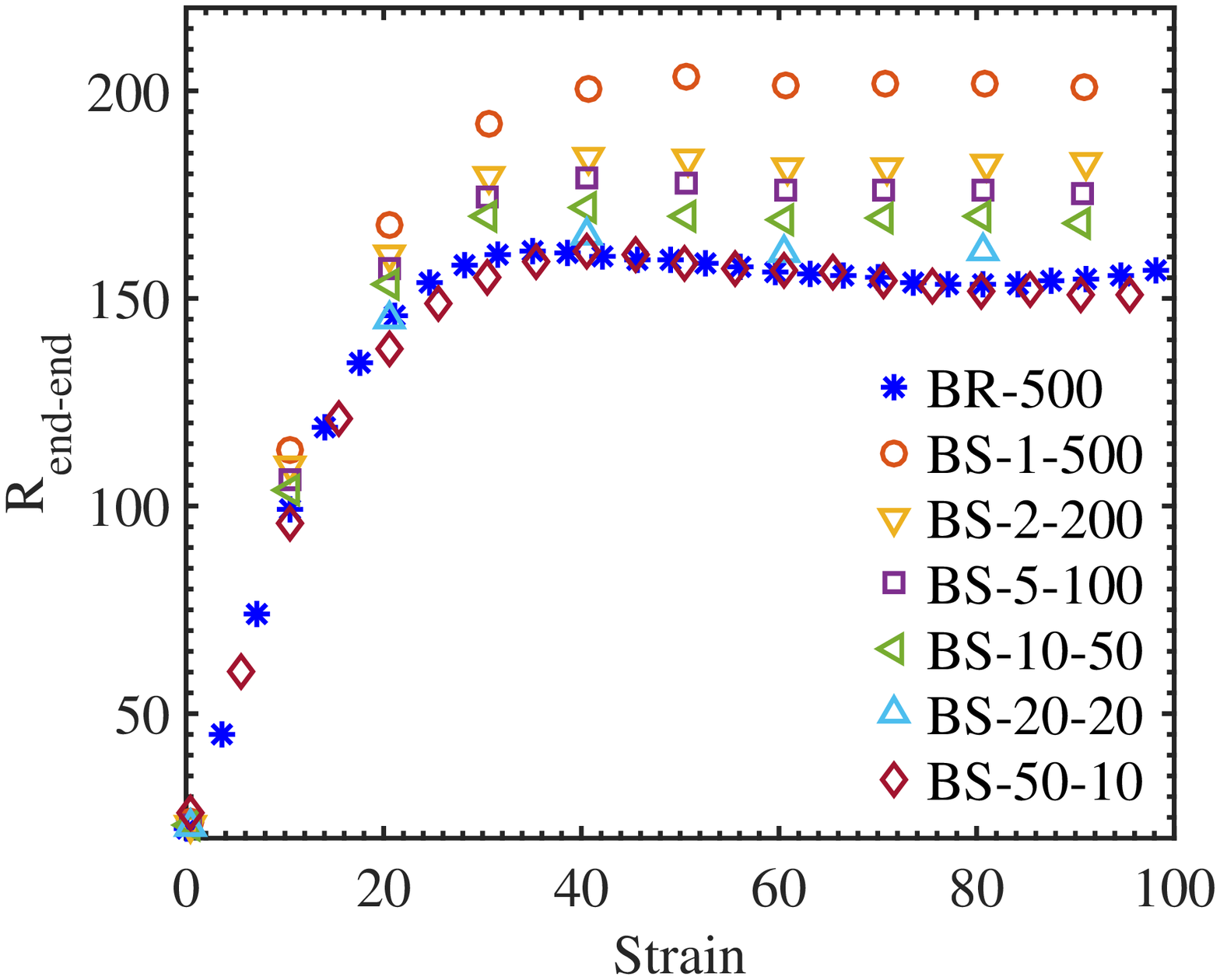}
	\put (-270,230){\LARGE($a$)}
	
	\includegraphics[width=0.7\textwidth]{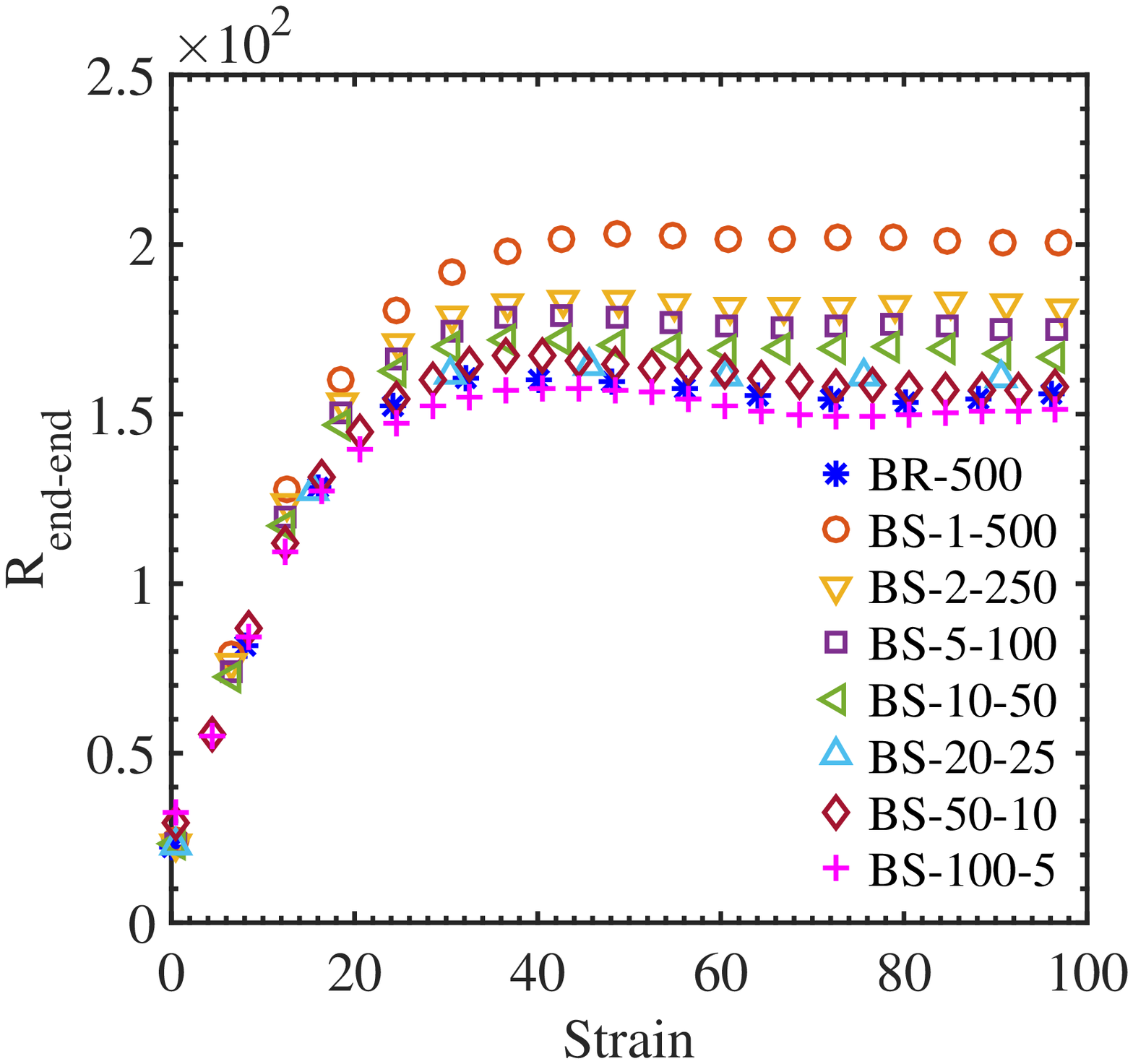}
	\put (-270,230){\LARGE($b$)}
	
	\caption {Variation of $R_{end-end}$ of a polymer chain with strain in shear flow for Wi = 30. The legends have the same meaning as in Fig. \ref{fig:1}. Results shown here are averaged over 2000 and 3000 cases for bead-rod and bead-spring model, respectively. Comparisons are shown for spring laws using the ($a$) Cohen-Padé approximation and ($b$) Underhill-Doyle spring law.  }
		\label{fig:12}
\end{figure}

\begin{figure}[hbt!]
\centering
	\includegraphics[width=0.7\textwidth]{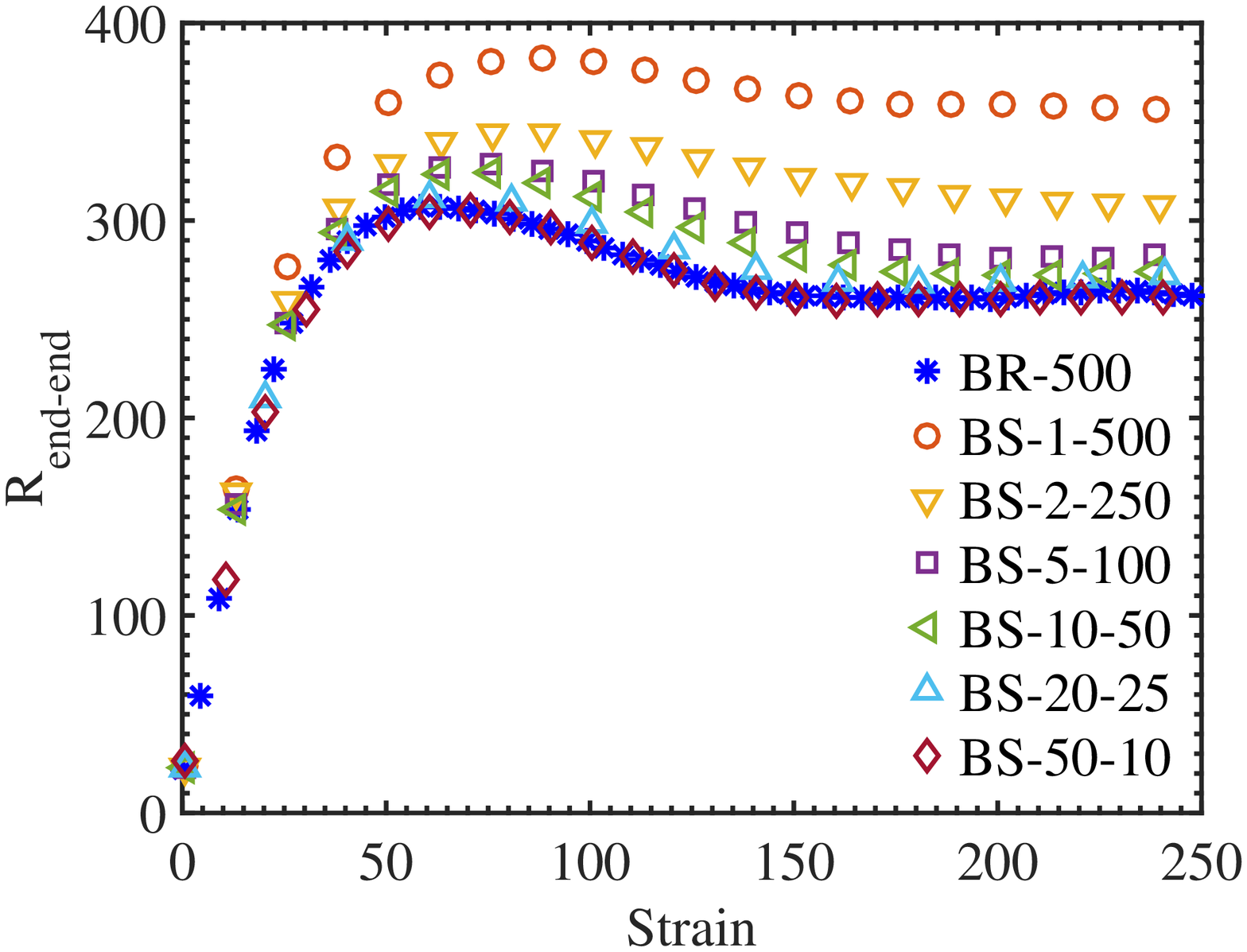}
	\put (-270,210){\LARGE($a$)}
	
\includegraphics[width=0.7\textwidth]{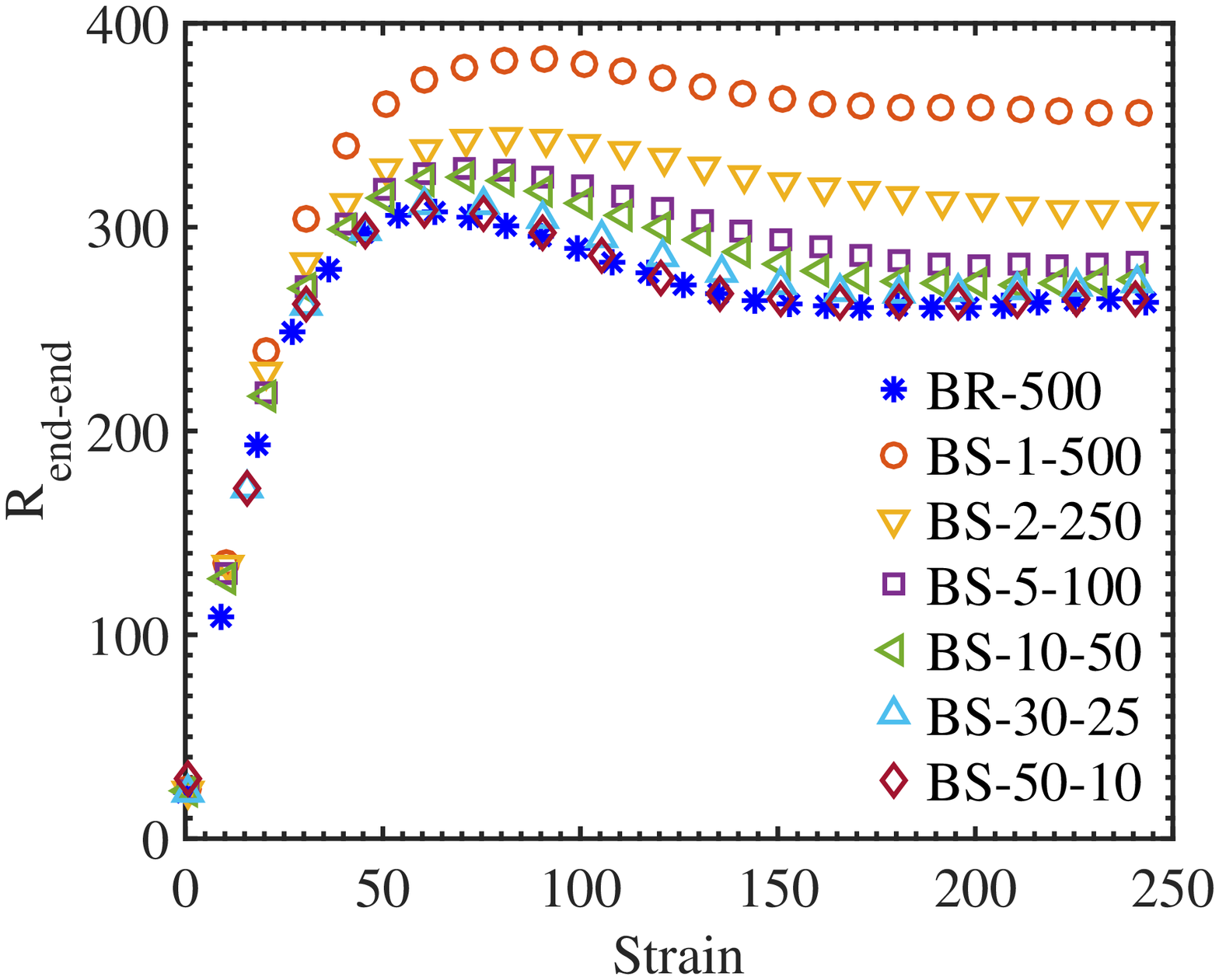}
	\put (-270,210){\LARGE($b$)}
	
	\caption {Variation of $R_{end-end}$ of a polymer chain with strain in shear flow for Wi = 300. The legends have the same meaning as in Fig. \ref{fig:1}. Results shown here are averaged over 2000 and 3000 cases for bead-rod and bead-spring model, respectively. Comparisons are shown for spring laws using the ($a$) Cohen-Padé approximation and ($b$) Underhill-Doyle spring law.  }
		\label{fig:13}
\end{figure}
\begin{figure}[hbt!]
\centering
	\includegraphics[width=0.7\textwidth]{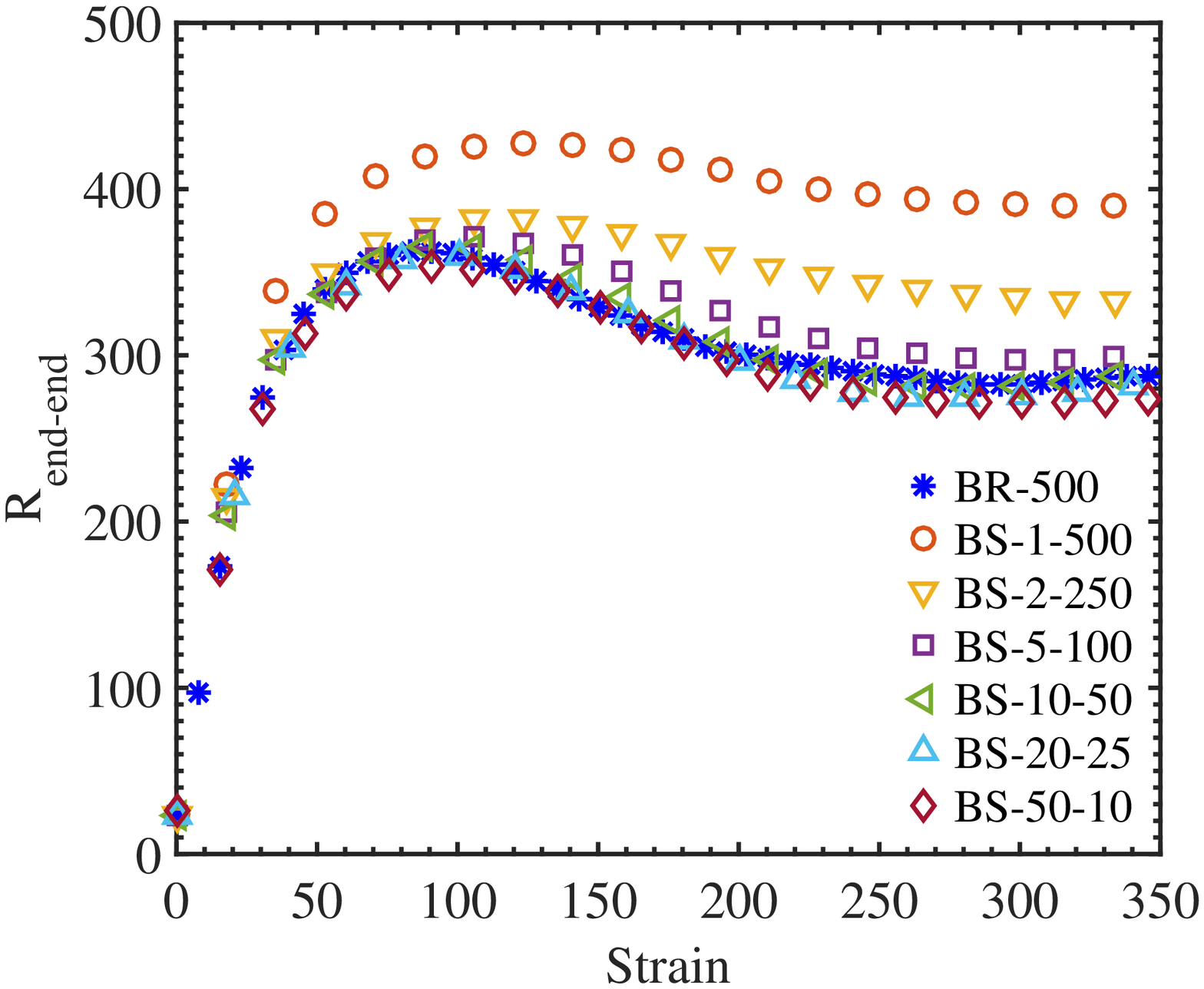}
	\put (-270,210){\LARGE($a$)}
	
	\includegraphics[width=0.7\textwidth]{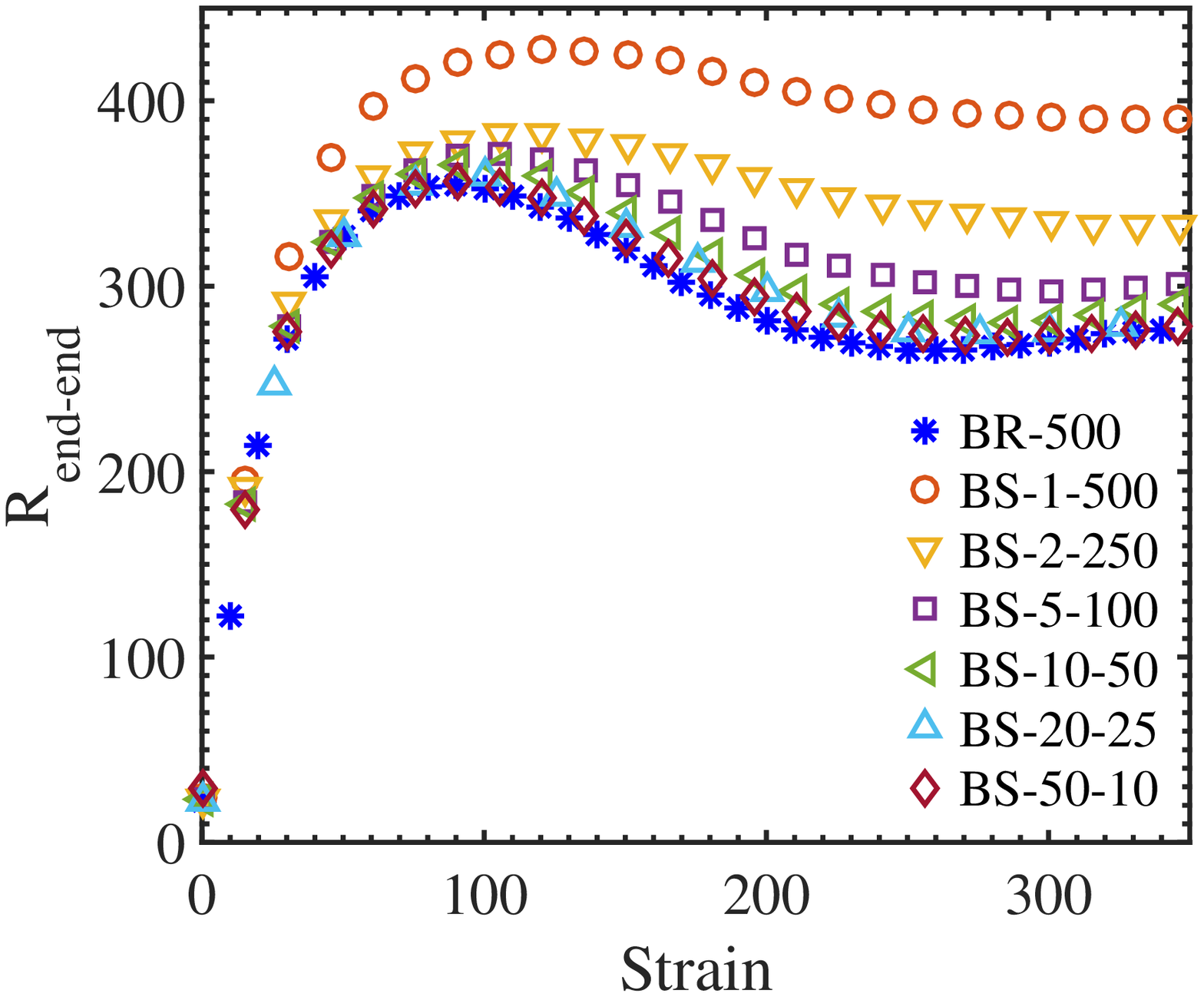}
	\put (-270,210){\LARGE($b$)}
	
	\caption {Variation of $R_{end-end}$ of a polymer chain with strain in shear flow for Wi = 1000. The legends have the same meaning as in Fig. \ref{fig:1}. Results shown here are averaged over 2000 and 3000 cases for bead-rod and bead-spring model, respectively. Comparisons are shown for spring laws using the ($a$) Cohen-Padé approximation and ($b$) Underhill-Doyle spring law  .}
		\label{fig:14}
\end{figure}

\begin{figure}[hbt!]
\centering
	\includegraphics[width=0.7\textwidth]{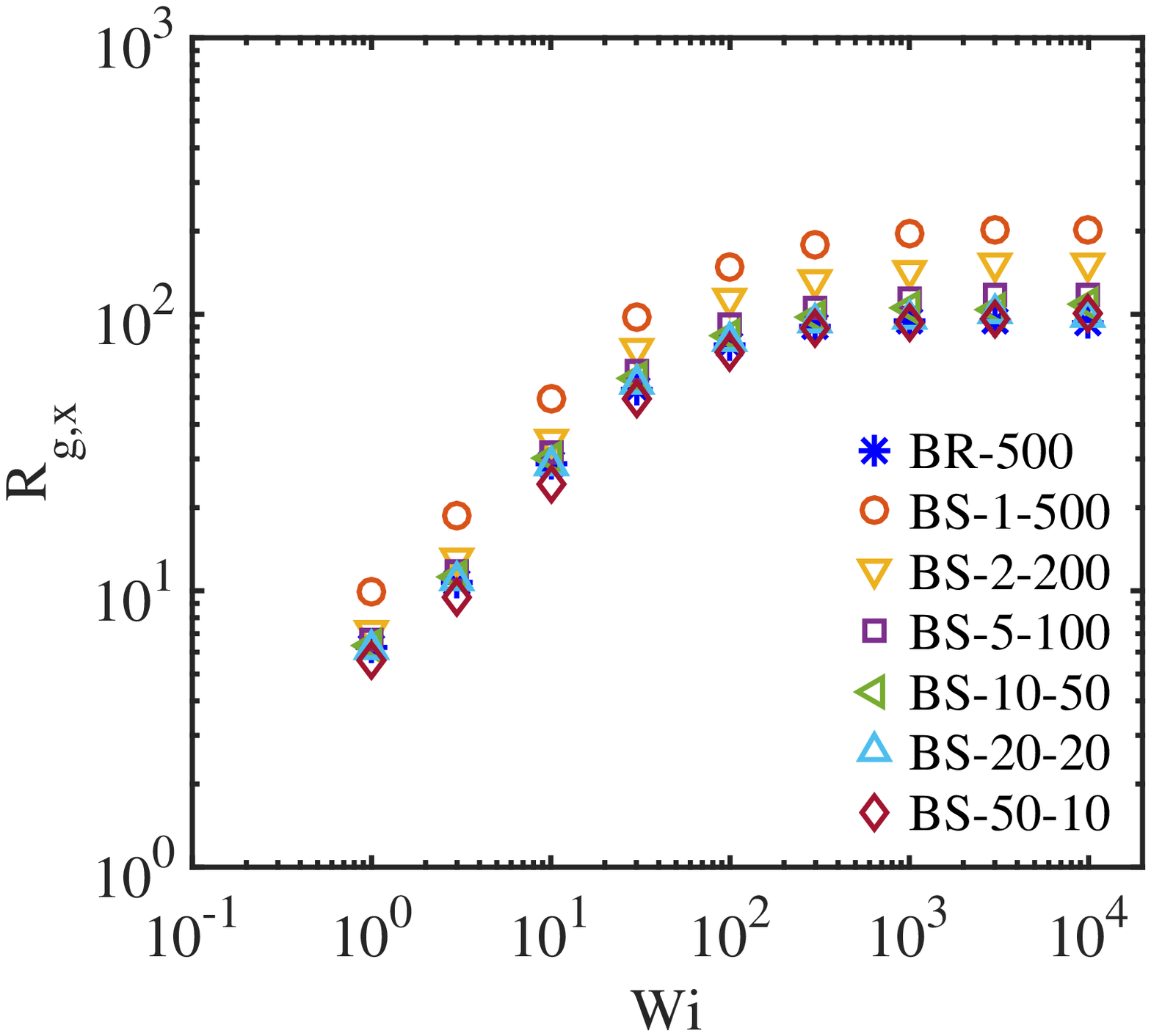}
	\put (-270,210){\LARGE($a$)}
	
	\includegraphics[width=0.7\textwidth]{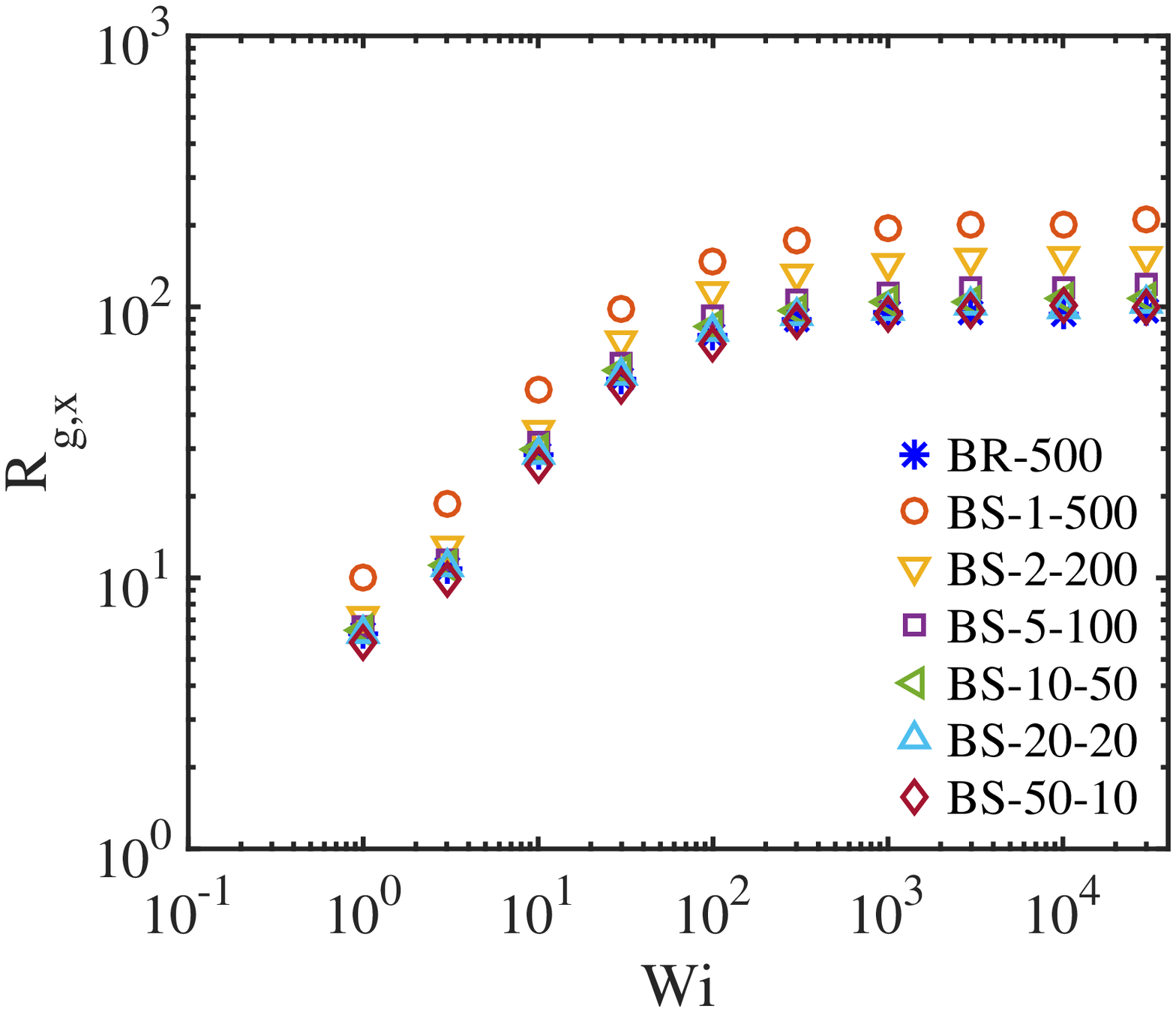}
	\put (-270,210){\LARGE($b$)}
	
	\caption {The variation of the final steady state value of $R_{g,x}$ with the flow rate (measured by Wi) in shear flow. The legends have similar meaning as in Fig. \ref{fig:1}. Comparisons are shown for springs using the ($a$) Cohen-Padé approximation and ($b$) Underhill-Doyle spring law. }
		\label{fig:15}
\end{figure}

\begin{figure}[hbt!]
\centering
	\includegraphics[width=0.7\textwidth]{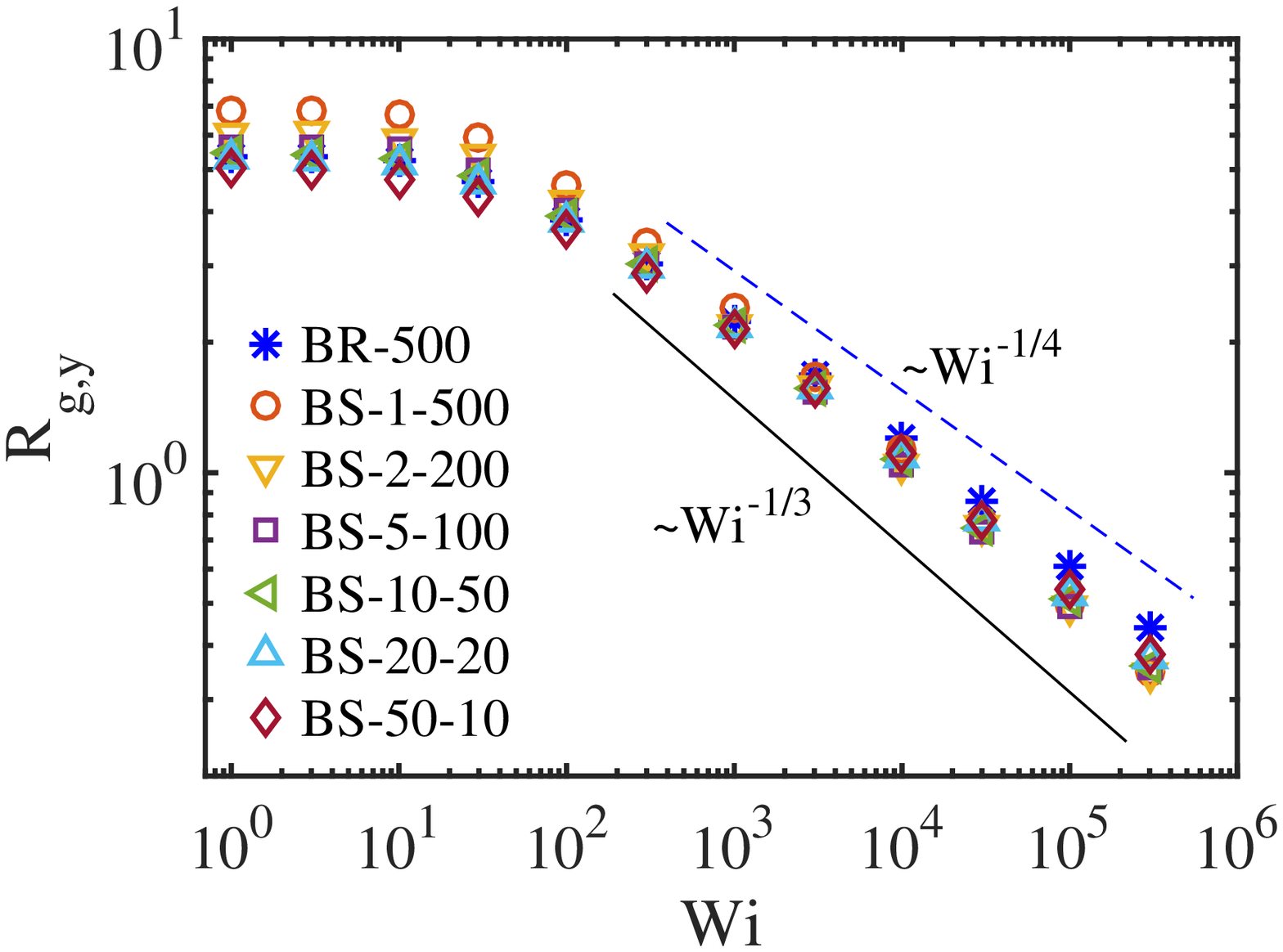}
	\put (-170,200){\LARGE($a$)}
	
	\includegraphics[width=0.7\textwidth]{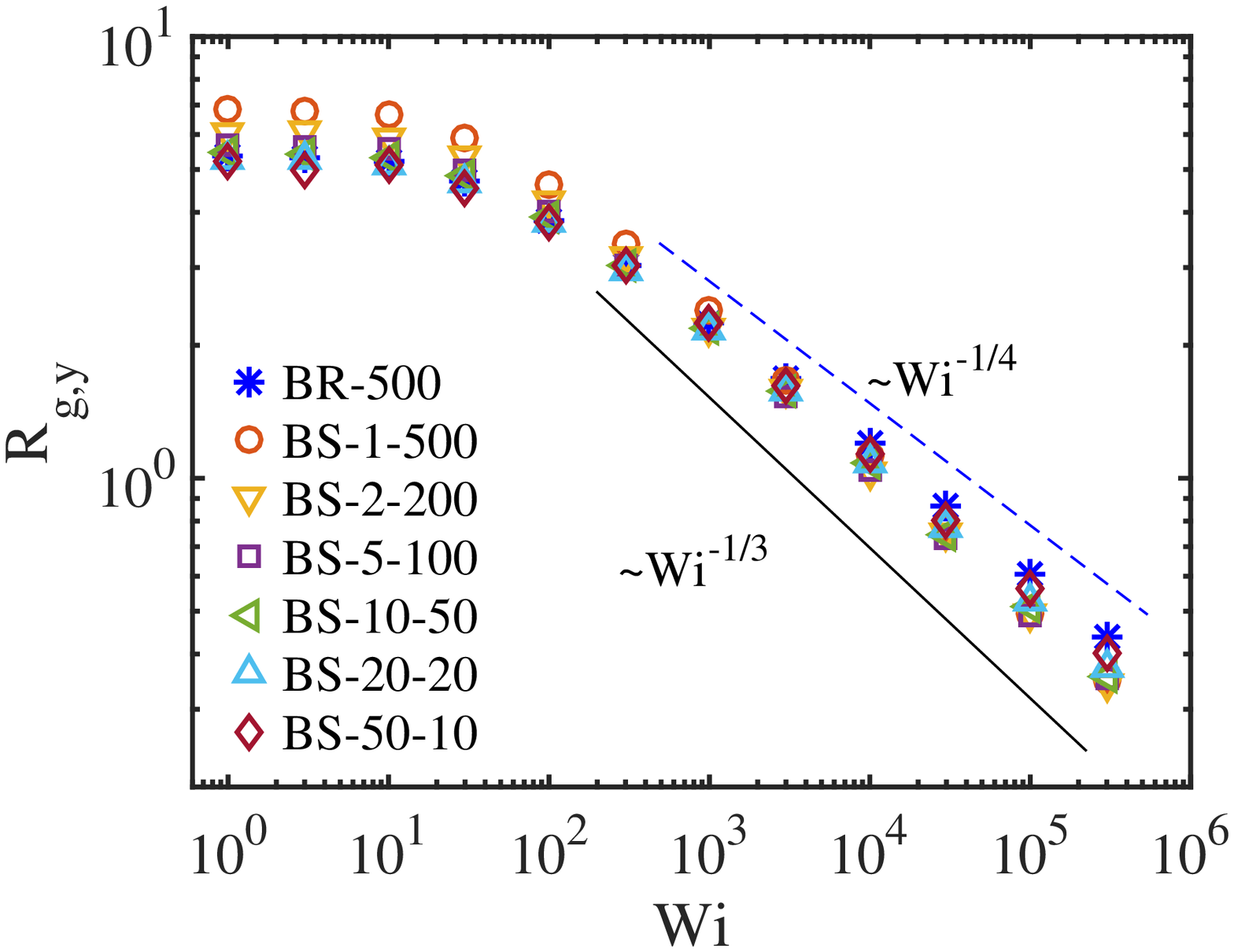}
	\put (-170,210){\LARGE($b$)}
		\caption {The variation of the final steady state value of $R_{g,y}$ with the flow rate (measured by Wi) in shear flow. The legends have similar meaning as in Fig. \ref{fig:1}. Comparisons are shown for springs using the ($a$) Cohen-Padé approximation and ($b$) Underhill-Doyle spring law. The dashed line shows the scaling for bead-KS and the solid line for bead-spring model, respectively.}
			\label{fig:16}
\end{figure}


\begin{figure}[hbt!]
\centering
	\includegraphics[width=0.7\textwidth]{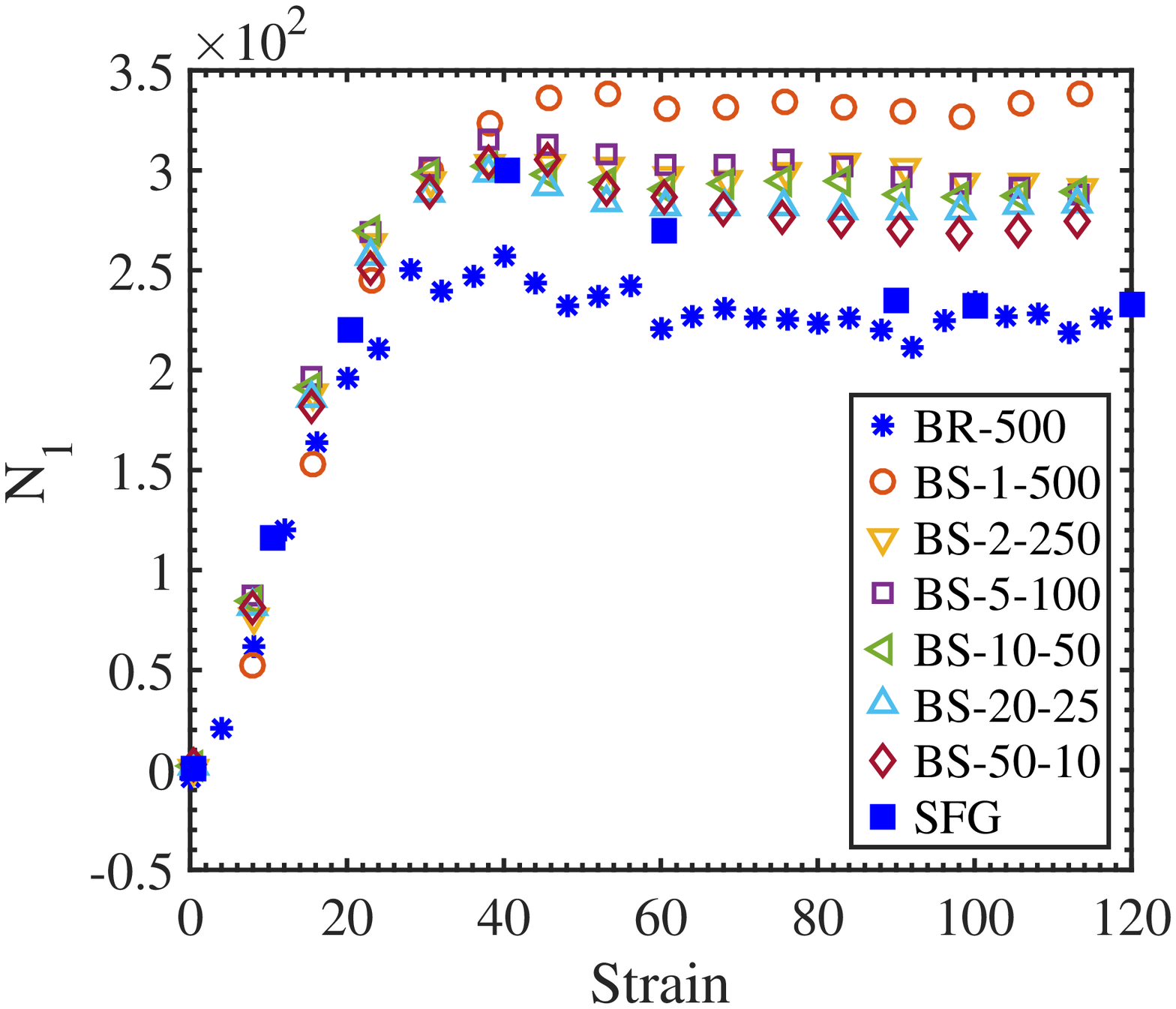}
	\put (-270,230){\LARGE($a$)}
	
	\includegraphics[width=0.7\textwidth]{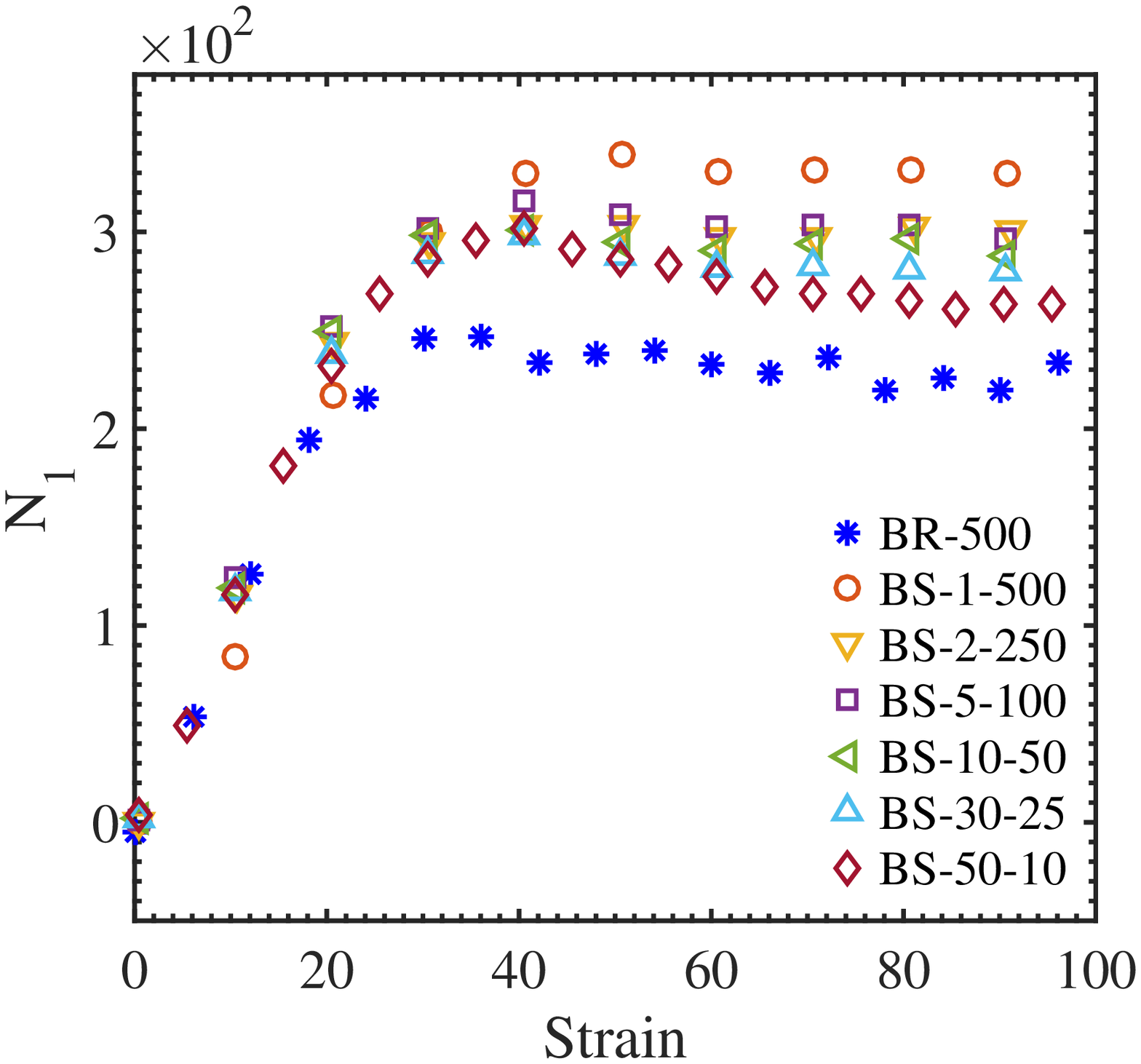}
	\put (-270,230){\LARGE($b$)}
	
	\caption {Variation of $N_1$ of a polymer chain with strain in shear flow for Wi = 30. The legends have the same meaning as in Fig. \ref{fig:1}. Results shown here are averaged over 2000 and 3000 cases for bead-rod and bead-spring model, respectively. Comparisons are shown for spring laws using the ($a$) Cohen-Padé approximation and ($b$) Underhill-Doyle spring law }
		\label{fig:17}
\end{figure}

\begin{figure}[hbt!]
\centering
	\includegraphics[width=0.7\textwidth]{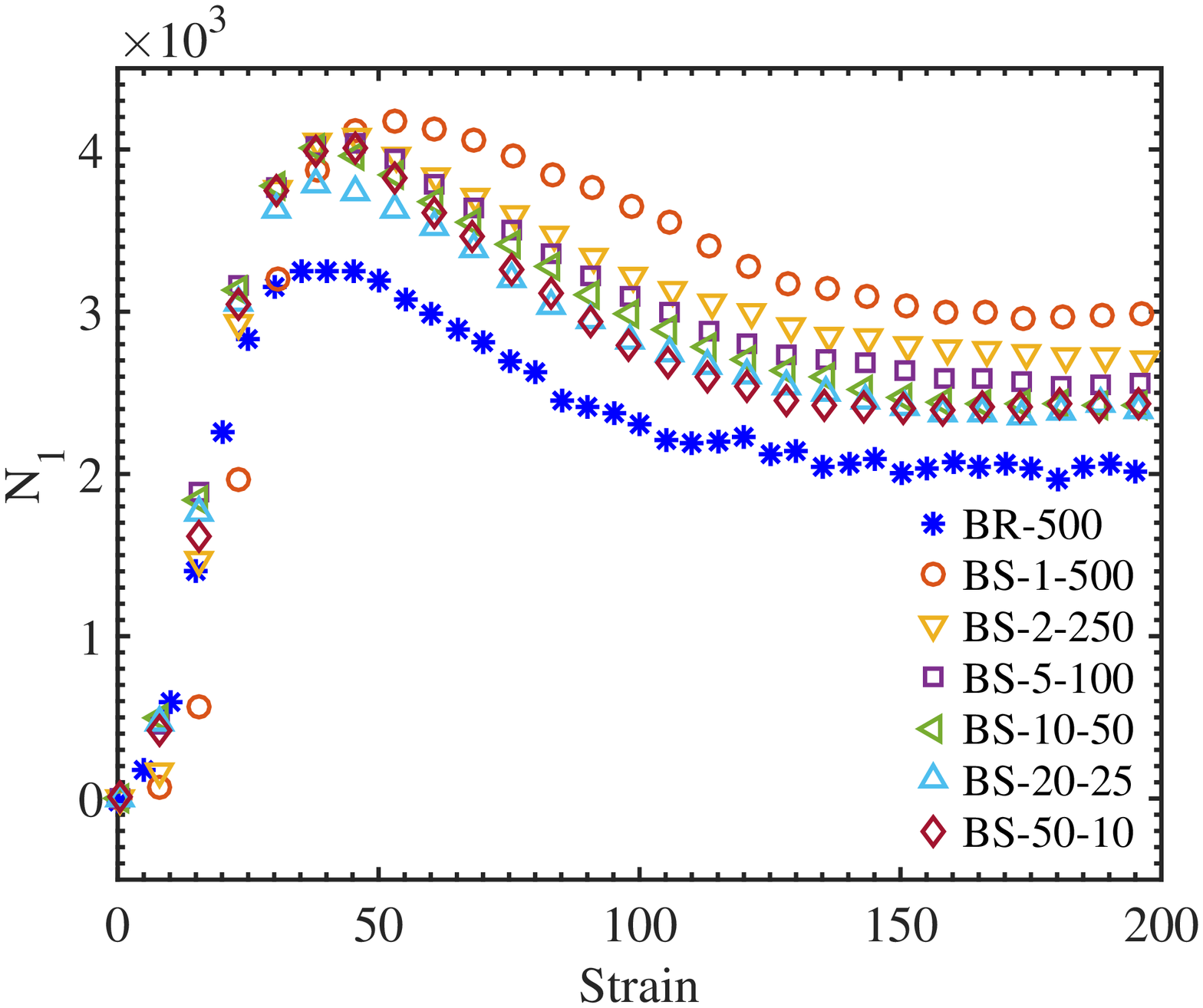}
	\put (-180,210){\LARGE($a$)}
	
	\includegraphics[width=0.7\textwidth]{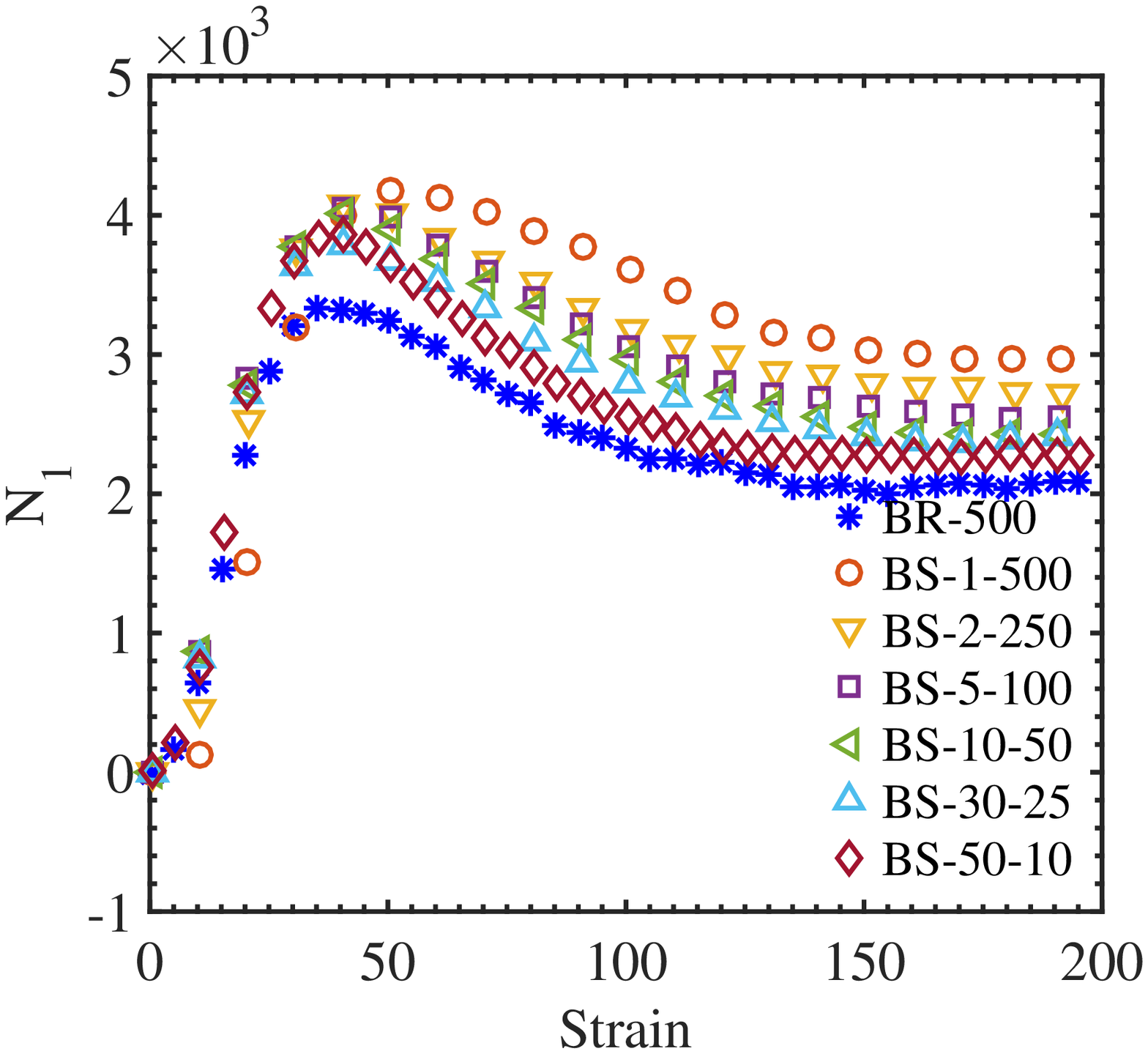}
	\put (-180,230){\LARGE($b$)}
	
	\caption {Variation of $N_1$ of a polymer chain with strain in shear flow for Wi = 300. The legends have the same meaning as in Fig. \ref{fig:1}. Results shown here are averaged over 2000 and 3000 cases for bead-rod and bead-spring model, respectively. Comparisons are shown for spring laws using the ($a$) Cohen-Padé approximation and ($b$) Underhill-Doyle spring law.}
		\label{fig:18}
\end{figure}

\begin{figure}[hbt!]
\centering
	\includegraphics[width=0.7\textwidth]{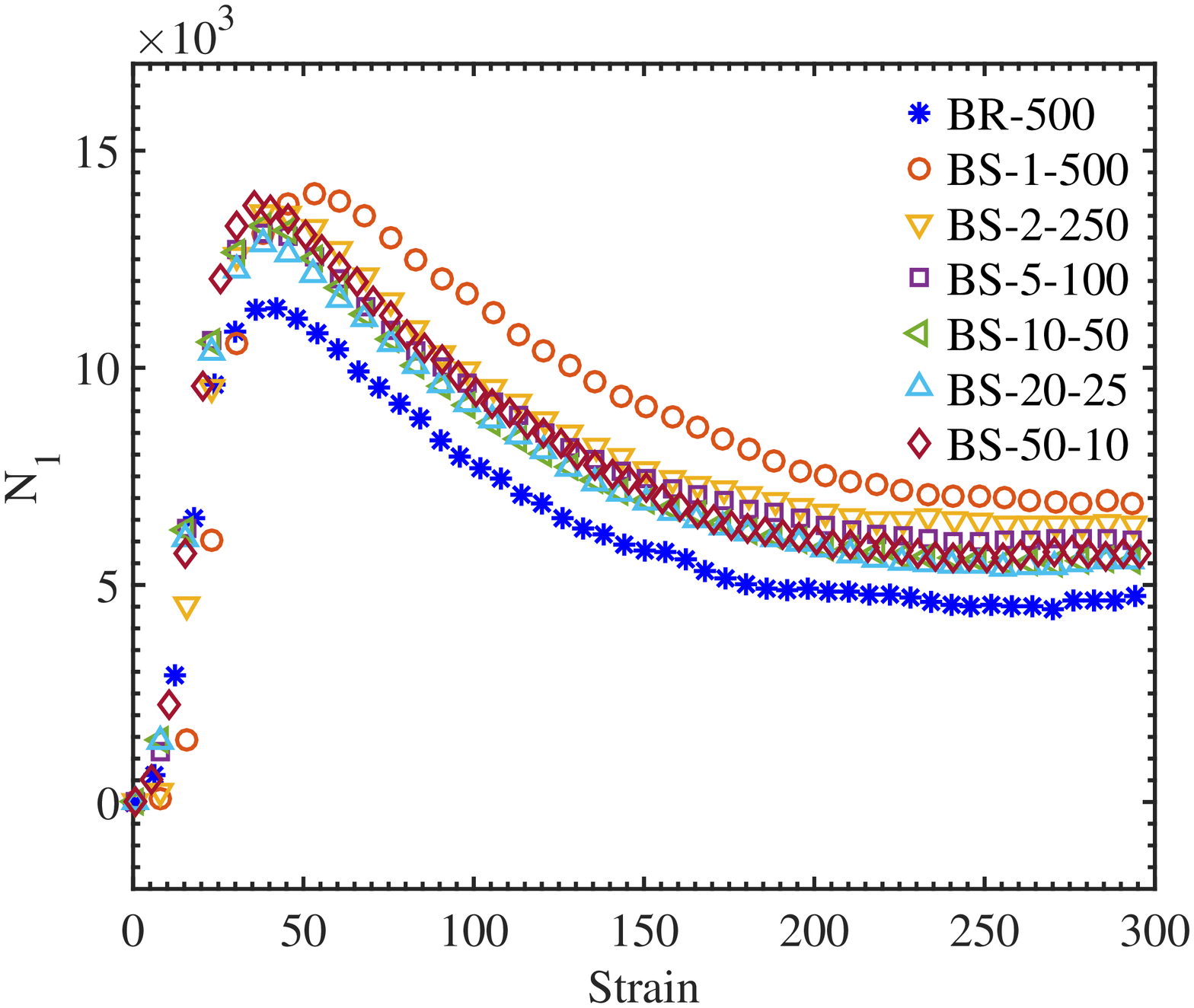}
	\put (-180,210){\LARGE($a$)}
	
	\includegraphics[width=0.7\textwidth]{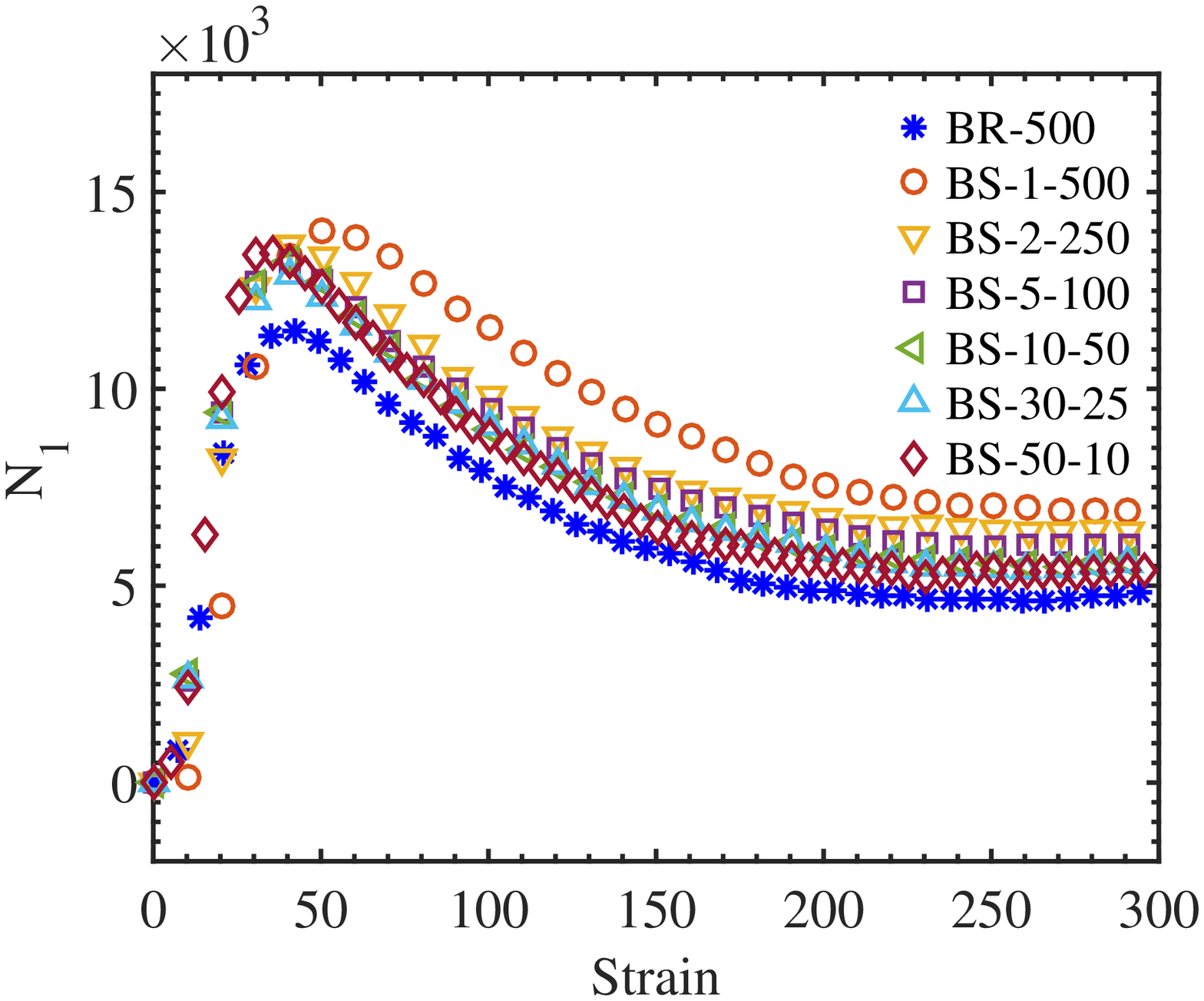}
	\put (-180,210){\LARGE($b$)}
	\caption {Variation of $N_1$ of a polymer chain with strain in shear flow for Wi = 1000. The legends have the same meaning as in Fig. \ref{fig:1}. Results shown here are averaged over 2000 and 3000 cases for bead-rod and bead-spring model, respectively. Comparisons are shown for spring laws using the ($a$) Cohen-Padé approximation and ($b$) Underhill-Doyle spring law.}
		\label{fig:19}
\end{figure}

\begin{figure}[hbt!]
\centering
	\includegraphics[width=0.7\textwidth]{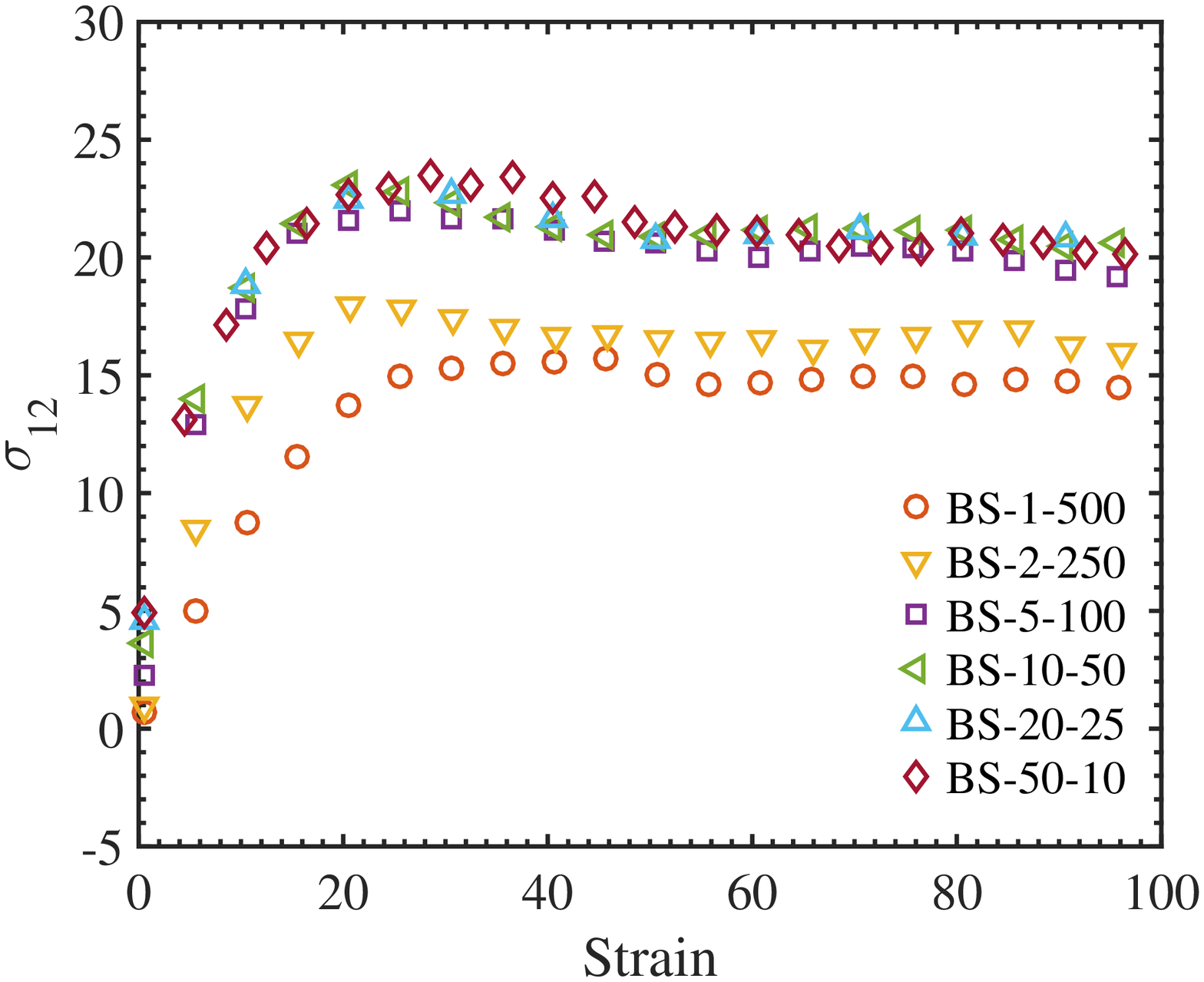}
	\put (-180,220){\LARGE($a$)}
	
	\includegraphics[width=0.7\textwidth]{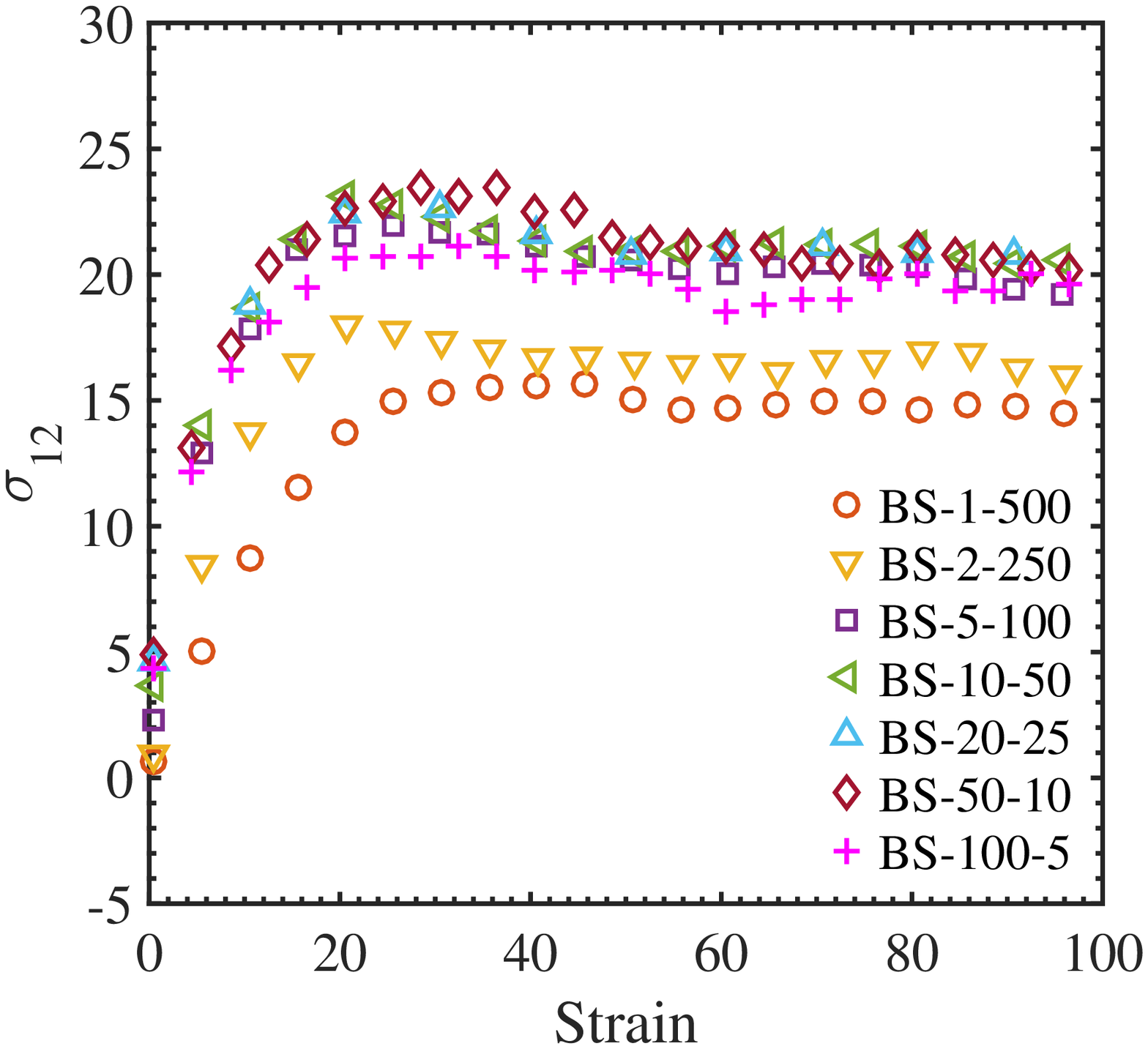}
	\put (-180,240){\LARGE($b$)}
	
	\caption {Variation of shear stress $\sigma_{12}$ of a polymer chain with strain in shear flow for Wi = 30. The legends have the same meaning as in Fig. \ref{fig:1}. Results shown here are averaged over 2000 and 3000 cases for bead-rod and bead-spring model, respectively. Comparisons are shown for spring laws using the ($a$) Cohen-Padé approximation and ($b$) Underhill-Doyle spring law.}
		\label{fig:20}
\end{figure}

\begin{figure}[hbt!]
\centering
	\includegraphics[width=0.7\textwidth]{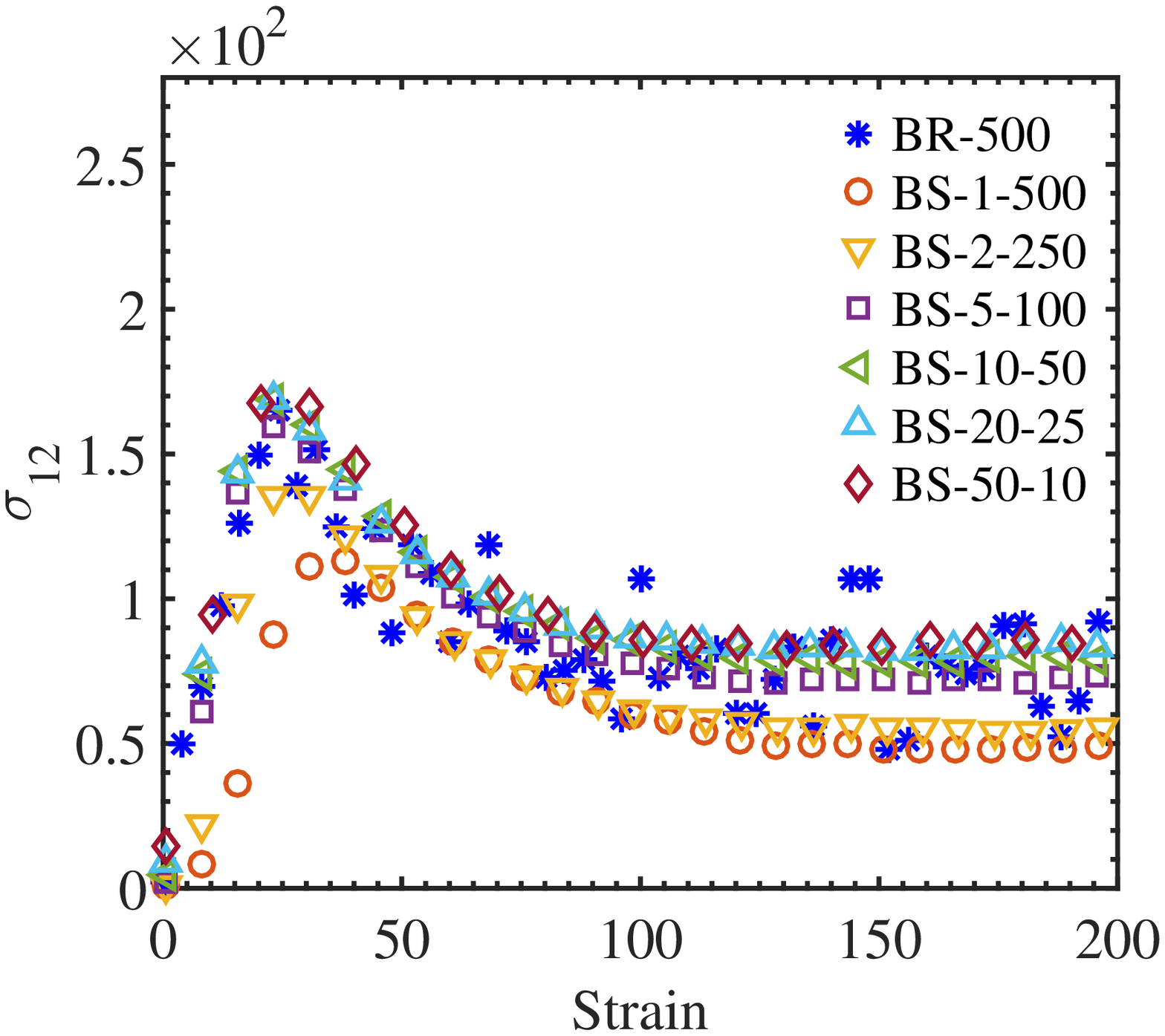}
	\put (-180,220){\LARGE($a$)}
	
	\includegraphics[width=0.7\textwidth]{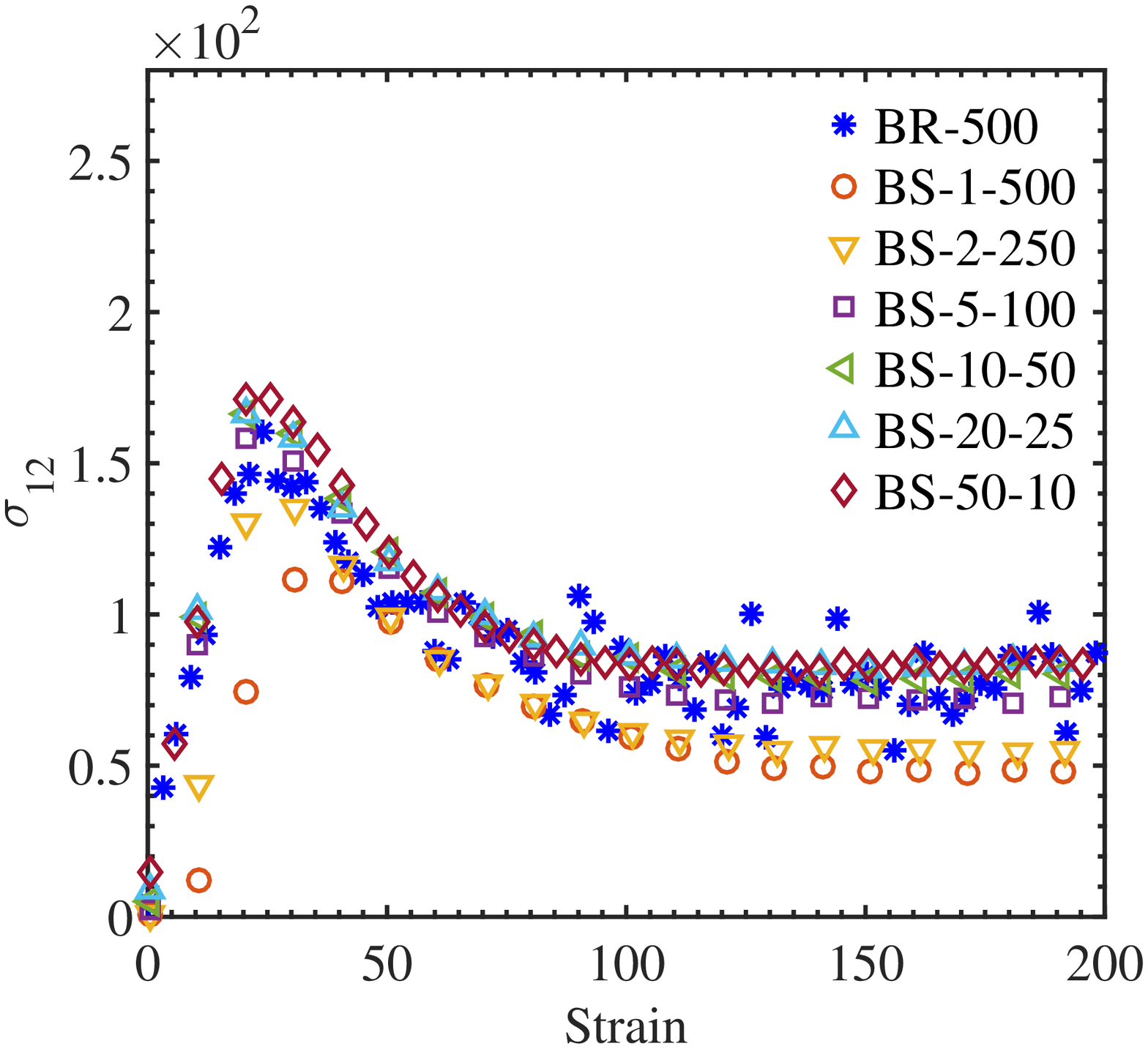}
	\put (-180,220){\LARGE($b$)}
	
	\caption {Variation of shear stress $\sigma_{12}$ of a polymer chain with strain in shear flow for Wi = 300. The legends have the same meaning as in Fig. \ref{fig:1}. Results shown here are averaged over 2000 and 3000 cases for bead-rod and bead-spring model, respectively. Comparisons are shown for spring laws using the ($a$) Cohen-Padé approximation and ($b$) Underhill-Doyle spring law.}
		\label{fig:21}
\end{figure}
\begin{figure}[hbt!]
\centering
	\includegraphics[width=0.7\textwidth]{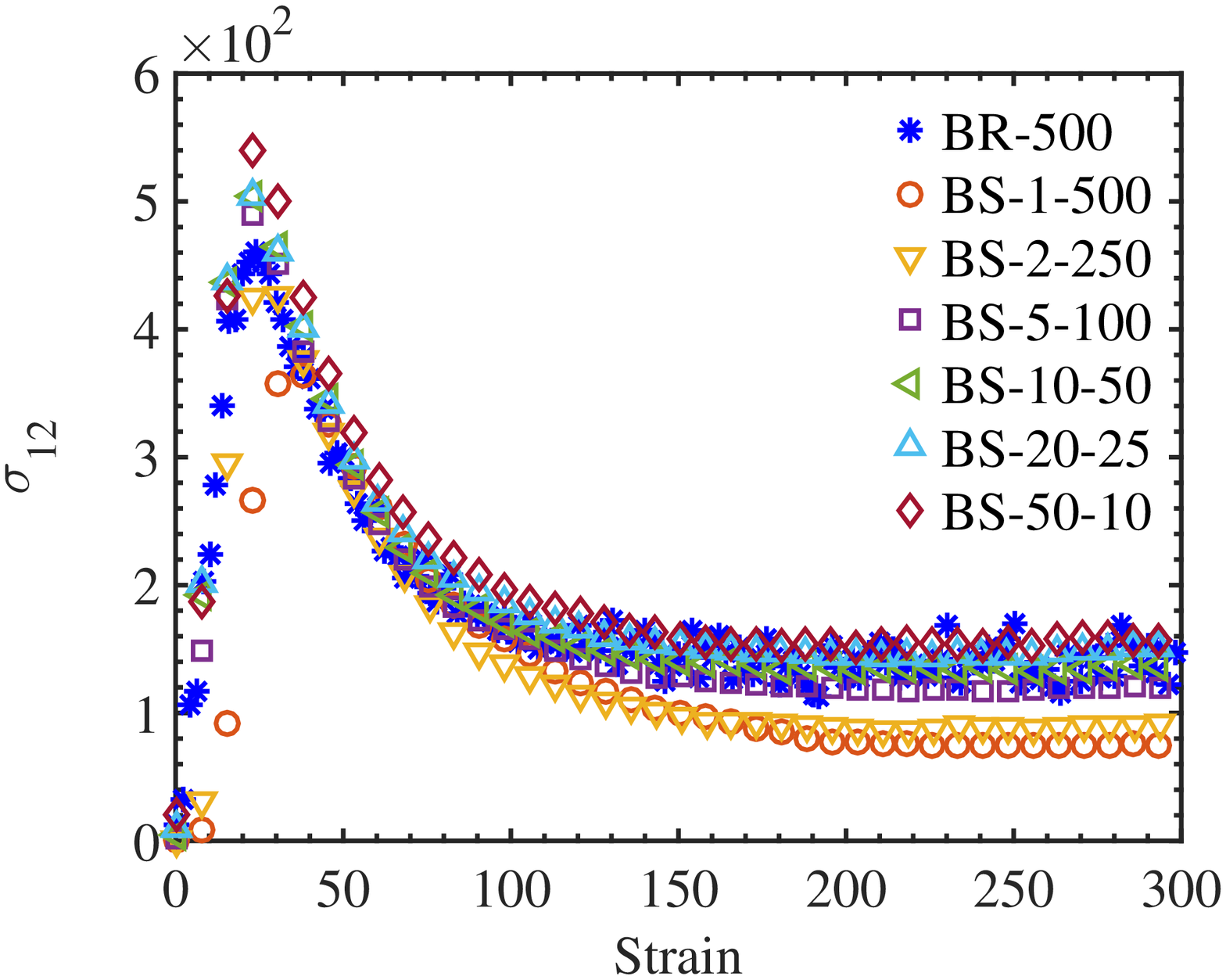}
	\put (-180,210){\LARGE($a$)}
	
	\includegraphics[width=0.7\textwidth]{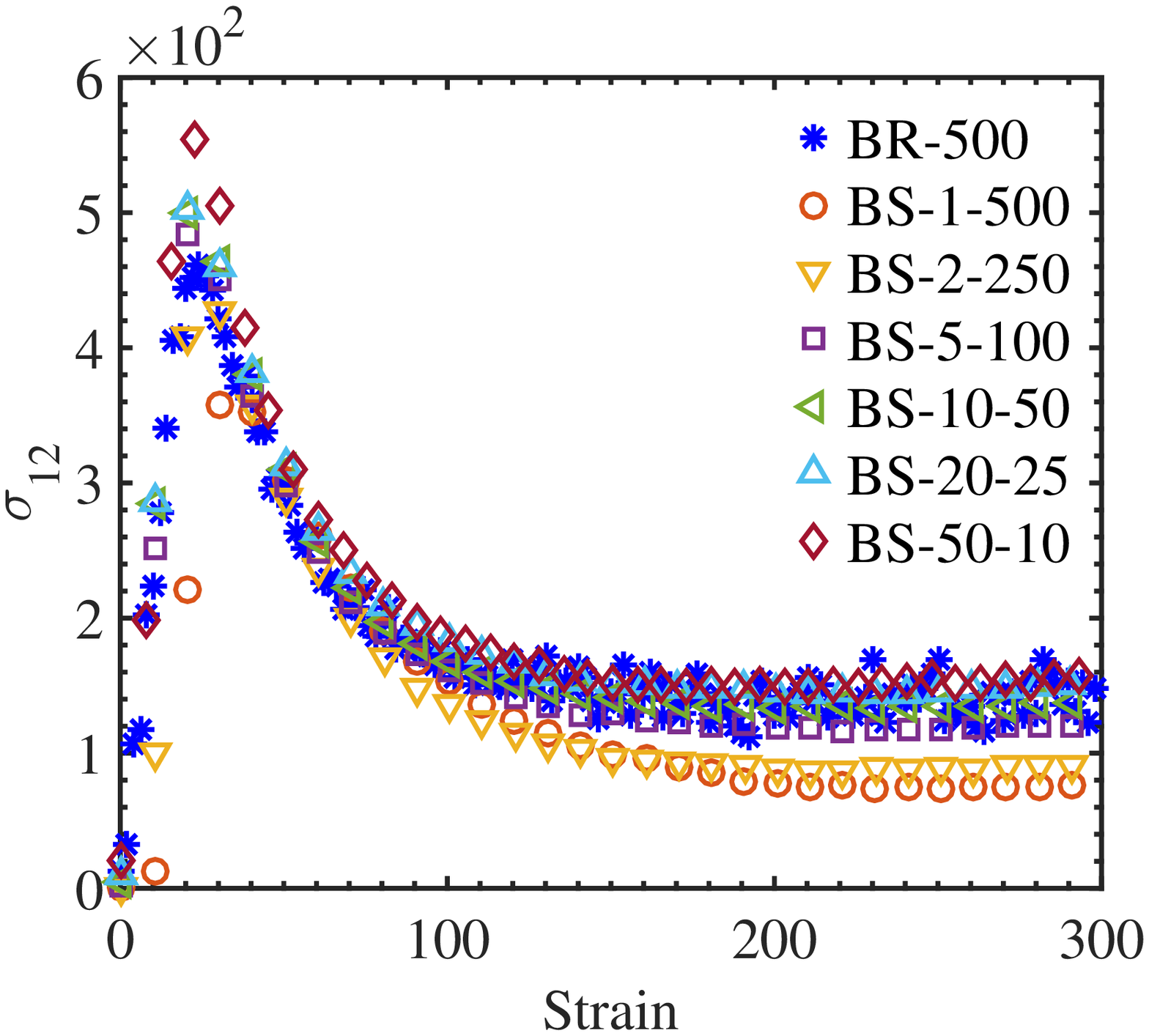}
	\put (-180,220){\LARGE($b$)}
	
	\caption {Variation of shear stress $\sigma_{12}$ of a polymer chain with strain in shear flow for Wi = 1000. The legends have the same meaning as in Fig. \ref{fig:1}. Results shown here are averaged over 2000 and 3000 cases for bead-rod and bead-spring model, respectively. Comparisons are shown for spring laws using the ($a$) Cohen-Padé approximation and ($b$) Underhill-Doyle spring law.}
		\label{fig:22}
\end{figure}


\begin{figure}[hbt!]
\centering
	\includegraphics[width=0.7\textwidth]{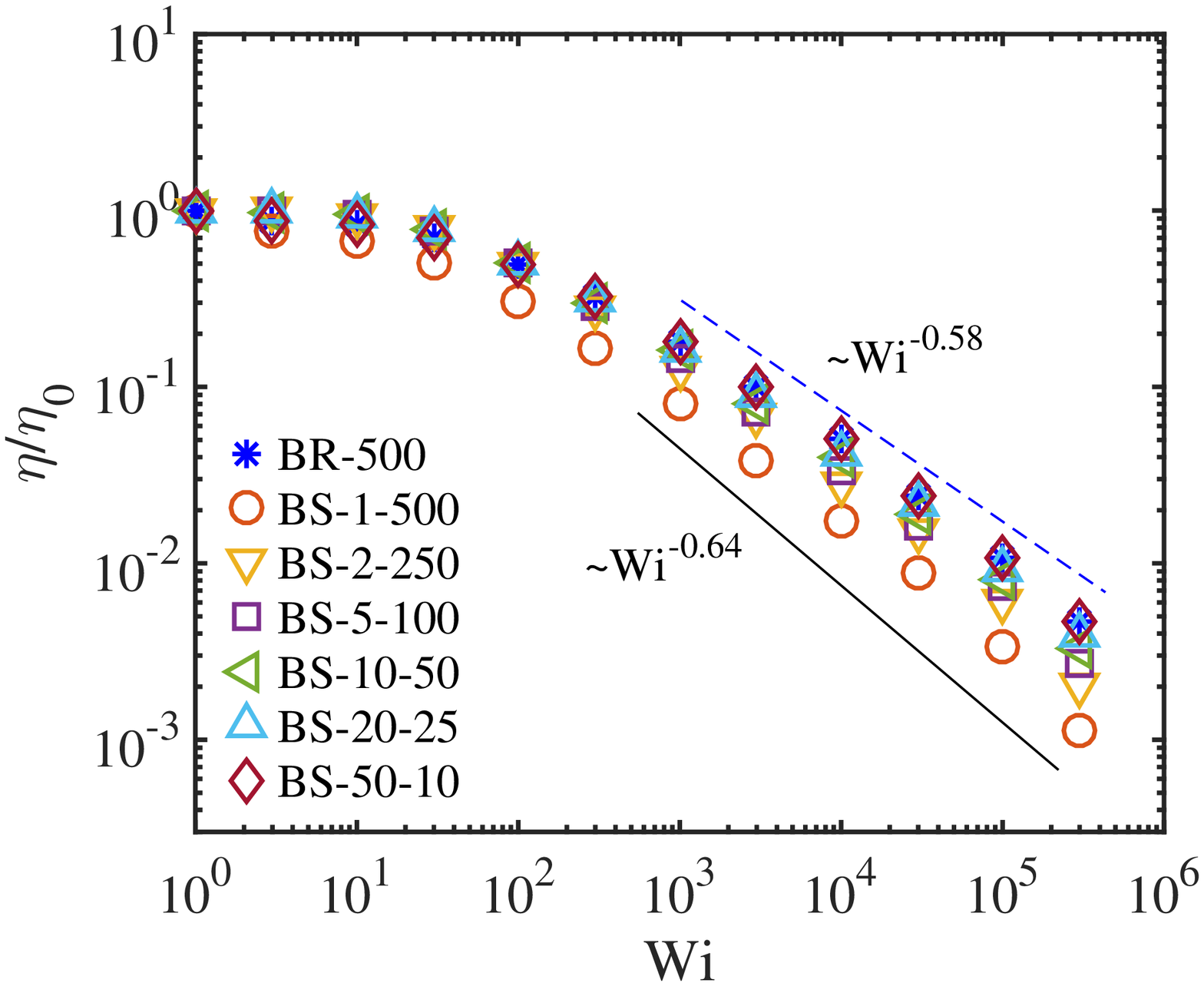}
	\put (-180,220){\LARGE($a$)}
	
	\includegraphics[width=0.7\textwidth]{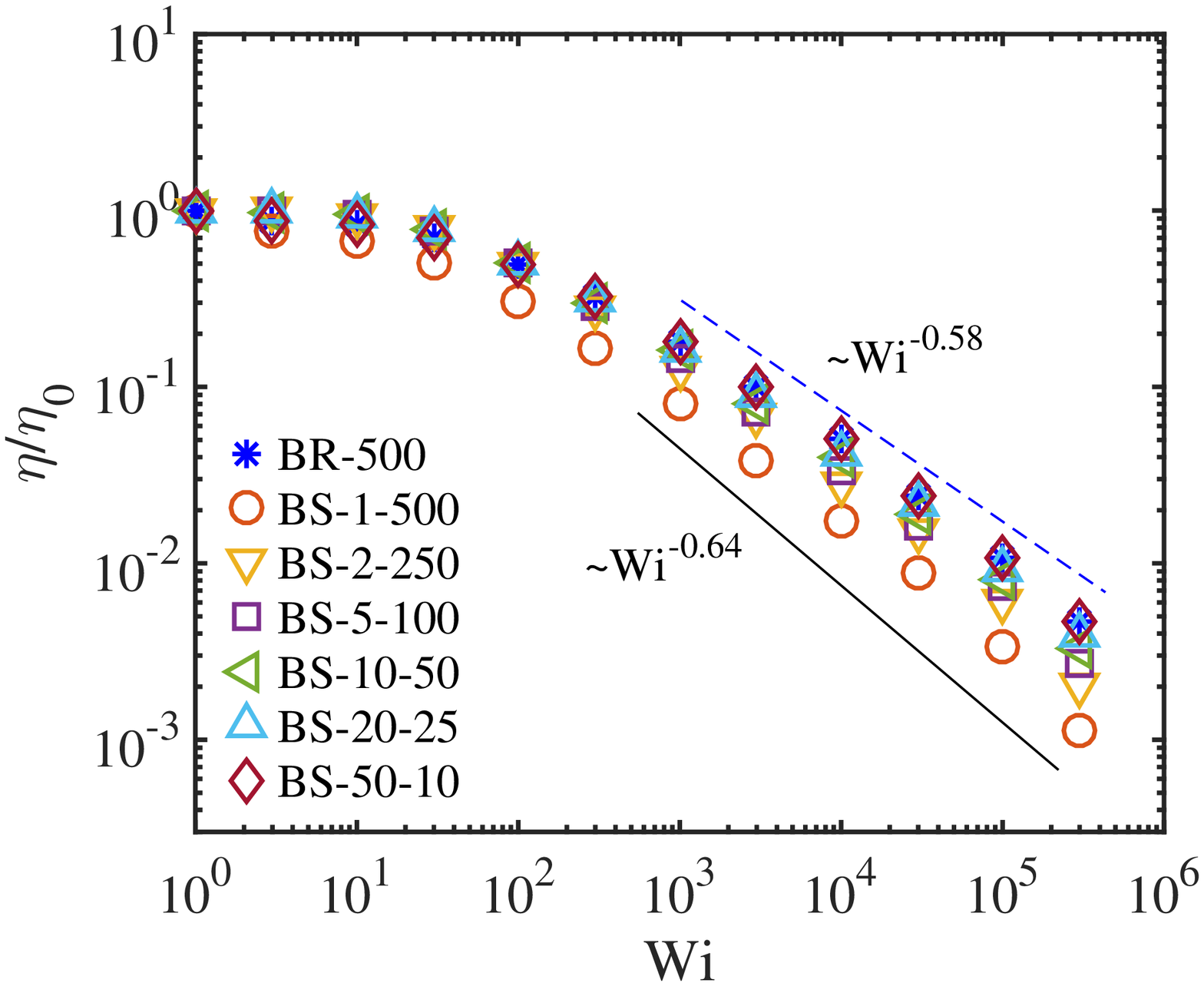}
	\put (-180,220){\LARGE($b$)}
	
	\caption {The variation of the relative steady shear viscosity with the flow rate (measured by Wi) in shear flow. The legends have similar meaning as in Fig. \ref{fig:1}. Comparisions are shown for springs using the ($a$) Cohen-Padé approximation and ($b$) Underhill-Doyle spring law. The dashed line shows the scaling law exponent for bead-rod and the solid line for the dumbbell. The scaling exponent varies from -0.64 to -0.58 with an increase in the number of springs.}
		\label{fig:23}
\end{figure}

Shear flow is really ubiquitous in scientific and industrial applications, since any flow near a boundary can be approximated by a shear flow [17]. When a polymer solution is subjected to a steady shear flow, the chain undergoes a repeated succession of stretching and tumbling events, owing to the presence of equal amounts of extension and rotation in shear flow. Here, we perform simulations of the start-up of shear flow for a wide range of Weissenberg numbers for various levels of chain discretization, similar to those for uniaxial extension.

The ensemble-averaged temporal variation of the chain size (measure by the end-to-end distance) is shown in Figs. \ref{fig:12}-\ref{fig:14}. As before, parts ($a$) and ($b$) provide results for the two different spring laws. For these simulations, we have considered 2000 and 3000 cases for the bead-KS and bead-spring models, respectively, to calculate the average. For start-up shear flows, for the selected flow rates in this study, an overshoot is observed before attaining the final steady state. This overshoot is significant for strong shear flows (Figs. \ref{fig:13} and \ref{fig:14}) and negligible for weaker flows (Fig. \ref{fig:12}). Some of the trends observed here are similar to those in extension. The differences between the bead-KS and bead-spring models reduce systematically as the number of springs (in the bead-spring model) is increased, with a near-perfect agreement obtained for 50 springs. The two spring laws agree well with each other for the same discretization level. Similar behaviour is observed for the variations of the steady state chain radii of gyration (flow and gradient direction) with flow rate (given by Wi). The chain stretch ($R_{g,x}$ – Fig. \ref{fig:15}) becomes consistent with the bead-KS model with progressively decreasing $\nu$ (i.e. increasing of number of springs). However, a difference in the scaling law exponents is observed between the bead-KS and bead-spring models for the coil thickness ($R_{g,y}$– Fig. \ref{fig:16}), which is consistent with earlier studies \cite{dalal2012multiple}.

The temporal behaviour of the first normal stress difference is shown in Figs.\ref{fig:17}-\ref{fig:19}, for different shear rates and chain discretization levels. Interestingly, the predictions are considerably different, especially with respect to the position and magnitude of the overshoot.  Similar to the observations for extensional flow, the bead-spring models are consistent with one another for the lower values of $\nu$ (higher number of springs) considered here. However, the bead-KS predictions are significantly different from this “converged” bead-KS predictions, with the difference getting systematically reduced as Wi is increased. However, even at Wi = 1000, a quantitative difference exists especially in the magnitude of the overshoot. Overall, the two spring laws agree well with one another, as with all earlier results.

Figs. \ref{fig:20}-\ref{fig:22} show the transient behaviour of the shear stress with strain for Wi = 30, 300 and 1000, respectively. Note that the bead-KS predictions are not included for Wi = 30 owing to large fluctuations. Interestingly, the overshoot predicted by the bead-KS model at Wi = 1000 is closer to the bead-spring model with few springs. However, the final steady state becomes more consistent with an increasing number of springs. Results obtained with the two spring laws for the coarse-grained bead-spring model are quite similar.

The normalized shear viscosity (normalized by the zero shear viscosity) is presented in Fig. \ref{fig:23}, for different chain discretization levels. As for other rheological properties, the behaviour becomes consistent with an increasing number of springs. The normalized viscosity scales approximately as -0.58 for the bead-KS chain. For the bead-spring model, the scaling varies from -0.64 to -0.58 as the number of springs are increased.

   \section{Summary and Conclusions}

In this study, we investigated the differences in the predictions of chain dynamics, under imposed flow fields, due to chain discretization levels. In particular, Brownian dynamics simulations of a polymer chain are performed with uniaxial extension and steady shear flows. The differing levels of chain discretization are obtained by using the bead-rod and bead-spring models, where each unit represents one Kuhn step and multiple Kuhn steps, respectively. The spring laws used in this work are the Cohen-Padé approximation and the Underhill-Doyle model \cite{cohen1991pade,underhill2004coarse,dalal2012multiple}. This detailed study is inspired by the differences observed in the configurational properties at steady state in shear flow in earlier studies \cite{dalal2012multiple,dalal2014effects}. The key observations are summarized below:

\begin{itemize}
	\item{The limiting values of $\nu$ of a bead-spring model, below which even the equilibrium chain properties are not predicted accurately, are ascertained as $\nu=10$ and $\nu=5$ for the Cohen-Padé approximation and the Underhill-Doyle spring law, respectively. Moreover, for flows with high Wi, inconsistencies were observed for $\nu=5$ with the Underhill-Doyle sping law.}
	
	\item{For uniaxial extension flow, the temporal variation of the chain sizes predicted by the bead-spring model agrees well with that of the bead-rod model as the chain resolution increases (i.e. $\nu$ decreases) and flow rate (i.e. Wi) increases. For large enough springs (50 – 100), the final steady state agrees well for all Wi.}
	
	\item{The first normal stress difference in uniaxial extension shows considerable differences between the bead-rod and the bead-spring models, which reduces as Wi is increased. Significantly, the bead-spring models with larger springs “converge” to a $N_1$ that is higher than that predicted by the bead-rod model. This is attributed to the much lower stretch experienced by the terminal segment in the bead-rod model, relative to the bead-spring models. Somewhat interestingly, the steady states obtained with a bead-spring model of 100 springs (Underhill-Doyle spring law) agrees well with that of the bead-rod model for all Wi considered, although the agreement is poor at intermediate strains leading to the steady state.}
	
	\item{The stress-conformation hysteresis loops agree well in general, with progressive increase in consistency as the number of springs are increased. However, the predictions of the chain with 50 springs is worse than the other representations, which may be due to the failure of the spring laws for very short springs.}

	\item{In steady shear, the temporal evolution of the chain size from the bead-spring models agrees well with the bead-rod model as the number of springs and Wi are increased.} 
	
	\item{Similar to uniaxial extension, the bead-spring models “converge” to a value of $N_1$ that is higher than that predicted by the bead-rod model. The difference decreases at larger Wi.}
	
	\item{The shear stress shows an overshoot followed by a steady state for the Wi values considered. The bead-spring models become consistent with the bead-rod model as $\nu$ is decreased and Wi is increased. However, differences in the overshoot exist between the bead-rod model and the bead-spring model with 50 springs.}
	
	\item{At the steady state, the coil thickness scales differently for the bead-rod (-1/4) and the bead-spring (-1/3) models with Wi, consistent with earlier studies \cite{dalal2012multiple}. For the shear viscosity, the exponent varies from -0.58 to -0.64 as $\nu$ decreases.} 

	\item{ The SFG technique provides predictions for the stress that agree reasonably well for the final steady state in extensional and shear flows. However, similar to other bead-spring representations, the transient startup behavior is significantly different than the bead-KS results. }

\end{itemize}

Overall, the bead-spring model predictions for both configurational and rheological properties are significantly different from those of the bead-rod model for low-to-intermediate number of springs. While the differences progressively reduce as the chain resolution is increased (i.e. number of springs are increased), the predictions are not identical even at the lowest possible $\nu$ before the spring laws break down. Consequently, whenever a bead-spring model is used for any application involving flow, care has to be taken to select the appropriate number of springs. We hope that the results of this study helps in this regard, by enabling the user to choose between various models for a given application by balancing the computational cost (adjusting $\nu$) and loss of accuracy.

Our results in this article clearly indicate that the bead-KS model should be used for any application, if computational resources permit. Regarding the usage of bead-spring models, we would like to conclude the following, based on our observations noted above:
\begin{enumerate}
    \item For any application, if bead-spring models need to be used with no constraint on the number of springs, the maximum possible number of springs should be used, such that $\nu>10$  for Cohen-Padé approximation and $\nu>5$  for the Underhill-Doyle spring law.
    \item For any application, if one needs to limit the number of springs to be used, a minimum of 10 springs should be used to retain a reasonable degree of accuracy.
    \item Alternatively, the SFG technique can provide a satisfactory prediction, if several bead-spring models can be simulated.
\end{enumerate}

   \section{Acknowledgement}
  The authors would like to thank the generous support from the IIT Kanpur initiation grant and the HPC center at IIT Kanpur.

\bibliographystyle{elsarticle-num} 
\bibliography{JNNFM}

\vskip3pt

\end{document}